\documentclass[nofootinbib,showpacs]{revtex4}
\usepackage{graphicx}% Include figure files
\usepackage{dcolumn}% Align table columns on decimal point
\usepackage{amsmath,amssymb,epsfig}
\usepackage{paralist}
\usepackage{comment}
\usepackage{graphicx}
\usepackage{multirow}
\usepackage{color,soul}

\allowdisplaybreaks

\renewcommand{\vec}[1]{\boldsymbol{\mathrm{#1}}}

\begin{document}

%\preprint{
%%APS/123-QED
%NOT FOR DISTRIBUTION}

\title{New perturbative method for solving \\ 
the gravitational $N$-body problem in the general theory of relativity}

\author{Slava G. Turyshev$^1$ and Viktor T. Toth$^2$
}

\affiliation{\vskip 3pt
$^1$Jet Propulsion Laboratory, California Institute of Technology,\\
4800 Oak Grove Drive, Pasadena, CA 91109-0899, USA
}%

\affiliation{\vskip 3pt
$^2$Ottawa, Ontario K1N 9H5, Canada
}%

\date{\today}% It is always \today, today,

\begin{abstract}

We present a new approach to describe the dynamics of an isolated, gravitationally bound astronomical $N$-body system in the weak field and slow-motion approximation of the general theory of relativity. Celestial bodies are described using an arbitrary energy-momentum tensor and assumed to possess any number of internal multipole moments. The solution of the gravitational field equations in any reference frame
is presented as a sum of three terms:
\begin{inparaenum}[i)]
\item the inertial flat spacetime in that frame,
\item unperturbed solutions for each body in the system that is covariantly transformed to the coordinates of this frame, and
\item the gravitational interaction term.
\end{inparaenum}
We use the harmonic gauge conditions that allow reconstruction of a significant part of the structure of the post-Galilean coordinate transformation functions relating global coordinates of the inertial reference frame to the local coordinates of the non-inertial frame associated with a particular body. The remaining parts of these functions are determined from dynamical conditions, obtained by constructing the relativistic proper reference frame associated with a particular body. In this frame, the effect of external forces acting on the body is balanced by the fictitious frame-reaction force that is needed to keep the body at rest with respect to the frame, conserving its relativistic three-momentum. We find that this is sufficient to determine explicitly all the terms of the coordinate transformation. The same method is then used to develop the inverse transformations. The resulting post-Galilean coordinate transformations have an approximate group structure that extends the Poincar\'e group of global transformations to the case of accelerating observers in a gravitational field of $N$-body system. We present and discuss the structure of the metric tensors corresponding to the reference frames involved, the rules for transforming relativistic gravitational potentials, the coordinate transformations between frames and the resulting relativistic equations of motion.

\end{abstract}

%\keywords{}

\maketitle

\section{Introduction}
\label{sec:introduction}

Recent experiments have successfully tested Einstein's general theory of relativity in a variety of ways and to remarkable precision \cite{Will_book93,Turyshev-2008ufn}. Diverse experimental techniques were used to test relativistic gravity in the solar system. These included spacecraft Doppler tracking, planetary radar ranging, lunar and satellite laser ranging, as well as a number of dedicated gravitational experiments in space and many ground-based efforts \cite{Will-lrr-2006-3,Turyshev:2007qy,Turyshev-2008ufn,Williams-etal-2009}. Given this phenomenological success, general relativity became the standard theory of gravitation, especially where the needs of astronomy, astrophysics, cosmology and fundamental physics are concerned \cite{Turyshev-2008ufn,Turyshev-PLR:2010}. The theory is used for many practical purposes that involve spacecraft navigation, geodesy and time transfer.  It is used to determine the orbits of planets and spacecraft and to describe the propagation of electromagnetic waves in our solar system \cite{Turyshev-2008ufn,Turyshev:2012nw,ref:Turyshev:2014dea}.

As we shall see, finding a solution to the Einstein's gravitational field equations in the case of an unperturbed one-body problem is quite a simple task. However, it turns out that a generalization of the resulting post-Newtonian solution to a system of $N$ extended arbitrary bodies is not so straightforward.

A neutral, pointlike test particle with no angular momentum follows a geodesic that is completely defined by the external gravitational field. However, the coupling of the intrinsic multipole moments of an extended body to the background gravitational field (present due to external gravitational sources), affects the equations of motion of such a body. Similarly, if a test particle is spinning, its equations of motion must account for the coupling of the body's angular momentum to the external gravitational field. As a result, one must be able to describe the interaction of a body's intrinsic multipole moments and angular momentum with the surrounding gravitational field. Multipole moments are well-defined in the local quasi-inertial reference frame, generalizing what was defined for the unperturbed one-body problem \cite{Soffel:2003cr}. While transforming these quantities from one coordinate frame to another, one should account for the fact that the gravitational interaction is non-linear and, therefore, these moments interact with gravitational fields, affecting the body's motion.

When the pseudo-Riemannian geometry of the general theory of relativity is concerned, it is well known that coordinate charts are merely labels. In fact, spacetime coordinates usually do not have direct physical meaning and one has to formulate the observables as coordinate-independent quantities \cite{Soffel:2003cr}. Thus, in order to interpret the results of observations or experiments, one picks a specific coordinate system, chosen for the sake of convenience and calculational expediency, formulates a coordinate representation of a particular measurement procedure, and then derives the corresponding observable needed to conduct and interpret the measurements. It is also known that an ill-defined reference frame may lead to the appearance of non-physical terms that may significantly complicate the interpretation of the data collected \cite{Brumberg-Kopeikin-1989}. Therefore, in practical problems involving relativistic reference frames, choosing the right coordinate system with clearly understood properties is of paramount importance, even as one recognizes that, in principle, all (non-degenerate) coordinate systems are created equal \cite{Soffel:2003cr}.

Before one can solve the {\it global} problem (the motion of the entire $N$-body system), the {\it local} gravitational problem (in the body's vicinity) must be solved first. The correspondence of the {\it global} and {\it local} problems is established using coordinate transformations by which representations of physical quantities are transformed from the coordinates of one reference frame to another. It is understood (see \cite{Brumberg-book-1991,Turyshev-96,DSX-I} for discussion) that, in order to present a complete solution for the $N$-body problem in the general theory of relativity, one must therefore find the solutions to the following three intertwined problems:

\begin{enumerate}[1).]
\item The {\it global} problem: We must construct a global inertial frame, such as a barycentric inertial reference frame, for the system under study. This frame can be used to describe the {\it global} translational motion of $N$ extended bodies comprising the system;
\item The {\it local} problem: For each body in the system, we must construct a local reference frame. This frame is used to study the gravitational field in the vicinity of a particular body, to establish its internal structure and to determine the features of its rotational motion; and finally,
\item A theory of {\it coordinate reference frames}: We must establish rules of coordinate transformations between the {\it global} and {\it local} frames and use these rules to describe physical processes in either of the two classes of reference frames.
\end{enumerate}

A prerequisite to solving the first two of these problems is knowing the transformation rules between {\it global} and {\it local} reference frames. Thus, establishing the theory of astronomical reference frames and solving the equations of motion for celestial bodies become inseparable.

Modern theories of relativistic reference frames \cite{Brumberg-Kopeikin-1989,Kopeikin-1988,Brumberg-Kopeikin-1989-2,DSX-I,DSX-II,DSX-III,DSX-IV,Kopeikin:2004ia}, dealing predominantly with the general theory of relativity, usually take the following approach: Before solving the gravitational field equations, coordinate or gauge conditions are introduced, which impose four restrictions on the components of the pseudo-Riemannian metric $g_{mn}$. This procedure extracts a particular subset from an infinite set of permissible spacetime coordinate charts. These coordinate charts in the resulting subset are related by smooth, differentiable transformations that do not change the imposed coordinate conditions \cite{Kopeikin-1988}. A set of differential coordinate conditions used in leading theories of relativistic reference systems, such as that recommended by the International Astronomical Union (see, for instance \cite{Soffel:2003cr}), are the harmonic gauge conditions. In addition, a set of specific conditions designed to fix a particular reference frame is added to eliminate most of the remaining degrees of freedom, yielding an explicit form for the coordinate system associated with either frame.

In a recent paper \cite{TMT:2011}, we studied accelerating relativistic reference frames in Minkowski spacetime under the harmonic gauge condition. We showed that the harmonic gauge, which has played a very prominent role in gravitational physics ever since the work of Fock \cite{Fock-book-1959}, allows us to present the accelerated metric in an elegant form that depends only on two harmonic potentials. It also allows reconstruction of a significant part of the structure of the post-Galilean coordinate transformation functions relating inertial and accelerating frames. In fact, using the harmonic gauge, we develop the structure of both the direct and inverse coordinate transformations between inertial and accelerated reference frames. Such a complete set of transformations cannot be found in the current literature, since usually either the direct \cite{DSX-I,Soffel:2003cr} or the inverse \cite{Brumberg-Kopeikin-1989-2} transformation is developed, but not both at the same time. A unique form for these coordinate transformation functions were determined from dynamical conditions, obtained by constructing the relativistic proper reference frame of an accelerated test particle. In this frame, the effect of external forces acting on the observer are balanced by the fictitious frame-reaction force that is needed to keep the test particle at rest with respect to the frame, conserving its relativistic three-momentum. We find that this is sufficient to determine explicitly all the terms of the coordinate transformations. The same method is then used to develop the inverse transformations. The resulting post-Galilean coordinate transformations exhibit an approximate group structure that extends the Poincar\'e group of global spacetime transformations to the case of arbitrarily accelerated observers moving in a gravitational field.

In the present paper we continue the development of this method and study the dynamics of a gravitationally bound astronomical $N$-body system.  To constrain the available degrees of freedom, we will again use the harmonic gauge. Using the harmonic gauge condition, we develop the structure of both the direct and inverse harmonic coordinate transformations between global and local reference frames. The method presented in this paper could help with the generalization of contemporary theories of relativistic reference frames and with dealing naturally with both transformations and the relevant equations of motion in a unified formalism. Finally, the method we present does not rely on a particular theory of gravitation; instead, it uses covariant coordinate transformations to explore the dynamics in spacetime from a general perspective. Thus, any description from a standpoint of a metric theory of gravity must have our results in the general relativistic limit.

The outline of this paper is as follows:

In Sec.~\ref{sec:direct-trans} we introduce notation and briefly study the trivial case of a single unperturbed body, followed by a description of the motion of a system of $N$ weakly interacting, self-gravitating, deformable bodies by generalizing the approximate Lorentz transformations. This is accomplished by introducing acceleration-dependent terms in the coordinate transformations. The resulting transformation is given in a general form that relies on a set of functions that are precisely determined in the subsequent sections.

The local coordinate system of an accelerated observer is not unique. We use the harmonic gauge to constrain the set of co-moving coordinate systems in the accelerated reference frame. We carry out this task and determine the metric tensor describing the accelerated reference frame and the structure of the coordinate transformations that satisfy the harmonic gauge conditions. We recover the well-known, elegant result that the metric in the accelerated frame has a form that depends only on two harmonic potentials, which yields a powerful tool that allows the reconstruction, presented in Sec.~\ref{sec:xformlocal}, of a significant part of the structure of the post-Galilean coordinate transformation functions between inertial and accelerating frames.

To fix the remaining degrees of freedom and to specify the proper reference frame of an accelerated observer, we introduce a set of dynamical conditions in Section \ref{sec:dyn}. Specifically, we require that the relativistic three-momentum in the coordinate frame of the accelerated observer must be conserved. This conservation introduces the necessary additional constraints to determine uniquely all remaining terms.

A similar approach can also be carried out in reverse, establishing coordinate transformation rules from an accelerating to an inertial reference frame, which is done in Sec.~\ref{sec:inverse-trans}. Key to the approach presented in this section is the use of the contravariant metric for the accelerating frame, which leads in a straightforward manner to the inverse Jacobian matrix, allowing us to present the inverse transformations in which the roles of the inertial and accelerating coordinates are reversed. Although the calculations are formally very similar to those presented in Secs.~\ref{sec:harm} and \ref{sec:dyn}, subtle differences exist; for this reason, and in order to keep Sec.~\ref{sec:inverse-trans} as self-contained as possible, we opted in favor of some repetitiveness.

We conclude by discussing these results and presenting our recommendations for future research in Section~\ref{sec:end}. We also show the correspondence of our results to those obtained previously by other authors.

\section{Motion of an isolated $N$-body system}
\label{sec:direct-trans}

Our first goal is to investigate the case of $N$ bodies, forming a gravitationally bound system that is isolated, i.e., free of external gravitational fields, described by a metric that is asymptotically flat. Thus, sufficiently far from the $N$-body system the structure of its gravitational field should resemble that of the one-body system.

\subsection{The unperturbed one-body system}

\label{sec:background}

From a theoretical standpoint, general relativity represents gravitation as a tensor field with universal coupling to the particles and fields of the Standard Model. It describes gravity as a universal deformation of the flat spacetime Minkowski metric, $\gamma_{mn}$:
\begin{equation}
g_{mn}(x^k)=\gamma_{mn}+h_{mn}(x^k).
\label{eq:gmn}
\end{equation}
Alternatively, it can also be defined as the unique, consistent, local theory of a massless spin-2 field $h_{mn}$, whose source is the total, conserved energy-momentum tensor (see \cite{MTW,Damour-2006-pdg} and references therein).\footnote{
The notational conventions employed here are those used by Landau and Lifshitz \cite{Landau-Lifshitz-1988}: Letters from the second half of the Latin alphabet, $m, n,...=0...3$ denote spacetime indices. Greek letters $\alpha, \beta,...=1...3$ denote spatial indices. The metric $\gamma_{mn}$ is that of Minkowski spacetime with $\gamma_{mn}={\rm diag}(+1,-1,-1,-1)$ in the Cartesian representation. The coordinates are formed such that $(ct,\vec{r})=(x^0,x^\alpha)$, where $c$ is the velocity of light. We employ the Einstein summation convention with indices being lowered or raised using $\gamma_{mn}$. Round brackets surrounding indices denote symmetrization and square brackets denote anti-symmetrization. A dot over any function means a differentiation with respect to time $t$, defined by $x^0=ct$. We use negative powers of $c$ as a bookkeeping device for order terms; $g^{[k]}_{mn}, k = 1,2,3...$ denotes the $k$-th term in the series expansion of the metric tensor. Other notations will be explained in the paper.}

Classically \cite{Einstein-1915,Hilbert:1915,Einstein-1916}, the general theory of relativity can be defined by postulating the action describing the gravitational field and its coupling to matter fields. Absent a cosmological constant, the propagation and self-interaction of the gravitational field is described by the action
\begin{eqnarray}
\label{eq:hilb-ens-action}
{\cal S}_{\rm G}[g_{mn}]
&=&\frac{c^4}{16\pi G}\int d^4x
\sqrt{-g} R,
\end{eqnarray}
where $G$ is Newton's constant, $g^{mn}$ is the inverse of $g_{mn}$, $g=\det g_{mn}$, $R$ is the trace of the Ricci tensor that in turn is given by $R_{mn}=\partial_k\Gamma^k_{mn}-\partial_m\Gamma^k_{nk}+\Gamma^k_{mn}\Gamma^l_{kl}-\Gamma^k_{ml}\Gamma^l_{nk}$, and $\Gamma^k_{mn}=\frac{1}{2}g^{kp}(\partial_m g_{pn}+\partial_n g_{pm}-\partial_p g_{mn})$ are the Christoffel symbols.

The coupling of $g_{mn}$ to all matter fields (this would generally mean all the fields of the Standard Model of particle physics, represented below symbolically by various gauge fields, fermions and the Higgs doublet as $\psi$, $A_m$ and $H$) is accomplished by using it to replace the Minkowski metric everywhere \cite{Turyshev-2008ufn}. Varying the total action
\begin{equation}
{\cal S}_{\rm tot}[\psi,A_m,H;g_{mn}]={\cal S}_{\rm G}[g_{mn}]+{\cal S}_{\rm SM}[\psi, A_m, H; g_{mn}],
\end{equation}
with respect to $g_{mn}$, we obtain Einstein's field equations for gravity:
\begin{equation}
R^{mn}= \frac{8\pi G}{c^4}\Big(T^{mn}-\frac{1}{2} g^{mn} T\Big),
\label{eq:GR-eq}
\end{equation}
where $T^{mn}=(2/\sqrt{-g})\delta {\cal L}_{\rm SM}/\delta g_{mn}$ is the (presumed to by symmetric) energy-momentum tensor of matter, represented by the Lagrangian density ${\cal L}_{\rm SM}$. As the density of the Einstein tensor is conserved, namely $\nabla_k[\sqrt{-g}(R^{mk}-\frac{1}{2} g^{mk} R)]=0 $, it follows that the density of the energy-momentum tensor of matter $\sqrt{-g}T^{mn}$ is also conserved, obeying the following covariant conservation equation:
{}
\begin{equation}
\nabla_k\big(\sqrt{-g}T^{mk}\big)=0.
\label{eq:tem-conserv}
\end{equation}

Einstein's equations~(\ref{eq:GR-eq}) connect the geometry of a four-dimensional pseudo-Riemannian manifold representing spacetime to the stress-energy-momentum of matter contained in that spacetime. The theory is invariant under arbitrary coordinate transformations: $x^{\prime m}=f^m(x^n)$. This freedom to choose coordinates allows us to introduce gauge conditions that may help with solving the field equations~(\ref{eq:GR-eq}). For instance, in analogy with the Lorenz gauge ($\partial_m A^m = 0$) of electromagnetism, the harmonic gauge corresponds to imposing the condition \cite{Fock-book-1959}:
\begin{equation}
\partial_n \big(\sqrt{-g}g^{mn}\big) = 0. \label{eq:gauge-c}
\end{equation}

We can now proceed with finding a solution to the Einstein's equations~(\ref{eq:GR-eq}) that satisfy the harmonic gauge (\ref{eq:gauge-c}).

\subsection{Solution to the unperturbed one-body problem}
\label{sec:unpert-1B}

Equations (\ref{eq:GR-eq}) are a set of non-linear hyperbolic-type differential equations of the second order with respect to the metric tensor of the pseudo-Riemannian spacetime. This non-linearity makes finding a solution to this set of equations in the general case a complicated problem. No analytical solution to this problem is known; therefore, a full numerical treatment is needed. Depending on a particular situation, one usually introduces relevant small parameters and develops a solution iteratively.

When studying a problem in the weak gravitational field and slow motion approximation, one often uses the ratio $v/c$ of the coordinate velocity $v$ of the bodies to the speed of light $c$, expressed, e.g., in the $N$-body system's barycentric coordinate frame. For the bodies of the solar system this ratio is a small parameter and typically is of the order $v/c\approx 10^{-4}$. Since we are considering a gravitationally bound $N$-body system in the weak field and slow motion approximation, the gravitational potential energy is linked to $v/c$ by the virial theorem: $v^2/c^2 \sim GM/(c^2R)$ \cite{Fock-book-1959,Chandrasekhar-1965}, with $M$ being the mass of the body, while $R$ is a typical distance between bodies. Under these assumptions, we can introduce a scheme of post-Newtonian expansions of various physical quantities, using the dimensioned parameter $c^{-1}$ as a bookkeeping device for order terms. Thus, when we write $y_0+c^{-2}{\cal K}$, for instance, this implies that $c^{-2}{\cal K}$ is of order $(v/c)^2y_0$, which remains small relative to $y_0$. The resulting post-Newtonian approximation (PNA) scheme deals with slowly moving bodies having weak gravitational fields, which makes it very appropriate for constructing the theory of the relativistic reference frames in the solar system.

To find the solution of the gravitational field equations in the first post-Newtonian approximation (1PNA) of the general theory of relativity (for details, consult \cite{Fock-book-1959,Will_book93}), we expand the metric tensor $g_{mn}$ given by Eq.~(\ref{eq:gmn}):
{}
\begin{eqnarray}
 g_{00} &=& 1 + c^{-2} g^{[2]}_{00} + c^{-4}g^{[4]}_{00} +
{\cal O}(c^{-6}), \label{(B1a+*)} \\
g_{0\alpha} &=& c^{-3}g^{[3]}_{0\alpha} +{\cal O}(c^{-5}), \label{(B1b+*)} \\
g_{\alpha\beta} &=& \gamma_{\alpha\beta} + c^{-2}g^{[2]}_{\alpha\beta} +
{\cal O}(c^{-4}),\label{eq:(B1+*)}
\end{eqnarray}

\noindent where  $\gamma_{\alpha\beta}=(-1,-1,-1)$ is the spatial part of the Minkowski metric $\gamma_{mn}$. Note that $g_{0\alpha}$ starts with a term of order $c^{-3}$, which is the lowest order gravitational contribution to this component. The $c^{-2}$ term  $ g_{00}$ in (\ref{(B1a+*)}) is the Newtonian term, while the remaining terms represent the 1PNA order \cite{Brumberg-Kopeikin-1989,Brumberg-book-1991,Turyshev-96}.

In our calculations we constrain the remaining coordinate freedom in the field equations by imposing the covariant harmonic gauge conditions in the form of Eq.~(\ref{eq:gauge-c}). Thus, for $n = 0$ and $n = \alpha$, we correspondingly have (see also Appendix B.2 in Ref.~\cite{Turyshev-96}):
{}
\begin{eqnarray}
{1\over2}c\partial_0 \Big(\gamma^{\epsilon\nu} g^{[2]}_{\epsilon\nu} -
g^{[2]}_{00}\Big) - \partial^\nu g^{[3]}_{0\nu}
&=& {\cal O}(c^{-5}), \label{eq:(DeDg)0}\\	
\frac{1}{2}\partial^\alpha \Big(g^{[2]}_{00} +
\gamma^{ \epsilon\nu} g^{[2]}_{\epsilon\nu}\Big) -
\gamma^{\alpha\mu}\partial^\nu g^{[2]}_{\mu\nu}
&=& {\cal O}(c^{-4}).
 \label{eq:(DeDg)a}
\end{eqnarray}

The metric tensor given by Eqs.~(\ref{(B1a+*)})--(\ref{eq:(B1+*)}) and the gauge conditions given by Eqs.~(\ref{eq:(DeDg)0})--(\ref{eq:(DeDg)a}) allow one to simplify the expressions for the Ricci tensor and to present its contravariant components $R^{mn}=g^{mk}g^{nl}R_{kl}$ in the following form:
{}
\begin{eqnarray}
R^{00} &=& {1\over2}
\Box \Big(c^{-2}g^{[2]00}+ c^{-4}\Big\{g^{[4]00}-
{\textstyle\frac{1}{2}}\big(g^{[2]00}\big)^2\Big\}\Big) +
c^{-4}{1\over2}\Big(g^{[2]\epsilon\lambda}+\gamma^{\epsilon\lambda}g^{[2]00}\Big)\partial_\epsilon\partial_\lambda g^{[2]00}+{\cal O}(c^{-6}), \label{eq:(1R00)}\\
{}
R^{0\alpha} &=& c^{-3}{1\over2}\Delta g^{[3]0\alpha} + {\cal O}(c^{-5}),
\label{eq:(1R0a)}\\
{}
R^{\alpha\beta} &=& c^{-2}{1\over2} \Delta g^{[2]\alpha\beta}  + {\cal O}(c^{-4}), \label{eq:(1Rab)}
\end{eqnarray}
where $\Box =\gamma^{mn}\partial_m\partial_n=\partial^2_{00}+\Delta$ and $\Delta=\partial_\epsilon\partial^\epsilon$  are the d'Alembert and Laplace operators of the Minkowski spacetime correspondingly.

To solve the gravitational field equations (\ref{eq:GR-eq}), we need to specify the form of the energy-momentum tensor. However,  it is sufficient to make only some very general assumptions on the form of this tensor that are valid within the post-Newtonian approximation. Indeed, we will only assume that the components of the energy-momentum tensor, $T^{mn}$, have the following form:
{}
\begin{eqnarray}
T^{00} &=& c^2\Big( T^{[0]00} + c^{-2}T^{[2]00}+{\cal O}(c^{-4})\Big),
\quad
T^{0\alpha} = c\Big(T^{[1]0\alpha}+{\cal O}(c^{-2})\Big),
\quad
T^{\alpha\beta} = T^{[2]\alpha\beta} + {\cal O}(c^{-2}).  \label{eq:(Tab+)}
\end{eqnarray}
This form is sufficient to construct an iterative solution scheme in the post-Newtonian approximation of the general theory of relativity leaving the precise form of the energy-momentum tensor $T^{mn}$ unspecified. These assumptions allow us to present the ``source term'', $S^{mn} = T^{mn} - {1\over2} g^{mn}T$, on the right-hand side of Eq.~(\ref{eq:GR-eq}) in the following form:
\begin{eqnarray}
S^{00} &=& \frac{1}{2}c^2\Big(\sigma + {\cal O}(c^{-4})\Big),
\label{eq:(S00-sig)}\\
{}
S^{0\alpha} &=& c\Big(\sigma^\alpha -c^{-2}{1\over2}\sigma g^{[3]0\alpha}+ {\cal O}(c^{-4})\Big),
\label{eq:(S0a-sig_a)}\\
{}
S^{\alpha\beta} &=& - {1\over2}c^2\Big(\gamma^{\alpha\beta}+c^{-2}\big(\gamma^{\alpha\beta}g_{00}^{[2]}+g^{[2]\alpha\beta}\big)\Big)\sigma+ \sigma^{\alpha\beta}+{\cal O}(c^{-2}), 
\label{eq:(Sab-sig)}
%\\
%{}
%S^{\alpha\beta} &=& - \gamma^{\alpha\beta}{1\over2}c^2\sigma \big(1+ c^{-2}2g_{00}^{[2]}\big)+ \sigma^{\alpha\beta}+{\cal O}(c^{-2}), \label{eq:(Sab-sig)}
\end{eqnarray}

\noindent where we have introduced the scalar, vector, and shear densities $\sigma$, $\sigma^\alpha$, and $\sigma^{\alpha\beta}$:
\begin{eqnarray}
\sigma &=& T^{[0] 00} + c^{-2}\big(T^{[2]00} -
\gamma_{\mu\lambda}T^{[2] \mu\lambda}\big) + {\cal O}(c^{-4})\equiv
\frac{1}{c^2}\Big(T^{00} - \gamma_{\mu\lambda}T^{\mu\lambda}\Big) + {\cal O}(c^{-4}),
\label{eq:(sig)}\\
{}
\sigma^\alpha &=& T^{[1] 0\alpha} + c^{-2}T^{[3] 0\alpha}+ {\cal O}(c^{-4})\hskip 60pt \equiv
\frac{1}{c}T^{0\alpha} + {\cal O}(c^{-4}),\\
\label{eq:(sig_a)}
\sigma^{\alpha\beta} &=& T^{[2]\alpha\beta} -
\gamma^{\alpha\beta}\gamma_{\mu\lambda}T^{[2] \mu\lambda} + {\cal O}(c^{-2})\hskip 43pt \equiv
T^{\alpha\beta} - \gamma^{\alpha\beta}\gamma_{\mu\lambda}T^{\mu\lambda} + {\cal O}(c^{-2}).
\label{eq:(sig_ab)}
\end{eqnarray}

We can now express the stress-energy tensor $T^{mn}$ in terms of $\sigma^{mn}$ introduced by Eq.~(\ref{eq:(sig)})--(\ref{eq:(sig_ab)}):
\begin{eqnarray}
T^{00} &=& c^2\sigma -\frac{1}{2}\gamma_{\mu\lambda}\sigma^{\mu\lambda}+ {\cal O}(c^{-2}),~~~~~
T^{0\alpha} = c\Big(\sigma^\alpha + {\cal O}(c^{-4})\Big),~~~~~
T^{\alpha\beta} = \sigma^{\alpha\beta}- \gamma^{\alpha\beta}{1\over2}\gamma_{\mu\lambda}\sigma^{\mu\lambda}+{\cal O}(c^{-2}).
\label{eq:(Tab-sig)}
\end{eqnarray}

Substituting the expressions (\ref{eq:(1R00)})--(\ref{eq:(1Rab)}) for the Ricci tensor together with the source term $S^{mn}$ given by Eqs.~(\ref{eq:(S00-sig)})--(\ref{eq:(sig_ab)}) into the field equations (\ref{eq:GR-eq}) of the general theory of relativity, we obtain the following equations to determine the components of the metric tensor in the post-Newtonian approximation:
{}
\begin{eqnarray}
\Box \Big\{\frac{1}{c^2}g^{[2]00}+ \frac{1}{c^4}\Big(g^{[4]00}-{\textstyle\frac{1}{2}}\big(g^{[2]00}\big)^2\Big)\Big\} + \frac{1}{c^4}\Big(g^{[2]\epsilon\lambda}+\gamma^{\epsilon\lambda}g^{[2]00}\Big)\partial_\epsilon\partial_\lambda g^{[2]00}&=&\frac{8\pi G}{c^2} \sigma + {\cal O}(c^{-6}), \label{eq:(1R00)S0}\\
{}
\frac{1}{c^3}\Delta g^{[3]0\alpha}&=& \frac{16\pi G}{c^3}\sigma^\alpha + {\cal O}(c^{-5}),
\label{eq:(1R0a)S0}\\
{}
\frac{1}{c^2} \Delta g^{[2]\alpha\beta} &=&-
\gamma^{\alpha\beta}\frac{8\pi G}{c^2}\sigma + {\cal O}(c^{-4}).
\label{eq:(1Rab)S0}
\end{eqnarray}

We assume that spacetime is asymptotically flat (no external gravitational field far from the body) and there is no gravitational radiation coming from outside the body. In terms of perturbations of the the Minkowski metric  $h_{mn}= g_{mn}-\gamma_{mn}$ introduced by Eq.~(\ref{eq:gmn}), the corresponding two boundary conditions \cite{Fock-book-1959} have the form
\begin{equation}
\lim_{\substack{r\rightarrow\infty\\t+r/c={\rm const}}} h_{mn}=0
~~~~~~{\rm and}~~~\lim_{\substack{r\rightarrow\infty\\t+r/c={\rm const}}}
\big[(rh_{mn})_{,r}+(rh_{mn})_{,0}\big]=0.
\label{eq:ass-flat}
\end{equation}
Multiplying Eq.~(\ref{eq:(1R00)S0}) by $\gamma^{\alpha\beta}$ and summing the result with Eq.~(\ref{eq:(1Rab)S0}) allows one to determine $g^{[2]\alpha\beta}+\gamma^{\alpha\beta}g^{[2]00} = {\cal O}(c^{-2})$, which was derived using the boundary conditions (\ref{eq:ass-flat}). As a result, the system of equations~(\ref{eq:(1R00)S0})--(\ref{eq:(1Rab)S0}) can equivalently be re-written in the following form:
{}
\begin{eqnarray}
\Box \Big(\frac{1}{c^2}g^{[2]00}+ \frac{1}{c^4}\Big\{g^{[4]00}-
{\textstyle\frac{1}{2}}\big(g^{[2]00}\big)^2\Big\}\Big)&=&\frac{8\pi G}{c^2} \sigma + {\cal O}(c^{-6}), \label{eq:(1R00)S1}\\
{}
\frac{1}{c^3}\Delta g^{[3]0\alpha}&=& \frac{16\pi G}{c^3}\sigma^\alpha + {\cal O}(c^{-5}),
\label{eq:(1R0a)S1}\\
{}
g^{[2]\alpha\beta}+\gamma^{\alpha\beta}g^{[2]00} &=&{\cal O}(c^{-2}).
\label{eq:(1Rab)S1}
\end{eqnarray}

The solution to this system of equations may be given as below (see also \cite{DSX-I} and references therein):
{}
\begin{eqnarray}
g^{00}&=&1+\frac{2w}{c^2}+\frac{2w^2}{c^4} + {\cal O}(c^{-6}),
\hskip 20pt
g^{0\alpha}=\frac{4w^\alpha}{c^3}+ {\cal O}(c^{-5}),
\hskip 20pt
g^{\alpha\beta}=\gamma^{\alpha\beta}-\gamma^{\alpha\beta}\frac{2w}{c^2} + {\cal O}(c^{-4}),\label{eq:gab1B*}
\end{eqnarray}
{}
where the scalar and vector gravitational potentials $w$ and $w^\alpha$ are determined from the following harmonic equations:
{}
\begin{eqnarray}
\Box w&=&4\pi G \sigma + {\cal O}(c^{-4}),
\hskip 20pt
\Delta w^\alpha= 4\pi G\sigma^\alpha + {\cal O}(c^{-2}),
\label{eq:eq-w^a*}
\end{eqnarray}
which, according to the harmonic gauge conditions (\ref{eq:(DeDg)0})--(\ref{eq:(DeDg)a}) and the energy-momentum conservation equation~(\ref{eq:tem-conserv}), satisfy the following two Newtonian continuity equations
{}
\begin{equation}
c\partial_0 w+\partial_\epsilon w^\epsilon=  {\cal O}(c^{-2})
~~~{\rm and} ~~~
c\partial_0 \sigma+\partial_\epsilon \sigma^\epsilon=  {\cal O}(c^{-2}).
\label{eq:eq-cont}
\end{equation}

Assuming that spacetime is asymptotically flat in the sense of Eq.~(\ref{eq:ass-flat}), one can write a solution for $w$ and $w^\alpha$ in terms of the advanced and retarded potentials. The recommended solution \cite{Soffel:2003cr}, half advanced and half retarded,  reads
{}
\begin{eqnarray}
w(t,\vec{x})&=&G\int d^3x'\frac{\sigma(t,\vec{x}')}{|\vec{x}-\vec{x}'|} + \frac{1}{2c^2}G\frac{\partial^2}{\partial t^2}\int d^3x'
{\sigma(t,\vec{x}')}{|\vec{x}-\vec{x}'|}+{\cal O}(c^{-3}),
\label{eq:w*}\\
{}
w^\alpha(t,\vec{x})&=& G\int d^3x'\frac{\sigma^\alpha(t,\vec{x}')}{|\vec{x}-\vec{x}'|}  + {\cal O}(c^{-2}).
\label{eq:w^a*}
\end{eqnarray}
These are symmetrical solutions for the post-Newtonian potentials, different from those resulting from the imposition of the no-incoming-wave condition (\ref{eq:ass-flat}). It is legitimate to resort to these solutions only because the difference between them and the retarded solutions is a pure gauge effect at this level of accuracy \cite{Soffel:2003cr}.

Finally, we can present the unperturbed solution for an isolated one-body problem  in terms of Minkowski metric perturbations $h_{mn}$ introduced by Eq.~(\ref{eq:gmn}) as below:
{}
\begin{eqnarray}
g_{mn}&=&\gamma_{mn}+h_{mn}^{(0)},\nonumber\\
h_{00}^{(0)}=-\frac{2w}{c^2}+\frac{2w^2}{c^4} + {\cal O}(c^{-6}), \qquad
h_{0\alpha}^{(0)}&=&-\gamma_{\alpha\lambda}\frac{4w^\lambda}{c^3}+ {\cal O}(c^{-5}), \qquad
h_{\alpha\beta}^{(0)}=\gamma_{\alpha\beta}\frac{2w}{c^2} + {\cal O}(c^{-4}).
\label{eq:g^mn1B}
\end{eqnarray}

Equations (\ref{eq:g^mn1B}) represent the well-known solution for the one-body problem in the general theory of relativity \cite{MTW,Will_book93,Turyshev-2008ufn}. The new method, presented here, relies on the properties of this solution in developing a perturbation theory needed to find a solution in the case of the $N$-body problem.

\subsection{Ansatz for the $N$-body problem}
\label{sec:ansatz-N-body}

To describe the dynamics of the $N$-body problem we introduce $N+1$ reference frames, each with its own coordinate chart. We need a {\it global} coordinate chart $\{x^m\}$ defined for an inertial reference frame that covers the entire system under consideration.  In the immediate vicinity of each body $b$ in the $N$-body system, we can also introduce a chart of {\it local} coordinates $\{y^m_b\}$ defined in the frame associated with this body. In the remainder of this paper, we use $\{x^m\}$ to represent the coordinates of the global inertial frame and $\{y^m_b\}$ to be the local coordinates of the frame associated with a particular body $b$.

\emph{First}, we assume that the studied $N$-body system is isolated and, similarly to the one-body problem, there exists a barycentric inertial system (with global coordinates denoted as $x^k\equiv(ct,\vec{x})$) that resembles the properties of the one-body problem discussed above. Namely, the metric tensor $g_{mn}(x^k)$, describing the $N$-body system in that reference frame, is asymptotically flat (no gravitational fields far from the system) and there is no gravitational radiation coming from outside the system.

Therefore, in terms of the perturbations of the Minkowski metric, $h^{mn}= g^{mn}-\gamma^{mn}$, the boundary conditions are identical to those given by (\ref{eq:ass-flat}) and, in the case of the $N$-body problem can be given as below:
\begin{equation}
\lim_{\substack{r\rightarrow\infty\\t+r/c={\rm const}}} h^{mn}=0
~~~~~~{\rm and}~~~\lim_{\substack{r\rightarrow\infty\\t+r/c={\rm const}}}
\big[(rh^{mn})_{,r}+(rh^{mn})_{,0}\big]=0,
\label{eq:bound-cond}
\end{equation}
which expresses the fact that no ``exterior'' wave is coming into the system from past null infinity (valid for all local frames introduced in the vicinity of the $N$ bodies constituting the system).

The boundary condition (\ref{eq:bound-cond}) above entails some subtleties. We note that, even if the post-Minkowskian metric perturbations are bound to decay at future null infinity, their post-Newtonian expansion may formally diverge as $r\rightarrow\infty$ (in an asymptotically inertial frame). At the 1PNA, this does not cause any trouble since $w_b$ and $w_b^\alpha$ have both compact support, but at higher order, some piece of the metric may actually blow up when $r$ increases. This simply reflects the fact that the post-Newtonian approximation is only valid in the near zone.

\emph{Second}, we assume that there exists a local reference frame (with coordinates denoted as $y^k_b\equiv(ct_b,\vec{y}_b)$) associated with each body $b$ in the $N$-body system. We further assume that there exists a smooth transformation connecting coordinates $y^k_b$ chosen in this local reference frame to coordinates $x^k$ in the global inertial frame. In other words, we assume that there exist both direct $x^k=g^k_b(y^l_b)$ and inverse $y^k_b=f^k_b(x^l)$ coordinate transformation functions between the two reference frames. Our objective is to construct  these transformations in explicit form.

The most general form of the post-Galilean coordinate transformations between the non-rotating reference frames defined by global coordinates $x^k\equiv (x^0,x^\alpha)$, and local coordinates $y^k_a\equiv(y^0_a,y^\alpha_a)$, introduced at the vicinity of a body $a$ from the $N$-body system, may be given in the following form \cite{Turyshev-96,TMT:2011}:
\begin{eqnarray}
x^0&=& y^0_a+c^{-2}{\cal K}_a(y^0_a,y^\epsilon_a)+c^{-4}{\cal L}_a(y^0_a,y^\epsilon_a)+{\cal O}(c^{-6})y^0_a,
\label{eq:trans-0}\\[3pt]
x^\alpha&=& y^\alpha_a+z^\alpha_{a_0}(y^0_a)+
c^{-2}{\cal Q}^\alpha_a(y^0_a,y^\epsilon_a)+{\cal O}(c^{-4}),
\label{eq:trans-a}
\end{eqnarray}
where $z^\mu_{a_0}\equiv z^\alpha_{a_0}(y^0_a)$ is the Galilean vector connecting the spatial origins of the two non-rotating frames, and we introduce the post-Galilean vector $x^\mu_{a_0}(y^0_a)$ connecting the origins of the two frames
\begin{equation}
x^\mu_{a_0}(y^0_a)=z^\mu_{a_0}+c^{-2}{\cal Q}^\mu_a(y^0_a,0)+{\cal O}(c^{-4}).\label{eq:totvec}
\end{equation}
The functions ${\cal K}_a, {\cal L}_a$ and ${\cal Q}^\mu_a$ are yet to be determined.

The local coordinates $\{y^m_a\}$ are expected to remain accurate in the neighborhood of the world-line of the body being considered. The functions ${\cal K}_a, {\cal L}_a$ and ${\cal Q}^\mu_a$ should contain the information about the specific physical properties of the reference frame chosen for the analysis and must depend only on the mutual dynamics between the reference frames, i.e., velocity, acceleration, etc. Furthermore, we would like these transformations to be smooth, which would warrant the existence of inverse transformations (introduced below and discussed in details in Sec.~\ref{sec:inverse-trans}).

The coordinate transformation rules for the general coordinate transformations above are easy to obtain and express in the form of the Jacobian matrix ${\partial x^m}/{\partial y^n_a}$. Using Eqs.~(\ref{eq:trans-0})--(\ref{eq:trans-a}), we get:
{}
\begin{eqnarray}
{\partial x^0\over\partial y^0_a}
&=& 1 + c^{-2}{\partial {\cal K}_{a}\over\partial y^0_{a}}  +
c^{-4}{\partial {\cal L}_{a}\over\partial y^0_{a}}  + {\cal O}(c^{-6}),
\qquad ~\,
{\partial x^0\over\partial y^\alpha_a} =
c^{-2} {\partial {\cal K}_{a}\over\partial y^\alpha_{a}}  +
c^{-4}{\partial {\cal L}_{a}\over\partial y^\alpha_{a}}  +
{\cal O}(c^{-5}),
\label{eq:(C1a)}\\
{}
{\partial x^\alpha\over\partial y^0_a} &=& \frac{v^\alpha_{a_0}}{c} +
c^{-2} {\partial {\cal Q}^\alpha_{a}\over\partial y^0_{a}} + {\cal O}(c^{-5}),
\qquad \qquad \qquad
{\partial x^\alpha\over \partial y^\mu_a}  =
\delta^\alpha_\mu  +
 c^{-2}{\partial {\cal Q}^\alpha_{a}\over\partial y^\mu_{a}}  + {\cal O}(c^{-4}), \label{eq:(C1d)}
\end{eqnarray}
where $v^\epsilon_{0a} =\dot z^\epsilon_{0a}\equiv cd z^\epsilon_{0a}/d y^0_a$ is the time-dependent velocity of the frame $a$ relative to the barycentric inertial frame.

The transformations  given by Eqs.~(\ref{eq:trans-0})--(\ref{eq:trans-a}) transform space and time coordinates from the global, inertial frame $\{x^m\}$ to space and time coordinates in the local, accelerating reference frame $\{y^m_a\}$. These transformations are not singular in those regions where the post-Galilean approximation remains valid. To check this, we can verify that the determinant of the Jacobian matrix does not vanish \cite{Brumberg-Kopeikin-1989-2}:
{}
\begin{equation}
\det\Big(\frac{\partial x^k}{\partial y^m_a}\Big) = 1+
 c^{-2}\Big\{{\partial {\cal K}_a\over\partial y^0_{a}}   -
v^\lambda_{a_0} {\partial {\cal K}_a\over\partial y^\lambda_{a}} +
{\partial {\cal Q}^\mu_{a}\over\partial y^\mu_{a}}\Big\} + {\cal O}(c^{-4}). \label{(C2)}
\end{equation}
This guarantees the invertibility of the Jacobian matrix and hence, the existence of the inverse transformations.

The corresponding inverse transformations between local, $\{y^k_a\}=(y^0_a,y^\epsilon_a)$, and global, $\{x^k\}=(x^0,x^\epsilon)$, coordinates (further discussed in Sec.~\ref{sec:inverse-trans}) are given by
\begin{eqnarray}
y^0_a&=& x^0+c^{-2}\hat{\cal K}_a(x^0,x^\epsilon)+c^{-4}\hat{\cal L}_a(x^0,x^\epsilon)+{\cal O}(c^{-6})x^0,
\label{eq:trans-0_inv}\\[3pt]
y^\alpha_a&=& x^\alpha-z^\alpha_{a_0}(x^0)+c^{-2}\hat{\cal Q}^\alpha_a(x^0,x^\epsilon)+{\cal O}(c^{-4}),
\label{eq:trans-a_inv}
\end{eqnarray}
where $z^\mu_{a_0}(x^0)$ is the Galilean position vector of the body $a$, expressed as a function of global time, $x^0$.

Although, the ``hatted'' functions ($\hat{\cal K}_a, \hat{\cal L}_a, \hat{\cal Q}^\alpha_a$) are yet unknown, we can verify that, in order for Eqs.~(\ref{eq:trans-0_inv})--(\ref{eq:trans-a_inv}) to be inverse to Eqs.~(\ref{eq:trans-0})--(\ref{eq:trans-a}), these functions must relate to the original set of (${\cal K}_a, {\cal L}_a, {\cal Q}^\alpha_a$) via the following expressions:
\begin{eqnarray}
\hat{\cal K}_a(x^k)&=& -{\cal K}_a(x^0,r^\epsilon_a)+{\cal O}(c^{-4}),\\[3pt]
\hat{\cal Q}^\alpha_a(x^k)&=& (v^\alpha_{a_0}/c)
\,{\cal K}_a(x^0,r^\epsilon_a)-{\cal Q}^\alpha_a(x^0,r^\epsilon_a)+{\cal O}(c^{-2}),\\[3pt]
\hat{\cal L}_a(x^k)&=&\frac{\partial{\cal K}_a(x^0,r^\epsilon_a)}{\partial x^0}{\cal K}_a(x^0,r^\epsilon_a)+
\frac{\partial{\cal K}_a(x^0,r^\epsilon_a)}{\partial x^\lambda}{\cal Q}^\lambda_a(x^0,r^\epsilon_a)-{\cal L}_a(x^0,r^\epsilon_a)+{\cal O}(c^{-2}),
\end{eqnarray}
with $r^\epsilon_a=x^\epsilon-x^\epsilon_{a_0}$, where $x^\mu_{a_0}(x^0)$ is the post-Galilean position vector of the body $a$ expressed as a function of the global time-like coordinate $x^0$ (as opposed to Eq.~(\ref{eq:totvec}), which is given in local time $y^0_a$) defined as
\begin{equation}
x^\mu_{a_0}(x^0)=z^\mu_{a_0}-c^{-2}\hat{\cal Q}^\mu_a(x^0,0)+{\cal O}(c^{-4}).
\label{eq:x_0Q}
\end{equation}
Also, $v^\alpha_{a_0}=\dot z^\alpha_{a_0}(x^0)$ is the velocity of the body $a$ as seen from the global inertial frame.

The inverse of the Jacobian matrix (\ref{eq:(C1a)})--(\ref{eq:(C1d)}), $\partial y^n_a/\partial x^m$, can be obtained directly from (\ref{eq:trans-0_inv})--(\ref{eq:trans-a_inv}):
\begin{eqnarray}
{\partial y^0_a \over\partial x^0}
&=& 1 + c^{-2}{\partial \hat{\cal K}_{a}\over\partial x^0}  +
c^{-4}{\partial \hat{\cal L}_{a}\over\partial x^0}  + {\cal O}(c^{-6}),
\qquad
{\partial  y^0_a \over\partial x^\alpha} =
c^{-2} {\partial \hat{\cal K}_{a}\over\partial x^\alpha}  +
c^{-4}{\partial \hat{\cal L}_{a}\over\partial x^\alpha}  +
{\cal O}(c^{-5}),
\label{eq:(C1a)_inv}\\
{}
{\partial y^\alpha_a  \over \partial x^0} &=&
-\frac{v^\alpha_{a_0}}{c} +
c^{-2} {\partial \hat{\cal Q}^\alpha_{a}\over\partial x^0} + {\cal O}(c^{-5}),
\qquad\qquad~~
{\partial y^\alpha_a\over \partial x^\mu}  =
\delta^\alpha_\mu  +
c^{-2} {\partial \hat{\cal Q}^\alpha_{a}\over\partial x^\mu}  + {\cal O}(c^{-4}). \label{eq:(C1d)_inv}
\end{eqnarray}
Similarly to (\ref{(C2)}), the condition of a non-vanishing  determinant of the inverse Jacobian matrix $\det\big(\partial y^n_a/\partial x^m\big)$ determines the area of applicability of the coordinate transformations Eqs.~(\ref{eq:trans-0_inv})--(\ref{eq:trans-a_inv}). As was shown in \cite{Brumberg-Kopeikin-1989-2} such area covers the entire solar system, which is more than adequate for our purposes.

\emph{Third}, we will search for the solution of the gravitational field equations, $g^{mn}$, in the case of $N$-body problem within the general theory of relativity in the global inertial (barycentric) coordinate frame in the following form:
\begin{align}
g^{mn}(x)&=\gamma^{mn}(x)+\sum^N_{b=1} \frac{\partial x^m}{\partial y^k_b}\frac{\partial x^n}{\partial y^l_b}~ h^{kl}_{b(0)}(y_b(x))+h^{mn}_{\rm int}(x),
\label{eq:ansatz}
\end{align}
where the terms on the right-hand side have the following meaning:
\begin{itemize}
  \item The first term, $\gamma^{mn}(x)=(1,-1,-1,-1)$, is the contravariant form of the Minkowski metric of flat spacetime;
  \item The second term is a summation term representing the superposition of the unperturbed one-body solutions $h^{kl}_{b(0)}$ given by Eqs.~(\ref{eq:g^mn1B}) for each body $b$ in the $N$-body system, which were covariantly transformed to the barycentric frame by coordinate transformations (\ref{eq:trans-0})--(\ref{eq:trans-a});
  \item Finally,  the last term, $h^{mn}_{\rm int}$, represents the gravitational interaction between the bodies in the $N$-body system. It is expected that the interaction term is at least of the second order in the gravitational constant, $h^{mn}_{\rm int}\propto {G^2}\approx c^{-4}$;
this assumption will be used in our approximation method.
\end{itemize}
Note that the contravariant components\footnote{Clearly, one can use the covariant form of the metric tensor, $g_{mn}$, instead of the contravariant form as in Eq.~(\ref{eq:ansatz}), together with the coordinate transformations given by Eqs.~(\ref{eq:trans-0_inv})-(\ref{eq:trans-a_inv}) (as was explored in \cite{Turyshev-96}), and obtain identical results. We choose this approach for computational expediency based on the convenience of using results for the one-body problem from Sec.~\ref{sec:unpert-1B} and the form of the direct coordinate transformations (\ref{eq:trans-0})--(\ref{eq:trans-a}).} of the metric tensor $g^{mn}$ allow us to rely directly on the form of the coordinate transformation $x^m=f^m(y_a^k)$ from Eqs.~(\ref{eq:trans-0})--(\ref{eq:trans-a}) and the associated Jacobian matrix ${\partial x^m}/{\partial y^n_a}$ given by (\ref{eq:(C1a)})--(\ref{eq:(C1d)}). Because of the metric nature of the general theory of relativity, Eq.~(\ref{eq:ansatz}) is valid for any admissible covariant coordinate transformations $x^m=f^m(y_b^k)$. As, at this stage, we already know the unperturbed one-body solution $h^{kl}_{b(0)}$, which is given for each body $b$ in the form of Eqs.~(\ref{eq:g^mn1B}), the only unknown part of the ansatz is the gravitational interaction term, $h^{mn}_{\rm int}$, which will be determined in Sec.~\ref{sec:interact-term}.

The form of the iterative solution  for $N$-body metric given by Eq.~(\ref{eq:ansatz}) was inspired by a successful perturbation theory approach of high-energy physics. In that approach, a solution for a system of interacting fields is sought in the form of a sum of the unperturbed solutions for each of the field in the system, plus the interaction term. The fact that the line element, formed using the metric tensor, represents a Lagrangian for test particles (see discussion in Sec.~\ref{sec:good-frame}), allows one to develop an iterative solution in the case of $N$ interacting bodies based on the principles of the perturbation theory approach.

\emph{Fourth}, we impose the harmonic gauge conditions given by Eq.~(\ref{eq:gauge-c}), $\partial_m(\sqrt{-g}g^{mn}\big)=0$, on the $N$-body metric tensor $g^{mn}$ given by Eq.~(\ref{eq:ansatz}), in both the global and local frames. As we will see in the following section, this condition that is imposed on the metric tensor leads to the requirement that the transformation functions (\ref{eq:trans-0})--(\ref{eq:trans-a}) must be harmonic. Below we investigate the impact of the harmonic gauge on the transformation functions ${\cal K}_a, {\cal L}_a$ and ${\cal Q}^\mu_a$ and the inverse transformation functions $\hat{\cal K}_a$, $\hat{\cal Q}_a$ and $\hat{\cal L}_a$.

\subsection{Imposing the harmonic gauge conditions}
\label{sec:harm}

The dynamical condition, i.e., the requirement that the spatial origin of the transformed system of coordinates $x^m=f(y^m)$ is to move along a specific world-line, does not uniquely define $y^m$. The existence of this coordinate freedom allows us to impose the harmonic gauge condition on the metric density $ \tilde{g}^{mn}={\sqrt{-g}}g^{mn}$ in the local frame:
\begin{equation}
\frac{\partial}{\partial y^m_a}\big({\sqrt{-g}}g^{mn}\big) = 0.
\label{eq:(DeDg)OK}
\end{equation}
The vanishing of the covariant derivative of the metric tensor (i.e., $({\sqrt{-g}}g^{mn})_{;m}=(\partial/\partial y^m_a)({\sqrt{-g}}g^{mn})+\sqrt{-g}g^{kl}\Gamma^n_{kl}=0$, where $\Gamma^n_{mk}$ are the Christoffel symbols associated with the metric tensor $g_{mn}$ in the moving frame $\{y^m_a\}$, and the semicolon represents covariant differentiation with respect to the coordinates, allows us to present Eq.~(\ref{eq:(DeDg)OK}) in the following equivalent form:
{}
\begin{equation}
g^{kl}\Gamma^n_{kl}=0.
\label{eq:(DeDg)CS}
\end{equation}
Furthermore, remembering the rules of transformation of the Christoffel symbols under general coordinate transformations, we can see that (\ref{eq:(DeDg)CS}) is equivalent to imposing the harmonic conditions on the transformation functions in Eqs.~(\ref{eq:trans-0})--(\ref{eq:trans-a}):
{}
\begin{equation}
\Box_y x^m= 0,
\label{eq:(DeDg)x0-box}
\end{equation}
where $\Box_y=(\sqrt{-g})^{-1}(\partial/\partial y_a^m)(\sqrt{-g}g^{mn}\partial/\partial y_a^n)$ denotes the d'Alembertian with respect to $\{y_a^n\}$ acting on $x^m$, which are treated as individual scalar functions.

Conversely, we also impose the harmonic gauge conditions on the inverse transformation:
{}
\begin{equation}
\frac{\partial}{\partial x^m}\big({\sqrt{-g}}g^{mn}\big) = 0 \qquad
{\rm and} \qquad \Box_x y^m_a= 0,
\label{eq:(DeDg)OK-ctv}
\end{equation}
where $\Box_x=(\sqrt{-g})^{-1}(\partial/\partial x^m)(\sqrt{-g}g^{mn}\partial/\partial x^n)$ denotes the d'Alembertian with respect to global coordinates $\{x_m\}$ acting on the scalar functions $y^m_a$.

Therefore, on the one hand, the harmonic gauge imposes restrictions on the partial derivatives of the metric tensor, as seen in Eq.~(\ref{eq:(DeDg)OK}). On the other hand, it restricts the choice of admissible coordinate transformations only to those that satisfy the harmonic equation (\ref{eq:(DeDg)x0-box}). These two consequences of imposing the harmonic gauge will be used to establish the structure that the metric tensor, which will be given by Eqs.~(\ref{eq:g00-loc_cov})--(\ref{eq:gab-loc_cov}), must satisfy under the harmonic coordinate transformations (\ref{eq:trans-0})--(\ref{eq:trans-a}). The harmonic gauge also constrains the form of the transformation functions (${\cal K}_a,{\cal L}_a,{\cal Q}^\alpha_a$).
Below we explore this important dual role of the harmonic gauge in more detail.

\subsubsection{Constraining degrees of freedom in the metric tensor}
\label{sec:harm_cond-metr}

To study the harmonic gauge conditions we will use a complete form of the metric tensor (\ref{(B1a+*)})--(\ref{eq:(B1+*)}), where the component $g_{0\alpha} = c^{-1}g^{[1]}_{0\alpha}+c^{-3}g^{[3]}_{0\alpha} +{\cal O}(c^{-5})$ has the leading term of $c^{-1}$ order that is of inertial nature \cite{TMT:2011}. As a result, the harmonic gauge conditions (\ref{eq:(DeDg)OK}) yield the following set of partial differential equations for $\tilde g^{mn}=\sqrt{-g}g^{mn}$:
{}
\begin{eqnarray}
\frac{\partial}{\partial y_a^0} \Big\{c^{-2}\tilde{g}^{[2]00}+{\cal O}(c^{-4})\Big\}+\frac{\partial}{\partial y_a^\epsilon} \Big\{c^{-1}\tilde{g}^{[1]\epsilon0}+c^{-3}\tilde{g}^{[3]\epsilon0}+{\cal O}(c^{-5})\Big\} &=& 0,
 \label{eq:(DeDg)OK0}\\
\frac{\partial}{\partial y_a^0} \Big\{c^{-1}\tilde{g}^{[1]0\alpha}+{\cal O}(c^{-3})\Big\}+\frac{\partial}{\partial y_a^\epsilon} \Big\{c^{-2}\tilde{g}^{[2]\epsilon\alpha}+{\cal O}(c^{-4})\Big\}&=&  0,
 \label{eq:(DeDg)OKa}
\end{eqnarray}
where we use bracketed superscript indices in the form $\tilde{g}^{[k]mn}$ to denote order terms with respect to inverse powers of $c^{-k}$, as discussed in the previous section. Grouping terms by order in these equations, we obtain the following set of equations on the metric components:
{}
\begin{eqnarray}
\frac{\partial}{\partial y_a^\epsilon}\tilde{g}^{[1]\epsilon0} &=& {\cal O}(c^{-4}),
 \label{eq:(DeDgGa)OK0a1}\\
c\frac{\partial}{\partial y_a^0} \tilde{g}^{[2]00}+\frac{\partial}{\partial y_a^\epsilon} \tilde{g}^{[3]\epsilon0} &=& {\cal O}(c^{-2}),
 \label{eq:(DeDgGa)OK0a3}\\
c\frac{\partial}{\partial y_a^0} \tilde{g}^{[1]0\alpha}+\frac{\partial}{\partial y_a^\epsilon}\tilde{g}^{[2]\epsilon\alpha} &=& {\cal O}(c^{-2}),
 \label{eq:(DeDgGa)OKab}
\end{eqnarray}
{}
where the metric density components in the expressions above can be expressed via the covariant metric as
\begin{eqnarray}
\tilde{g}^{[1]0\alpha} &=& -\gamma^{\alpha\nu}g^{[1]}_{0\nu},
\label{eq:GC_metr_10a}\\
\tilde{g}^{[2]00} &=& -\frac{1}{2}\Big\{g^{[2]}_{00} -\gamma^{\epsilon\lambda}\Big(g^{[2]}_{\epsilon\lambda}
+g^{[1]}_{0\epsilon}g^{[1]}_{0\lambda}\Big)\Big\},
\label{eq:GC_metr_200}\\
\tilde{g}^{[3]0\alpha} &=& -\gamma^{\alpha\nu}\Big\{g^{[3]}_{0\nu}
+\frac{1}{2}g^{[1]}_{0\nu}\Big(\gamma^{\epsilon\lambda}\big(g^{[2]}_{\epsilon\lambda}+g^{[1]}_{0\epsilon}g^{[1]}_{0\lambda}\big)- g^{[2]}_{00}\Big)-
\gamma^{\epsilon\lambda}
g^{[1]}_{0\epsilon}g^{[2]}_{\nu\lambda}\Big\},
\label{eq:GC_metr_30a}\\
\tilde{g}^{[2]\alpha\beta} &=& -
\gamma^{\alpha\epsilon}\gamma^{\beta\lambda}\Big(
g^{[2]}_{\epsilon\lambda}-g^{[1]}_{0\epsilon}g^{[1]}_{0\lambda}\Big)+
 {1\over2} \gamma^{\alpha\beta}\Big(g^{[2]}_{00} +\gamma^{\epsilon\lambda} \big(g^{[2]}_{\epsilon\lambda}-g^{[1]}_{0\epsilon}g^{[1]}_{0\lambda}\big)\Big).
\label{eq:GC_metr_ab}
\end{eqnarray}

As a result, the harmonic conditions Eqs.~(\ref{eq:(DeDgGa)OK0a1})--(\ref{eq:(DeDgGa)OKab}) take the following form:
{}
\begin{eqnarray}
\frac{\partial}{\partial y_{a\epsilon}} g^{[1]}_{0\epsilon} &=& {\cal O}(c^{-4}),
 \label{eq:(DeDgGa)OK01}\\
\frac{1}{2}c\frac{\partial}{\partial y_a^0} \Big\{g^{[2]}_{00}
-\gamma^{\epsilon\lambda}\Big(g^{[2]}_{\epsilon\lambda}
+g^{[1]}_{0\epsilon}g^{[1]}_{0\lambda}\Big)\Big\} +\frac{\partial}{\partial y_{a\nu}} \Big\{g^{[3]}_{0\nu}
+\frac{1}{2}g^{[1]}_{0\nu}\Big(\gamma^{\epsilon\lambda}\big(g^{[2]}_{\epsilon\lambda}+g^{[1]}_{0\epsilon}g^{[1]}_{0\lambda}\big)- g^{[2]}_{00}\Big)-
\gamma^{\epsilon\lambda}
g^{[1]}_{0\epsilon}g^{[2]}_{\nu\lambda}\Big\} &=& {\cal O}(c^{-2}), \hskip 20pt
 \label{eq:(DeDgGa)OK0}\\
c\frac{\partial}{\partial y_a^0} g^{[1]}_{0\alpha}+\frac{\partial}{\partial y_{a\beta}}
\Big\{g^{[2]}_{\alpha\beta}
-g^{[1]}_{0\alpha}g^{[1]}_{0\beta}
-{1\over2} \gamma_{\alpha\beta}\Big(g^{[2]}_{00} +\gamma^{\epsilon\lambda} \big(g^{[2]}_{\epsilon\lambda}-g^{[1]}_{0\epsilon}g^{[1]}_{0\lambda}\big)\Big)\Big\} &=& {\cal O}(c^{-2}).
 \label{eq:(DeDgGa)OKa}
\end{eqnarray}

A general solution to Eq.~(\ref{eq:(DeDgGa)OK01}) may be presented in the form below:
{}
\begin{equation}
g^{[1]}_{0\alpha}(y_a)=\mu_\alpha(y^0_a)+\epsilon_{\alpha\nu}(y^0_a)y^\nu_a+{\cal O}(c^{-4}),
\label{eq:form-inv_eta_0a}
\end{equation}
where $\mu_\alpha$ and $\epsilon_{\alpha\nu}$ (with $\epsilon_{\alpha\nu}=-\epsilon_{\nu\alpha}$) are arbitrary functions of the time-like coordinate $y^0_a$. The function $\mu_\alpha$ may be interpreted as a rate of time shift in the chosen reference frame, while $\epsilon_{\alpha\nu}$ represents a rotation and a uniform shear of the coordinate axis. By an additional gauge transformation of the coordinates to a co-moving reference frame, we can always eliminate $\mu_\alpha$ without loss of generality. We can also require the proper reference frame of a moving observer to exhibit no Newtonian rotation or shear of its coordinate axes. In addition to the fact that the leading term in the mixed components of the energy-momentum tensor is of the order $S_{0\alpha}\propto c$ (see Eq.~(\ref{eq:S0a_tranf-loc-ctv-2}) below), in terms of the metric tensor, these two physical conditions are equivalent to a requirement that all mixed components of the metric tensor of the proper reference frame at the $c^{-1}$ order vanish. This represents the anticipated  behavior of the metric as $r\rightarrow\infty$, allowing us to derive the conditions on $g_{0\alpha}$ and $g_{\alpha\beta}$ from the no-incoming-wave condition (\ref{eq:bound-cond}). In other words, $g^{[1]}_{0\alpha}$ has the form:
\begin{equation}
g^{[1]}_{0\alpha}={\cal O}(c^{-4}).
\label{eq:form-inv_eta_0a+}
\end{equation}
This condition leads to a significant simplification of the system of equations~(\ref{eq:(DeDgGa)OK01})--(\ref{eq:(DeDgGa)OKa}):
\begin{eqnarray}
g^{[1]}_{0\alpha}&=&{\cal O}(c^{-4}),
 \label{eq:(DeDgGa)OKs01+}\\
\frac{1}{2}c\frac{\partial}{\partial y_a^0} \Big\{g^{[2]}_{00}
-\gamma^{\epsilon\lambda}g^{[2]}_{\epsilon\lambda}\Big\} +\frac{\partial}{\partial y_{a\nu}} g^{[3]}_{0\nu} &=& {\cal O}(c^{-2}), \hskip 20pt
 \label{eq:(DeDgGa)OKs0+}\\
\frac{\partial}{\partial y_{a\beta}} \Big\{g^{[2]}_{\alpha\beta}
-{1\over2} \gamma_{\alpha\beta}\Big(g^{[2]}_{00} +\gamma^{\epsilon\lambda}g^{[2]}_{\epsilon\lambda}\Big)\Big\} &=& {\cal O}(c^{-2}).
 \label{eq:(DeDgGa)OKsa+}
\end{eqnarray}

We note that, typically, researchers use only the last two of these equations. As we discussed above, the gauge conditions are used to constrain the Ricci tensor's degrees of freedom. For these purposes the harmonic gauge conditions (\ref{eq:(DeDg)OK}) reduce to the following two equations (identical to Eqs.~(\ref{eq:(DeDg)0})--(\ref{eq:(DeDg)a})):
{}
\begin{eqnarray}
{1\over2}c\frac{\partial}{\partial y_a^0} \Big(\gamma^{\epsilon\nu} g^{[2]}_{\epsilon\nu} -
g^{[2]}_{00}\Big) - \frac{\partial}{\partial y_{a\nu}} g^{[3]}_{0\nu}
&=& {\cal O}(c^{-2}),
 \label{eq:harm-gauge-0}	\\
\frac{1}{2}\frac{\partial}{\partial y_{a\alpha}} \Big(g^{[2]}_{00} +
\gamma^{ \epsilon\nu} g^{[2]}_{\epsilon\nu}\Big) -
\gamma^{\alpha\mu}\frac{\partial}{\partial y_{a\nu}} g^{[2]}_{\mu\nu}
&=& {\cal O}(c^{-2}).
 \label{eq:harm-gauge-a}
\end{eqnarray}
However, as we shall see shortly, the harmonic gauge conditions offer more than is typically used. Below we shall explore these additional possibilities. For this, by formally integrating Eq.~(\ref{eq:(DeDgGa)OKsa+}), we get
\begin{equation}
g^{[2]}_{\alpha\beta}
-{1\over2} \gamma_{\alpha\beta}\Big(g^{[2]}_{00} +\gamma^{\epsilon\lambda}g^{[2]}_{\epsilon\lambda}\Big) = \mu_{\alpha\beta}(y^0)+{\cal O}(c^{-2}),
 \label{eq:(DeDgGa)OKsaf}
\end{equation}
where the symmetric tensor $\mu_{\alpha\beta}$ is an arbitrary function of the time-like coordinate $y^0$ only, representing a uniform scale expansion of the spatial part of the symmetric metric $g_{mn}$. After some algebra,  Eq.~(\ref{eq:(DeDgGa)OKsaf}) may be rewritten in the following form
\begin{equation}
g^{[2]}_{\alpha\beta}+\gamma_{\alpha\beta}g^{[2]}_{00}
=\mu_{\alpha\beta}(y^0_a)-\gamma_{\alpha\beta}\gamma^{\epsilon\lambda}\mu_{\epsilon\lambda}(y^0_a)+{\cal O}(c^{-2}).
 \label{eq:(DeDgGa)OKsaf+}
\end{equation}

The asymptotic boundary conditions (\ref{eq:bound-cond}), representing the assumption that no incoming waves fall on the system, require that the arbitrary function of time $\mu_{\alpha\beta}$ that enters Eq.~(\ref{eq:(DeDgGa)OKsaf+}) must be zero, namely
$\mu_{\alpha\beta}(y^0_a)=0$.
Therefore, the chosen coordinates are isotropic, and we are led to the following form of the gauge conditions (\ref{eq:(DeDgGa)OKs01+})--(\ref{eq:(DeDgGa)OKsa+}):
{}
\begin{eqnarray}
g^{[1]}_{0\alpha}&=&{\cal O}(c^{-4}),
 \label{eq:(DeDgGa)OKs01}\\
2c\frac{\partial}{\partial y_a^0} g^{[2]}_{00} +\frac{\partial}{\partial y_{a\nu}} g^{[3]}_{0\nu} &=&
{\cal O}(c^{-2}), \hskip 20pt
 \label{eq:(DeDgGa)OKs0}\\
g^{[2]}_{\alpha\beta}+\gamma_{\alpha\beta}g^{[2]}_{00}&=& {\cal O}(c^{-2}).
 \label{eq:(DeDgGa)OKsa}
\end{eqnarray}
Compare these new expressions for the harmonic gauge conditions in the 1PNA to those given by Eqs.~(\ref{eq:(DeDg)0})--(\ref{eq:(DeDg)a}). Although both forms are similar, Eqs.~(\ref{eq:(DeDgGa)OKs01})--(\ref{eq:(DeDgGa)OKsa}) were obtained relying on the asymptotic boundary conditions (\ref{eq:bound-cond}) and, thus, offer more insight regarding the properties of the metric tensor.

Equivalently, for the contravariant form of the metric from Eq.~(\ref{eq:(DeDg)OK-ctv}), we obtain:
\begin{eqnarray}
g^{[1]0\alpha}&=&{\cal O}(c^{-4}),
 \label{eq:(DeDgGa)OKs01-ctv}\\
2c\frac{\partial}{\partial x^0} g^{[2]00} +\frac{\partial}{\partial x^\nu} g^{[3]0\nu} &=&
{\cal O}(c^{-2}), \hskip 20pt
 \label{eq:(DeDgGa)OKs0-ctv}\\
g^{[2]\alpha\beta}+\gamma^{\alpha\beta}g^{[2]00}&=& {\cal O}(c^{-2}).
 \label{eq:(DeDgGa)OKsa-ctv}
\end{eqnarray}
The two sets of gauge conditions presented above restrict the components of the same metric tensor in coordinates introduced in local and global frames. The gauge conditions in this form establish the foundation of our method of constructing a proper reference frame of a moving observer in the gravitational $N$-body problem.

\subsubsection{Constraining the form of the admissible coordinate transformations}
\label{seq:harmform}

We now explore the alternative form of the harmonic gauge given by Eq.~(\ref{eq:(DeDg)x0-box}). Substituting the coordinate transformations (\ref{eq:trans-0})--(\ref{eq:trans-a}) into Eq.~(\ref{eq:(DeDg)x0-box}), we can see that the harmonic gauge conditions restrict the coordinate transformation functions ${\cal K}_a, {\cal L}_a$ and ${\cal Q}^\alpha_a$ (as opposed to restricting the components of the metric tensor, discussed in Sec.~\ref{sec:harm_cond-metr}) such that they must satisfy the following set of second order partial differential equations:
{}
\begin{eqnarray}
\gamma^{\epsilon\lambda}\frac{\partial^2 {\cal K}_a}{\partial y^\epsilon_a \partial y^\lambda_a}
&=& {\cal O}(c^{-4}), \label{eq:DeDoGa-K}\\
c^2\frac{\partial^2 {\cal K}_a}{\partial {y^0_a}^2}+\gamma^{\epsilon\lambda}\frac{\partial^2 {\cal L}_a}{\partial y^\epsilon_a \partial y^\lambda_a}
&=& {\cal O}(c^{-2}), \label{eq:DeDoGa-L}\\
a_{a_0}^\alpha+\gamma^{\epsilon\lambda}\frac{\partial^2 {\cal Q}^\alpha_a}{\partial y^\epsilon_a \partial y^\lambda_a} &=& {\cal O}(c^{-2})
\label{eq:DeDoGa-Q}.
\end{eqnarray}

Analogously, for the inverse transformation functions from Eq.~(\ref{eq:(DeDg)OK-ctv}) we obtain:
\begin{eqnarray}
\gamma^{\epsilon\lambda}\frac{\partial^2 \hat{\cal K}_a}{\partial x^\epsilon \partial x^\lambda}&=&{\cal O}(c^{-4}), \label{eq:DeDoGa-K-hat}\\
c^2\frac{\partial^2 \hat{\cal K}_a}{\partial {x^0}^2}+\gamma^{\epsilon\lambda}\frac{\partial^2 \hat{\cal L}_a}{\partial x^\epsilon \partial x^\lambda}&=&{\cal O}(c^{-2}), \label{eq:DeDoGa-L-hat}\\
{}
-a_{a_0}^\alpha+\gamma^{\epsilon\lambda}\frac{\partial^2 \hat{\cal Q}^\alpha_a}{\partial x^\epsilon \partial x^\lambda}&=&{\cal O}(c^{-2})
\label{eq:DeDoGa-Q-hat}.
\end{eqnarray}

As discussed at the beginning of this section, Eqs.~(\ref{eq:(DeDgGa)OKs01})--(\ref{eq:(DeDgGa)OKsa}) provide valuable constraints on the form of the metric tensor. As a matter of fact, these equations provide two additional conditions on ${\cal K}_a$ and ${\cal Q}^\alpha_a$. It follows from Eqs.~(\ref{eq:(DeDgGa)OKs01}), (\ref{eq:(DeDgGa)OKsa}) and the form of the metric tensor $g_{mn}$ in the local frame given by Eqs.~(\ref{eq:g00-loc_cov})--(\ref{eq:gab-loc_cov}) (which are yet to be discussed in Sec.~\ref{sec:grav-field-eqs}) that these two functions must also satisfy the following two first order partial differential equations:
{}
\begin{eqnarray}
\frac{1}{c}{\partial {\cal K}_a\over\partial y^\alpha_a} +
v^{}_{a_0\alpha}
&=&{\cal O}(c^{-4}),
\label{eq:form-inv_0a} \\
\frac{1}{c}{\partial {\cal K}_a\over\partial y^\alpha_a}
\frac{1}{c}{\partial {\cal K}_a\over\partial y^\beta_a}+\gamma_{\alpha\lambda}\frac{\partial {\cal Q}^\lambda_a}{\partial y^\beta_a}+\gamma_{\beta\lambda}\frac{\partial {\cal Q}^\lambda_a}{\partial y^\alpha_a}+2\gamma_{\alpha\beta}\Big(\frac{\partial {\cal K}_a}{\partial y^0_a}+
{\textstyle\frac{1}{2}}v^{}_{a_0\epsilon} v_{a_0}^\epsilon\Big) &=& {\cal O}(c^{-2}),
\label{eq:form-inv_ab}
\end{eqnarray}
where Eq.~(\ref{eq:form-inv_0a}) signifies the asymptotic absence of Newtonian rotation and shear of coordinate axis in the proper reference frame of an accelerated observer, while Eq.~(\ref{eq:form-inv_ab}) represents asymptotic isotropy of the chosen coordinates.

Analogously, the inverse functions $\hat{\cal K}_a$ and $\hat{\cal Q}^\alpha_a$ satisfy
\begin{eqnarray}
\gamma^{\alpha\epsilon}\frac{1}{c}{\partial \hat{\cal K}_a\over\partial x^\epsilon}-v_{a_0}^\alpha &=&{\cal O}(c^{-4}),
\label{eq:form-inv_0a+} \\
v_{a_0}^\alpha v_{a_0}^\beta+\gamma^{\alpha\lambda}
\frac{\partial \hat {\cal Q}^\beta_a}{\partial x^\lambda}+\gamma^{\beta\lambda}\frac{\partial \hat {\cal Q}^\alpha_a}{\partial x^\lambda}
+2\gamma_{\alpha\beta}\Big(\frac{\partial \hat {\cal K}_a}{\partial x^0}+ {\textstyle\frac{1}{2}}\gamma^{\epsilon\lambda}\frac{1}{c}
{\partial \hat{\cal K}_a\over\partial x^\epsilon}\frac{1}{c}
{\partial \hat{\cal K}_a\over\partial x^\lambda}\Big) &=& {\cal O}(c^{-2}).
\label{eq:form-inv_ab+}
\end{eqnarray}

The two sets of partial differential equations on ${\cal K}_a, {\cal L}_a$ and ${\cal Q}^\alpha_a$ given by Eqs.~(\ref{eq:DeDoGa-K})--(\ref{eq:DeDoGa-Q}) and (\ref{eq:form-inv_0a})--(\ref{eq:form-inv_ab}) can be used to determine the general structure of ${\cal K}_a$ and ${\cal Q}_a^\alpha$. Similarly, Eqs.~(\ref{eq:DeDoGa-K-hat})--(\ref{eq:DeDoGa-Q-hat}) and (\ref{eq:form-inv_0a+})--(\ref{eq:form-inv_ab+}) determine the general structure of $\hat{\cal K}_a$ and $\hat{\cal Q}^\alpha_a$. These will be required in order to solve the gravitational field equations. However, we must first develop the solution to the $N$-body problem in the quasi-inertial barycentric coordinate reference system.

\subsection{Solution for the gravitational interaction term in the global frame}
\label{sec:interact-term}

In order to find the gravitational interaction term $h^{mn}_{\rm int}(x)$ in Eq.~(\ref{eq:ansatz}) and, thus, to determine the solution for the $N$-body problem in the barycentric frame, we will use the field equations of the general theory of relativity (\ref{eq:GR-eq}). To do this, first of all we need to obtain the components of the metric tensor (\ref{eq:ansatz}) in the barycentric reference system. This could be done by using the solution for the unperturbed one-body problem presented by Eq.~(\ref{eq:g^mn1B}) (or, equivalently, Eq.~(\ref{eq:gab1B*})), the transformation rules for coordinate base vectors given by Eqs.(\ref{eq:(C1a)})--(\ref{eq:(C1d)}). Using all these expressions in Eq.~(\ref{eq:ansatz}) results in the following structure of the metric components in the global (barycentric) reference frame:
{}
\begin{eqnarray}
g^{00}(x)&=& 1+\frac{2}{c^2}\sum_b\Big\{w_b(y_b(x))
\big(1-\frac{2}{c^2}v^{}_{b_0\epsilon}v^\epsilon_{b_0}\big)-
\frac{4}{c^2}v^{}_{b_0\epsilon}w^\epsilon_b(y_b(x))+\nonumber\\
&&\hskip 40pt +
\frac{2}{c^2}w_b(y_b(x))
\Big(\frac{\partial {\cal K}_b}{\partial y^0_b}+
{\textstyle\frac{1}{2}}v^{}_{b_0\epsilon}v^\epsilon_{b_0}\Big)+
\frac{1}{c^2}w^2_b(y_b(x))\Big\}+\frac{1}{c^4}h^{[4]00}_{\rm int}(x)+{\cal O}(c^{-6}),
\label{eq:g00-trans-brf}\\[3pt]
{}
g^{0\alpha}(x)&=& \frac{4}{c^3}\sum_b \Big\{w^\alpha(y_b(x))+
v^\alpha_{b_0}w_b(y_b(x))\Big\}+{\cal O}(c^{-5}),
\label{eq:g0a-trans-brf}\\[3pt]
{}
g^{\alpha\beta}(x)&=& \gamma^{\alpha\beta}-
\gamma^{\alpha\beta}\frac{2}{c^2}\sum_b w_b(y_b(x))+{\cal O}(c^{-4}).
\label{eq:gab-trans-brf}
\end{eqnarray}

Next, we need to specify the ``source term'' for the $N$-body problem on the right-hand side of Eq.~(\ref{eq:GR-eq}). Similarly to (\ref{eq:ansatz}), this term may be written in the following form:
\begin{equation}
S^{mn}(x)=\sum_b \frac{\partial x^m}{\partial y^k_b}\frac{\partial x^n}{\partial y^l_b}~ S^{kl}_{b}(y_b(x)),
\label{eq:N-source^mn}
\end{equation}
where $S^{kl}_{b}$ is the ``source term'' for a particular body $b$, given by Eqs.~(\ref{eq:(S00-sig)})--(\ref{eq:(sig_a)}), which was used to derive the one-body solution (\ref{eq:g^mn1B}).

By using the source term (\ref{eq:N-source^mn}) and the transformation rules for coordinate base vectors given by Eqs.~(\ref{eq:(C1a)})--(\ref{eq:(C1d)}), and definitions Eqs.~(\ref{eq:(S00-sig)})--(\ref{eq:(sig_a)}), we obtain the components of the source term $S^{mn}$ in the barycentric reference frame:
{}
\begin{eqnarray}
S^{00}(x)&=& \frac{1}{2}c^2\sum_b\Big\{\Big(1 -
\frac{2}{c^2}v^{}_{b_0\epsilon}v^\epsilon_{b_0}\Big)\sigma_b(y_b(x))-
\frac{4}{c^2}v^{}_{b_0\epsilon} \sigma^\epsilon_b(y_b(x))
+\frac{2}{c^2}
\big(\frac{\partial {\cal K}_b}{\partial y^0_b}+
{\textstyle\frac{1}{2}}v^{}_{b_0\epsilon}v^\epsilon_{b_0}\big)
\sigma_b(y_b(x))+{\cal O}(c^{-4})\Big\},
\label{eq:S00_tranf-brf}\\[3pt]
{}
S^{0\alpha}(x)&=& c\sum_b\Big\{\sigma^{\alpha}_b(y^k_b(x))+
v^\alpha_{b_0}\sigma_b(y_b(x))+{\cal O}(c^{-2})\Big\},
\label{eq:S0a_tranf-brf}\\[3pt]
{}
S^{\alpha\beta}(x)&=& -\gamma^{\alpha\beta}\frac{1}{2}c^2\sum_b
\Big\{\sigma_b(y_b(x))+{\cal O}(c^{-2})\Big\}.
\label{eq:Sab_tranf-brf}
\end{eqnarray}
Comparing with Eqs.~(\ref{eq:(S00-sig)})--(\ref{eq:(Sab-sig)}), these expressions introduce transformation rules for the mass (scalar) and current (vector) densities of a particular body $b$ under a coordinate transformation from local, $\{y^m_b\}$, to global, $\{x^m\}$,  coordinates. Thus, defining the scalar, $\hat\sigma_b(x)$, and vector, $\hat\sigma^\alpha_b(x)$, densities for a particular body $b$ in the coordinates of the global
frame\footnote{The ``hat'' above any function, for instance $\hat\sigma_b(x)$ and $\hat\sigma^\alpha_b(x)$, indicates the fact that this quantity is defined in the global frame.} the corresponding transformation rules for the relativistic densities are given as
\begin{eqnarray}
\hat\sigma_b(x)&=& \Big(1 -
\frac{2}{c^2}v^{}_{b_0\epsilon}v^\epsilon_{b_0}\Big)\sigma_b(y_b(x))-
\frac{4}{c^2}v^{}_{b_0\epsilon} \sigma^\epsilon_b(y_b(x))
+\frac{2}{c^2}
\big(\frac{\partial {\cal K}_b}{\partial y^0_b}+
{\textstyle\frac{1}{2}}v^{}_{b_0\epsilon}v^\epsilon_{b_0}\big)
\sigma_b(y_b(x))+{\cal O}(c^{-4}),
\label{eq:sig0_tranf}\\[3pt]
{}
\hat\sigma^{\alpha}_b(x)&=& \sigma^{\alpha}_b(y^k_b(x))+
v^\alpha_{b_0}\sigma_b(y_b(x))+{\cal O}(c^{-2}).
\label{eq:siga_tranf}
\end{eqnarray}
The results (\ref{eq:sig0_tranf})--(\ref{eq:siga_tranf}), allow
presenting the source term $S^{mn}(x)$ in the barycentric reference frame (\ref{eq:S00_tranf-brf})--(\ref{eq:Sab_tranf-brf}) as:
\begin{eqnarray}
S^{00}(x)&=& \frac{1}{2}c^2\sum_b\Big\{\hat\sigma_b(x)+{\cal O}(c^{-4})\Big\}, \quad
S^{0\alpha}(x)= c\sum_b\Big\{\hat\sigma^{\alpha}_b(x)+{\cal O}(c^{-2})\Big\}, \quad
S^{\alpha\beta}(x)= -\gamma^{\alpha\beta}S^{00}(x)+{\cal O}(c^{-2}).~~~~~~~
\label{eq:Sab_tranf-glob}
\end{eqnarray}

The components of the Ricci tensor in the global frame can be directly calculated from Eqs.~(\ref{eq:(1R00)})--(\ref{eq:(1Rab)}):
\begin{eqnarray}
R^{00}(x) &=& {1\over2} \Box_x
\Big(c^{-2}g^{[2]00}+ c^{-4}\Big\{g^{[4]00}-
{\textstyle\frac{1}{2}}\big(g^{[2]00}\big)^2\Big\}\Big) +
c^{-4}\frac{1}{2}\Big(g^{[2]\epsilon\lambda}+\gamma^{\epsilon\lambda}g^{[2]00}\Big)\frac{\partial^2 g^{[2]00}}{\partial x^\epsilon\partial x^\lambda}+{\cal O}(c^{-6}), \label{eq:(1R_00)}\\
{}
R^{0\alpha}(x) &=& c^{-3}{1\over2}\Delta_x g^{[3]0\alpha} + {\cal O}(c^{-5}),
\label{eq:(1R_0a)}\\
{}
R^{\alpha\beta}(x) &=& c^{-2}{1\over2} \Delta_x g^{[2]\alpha\beta}  + {\cal O}(c^{-4}), \label{eq:(1R_ab)}
\end{eqnarray}
where $\Box_x=\partial^2/(\partial x^0)^2+\Delta_x$ and $\Delta_x=\gamma^{\epsilon\lambda}\partial^2/\partial x^\epsilon\partial x^\lambda$ are the d'Alembert and Laplace operators with respect to $\{x^k\}$, correspondingly.

Substituting the components (\ref{eq:g00-trans-brf})--(\ref{eq:gab-trans-brf}) of the metric tensor  and the components (\ref{eq:S00_tranf-brf})--(\ref{eq:Sab_tranf-brf}) of the source term into the field equations (\ref{eq:GR-eq}) of general relativity with the Ricci tensor (\ref{eq:(1R_00)})--(\ref{eq:(1R_ab)}), we obtain the gravitational field equations of the general theory of relativity in the barycentric reference frame:
{}
\begin{eqnarray}
 \Box_x \Big\{\sum_b\Big[\big(1 -
\frac{2}{c^2}v^{}_{b_0\epsilon}v^\epsilon_{b_0}\big)w_b(y_b(x))
-\frac{4}{c^2}v^{}_{b_0\epsilon}w^\epsilon_b(y_b(x))
&+&\frac{2}{c^2}w_b(y_b(x))
\Big(\frac{\partial {\cal K}_b}{\partial y^0_b}+
{\textstyle\frac{1}{2}}v^{}_{b_0\epsilon}v^\epsilon_{b_0}\Big)+
\frac{1}{c^2}w^2_b(y_b(x))\Big]-~~~\nonumber\\
{}
- \frac{1}{c^2}\Big(\sum_bw_b(y_b(x))\Big)^2+
\frac{1}{2c^2}h^{[4]00}_{\rm int}(x)\Big\}
&=&
{4\pi G}\sum_b\Big\{\Big(1 -\frac{2}{c^2}v^{}_{b_0\epsilon}v^\epsilon_{b_0}\Big)\sigma_b(y_b(x))-\frac{4}{c^2}v^{}_{b_0\epsilon} \sigma^\epsilon_b(y_b(x))+
\nonumber\\
&&\hskip 20pt +~
\frac{2}{c^2}
\big(\frac{\partial {\cal K}_b}{\partial y^0_b}+
{\textstyle\frac{1}{2}}v^{}_{b_0\epsilon}v^\epsilon_{b_0}\big)
\sigma_b(y_b(x))\Big\}+{\cal O}(c^{-4}),
\label{eq:(1R00)-N}\\
{}
\Delta_x \sum_b
\Big\{w^\alpha_b(y_b(x))+v^\alpha_{b_0}w_b(y_b(x))\Big\}
&=& {4\pi G}\sum_b\Big\{\sigma^{\alpha}_b(y_b(x))+
v^\alpha_{b_0}\sigma_b(y_b(x))\Big\}+{\cal O}(c^{-2}),
\label{eq:(1R0a)-N}\\
{}
\Delta_x \sum_b w_b(y_b(x)) &=& {4\pi G}\sum_b
\sigma_b(y_b(x))+{\cal O}(c^{-2}). \label{eq:(1Rab)-N}
\end{eqnarray}

We observe that Eqs.~(\ref{eq:(1R0a)-N}) and (\ref{eq:(1Rab)-N}) are satisfied identically. We will focus on Eq.~(\ref{eq:(1R00)-N}) with the aim to determine the interaction term. For these purposes,  we use Eqs.~(\ref{eq:(1R0a)-N}) and (\ref{eq:(1Rab)-N}) and rewrite (\ref{eq:(1R00)-N}) in the following form:
{}
\begin{eqnarray}
 \Box_x \Big\{\sum_b\Big[w_b(y_b(x))
&+&\frac{2}{c^2}\Big(\frac{\partial {\cal K}_b}{\partial y^0_b}+
{\textstyle\frac{1}{2}}
v^{}_{b_0\epsilon}v^\epsilon_{b_0}\Big)w_b(y_b(x))+
\frac{1}{c^2}w^2_b(y_b(x))\Big]-
\frac{1}{c^2}\Big(\sum_bw_b(y_b(x))\Big)^2+
\frac{1}{2c^2}h^{[4]00}_{\rm int}(x)\Big\}=\nonumber\\
{}
&=&
{4\pi G}\sum_b\Big\{\sigma_b(y_b(x))+\frac{2}{c^2}\Big(
\frac{\partial {\cal K}_b}{\partial y^0_b}+
{\textstyle\frac{1}{2}}v^{}_{b_0\epsilon}v^\epsilon_{b_0}\Big)
\sigma_b(y_b(x))\Big\}+{\cal O}(c^{-4}).
\label{eq:(1R00)-N1}
\end{eqnarray}

To analyze this equation, it is helpful to express the d'Alembert operator $\Box_x$ via its counterpart $\Box_{y_a}$ expressed in coordinates $y_a$ associated with body $a$. Using Eqs.~(\ref{eq:trans-0})--(\ref{eq:trans-a}), one easily obtains the following relation:
{}
\begin{eqnarray}
\frac{\partial^2}{\partial {x^0}^2}+\gamma^{\epsilon\lambda}
\frac{\partial^2}{\partial x^\epsilon \partial x^\lambda}
=
\frac{\partial^2}{\partial {y^0_a}^2}+\gamma^{\epsilon\lambda}
\frac{\partial^2}{\partial y^\epsilon_a \partial y^\lambda_a}&-&
\frac{1}{c^2}\Big\{a_{a_0}^\mu+\gamma^{\epsilon\lambda}
\frac{\partial^2 {\cal Q}^\mu_a}{\partial y^\epsilon_a \partial y^\lambda_a}\Big\}\frac{\partial}{\partial y^\mu_a}-\nonumber\\
&-&
\frac{1}{c^2}\Big\{v_{a_0}^\epsilon v_{a_0}^\lambda+\gamma^{\epsilon\mu}\frac{\partial {\cal Q}^\lambda_a}{\partial y^\mu_a}+\gamma^{\lambda\mu}\frac{\partial {\cal Q}^\epsilon_a}{\partial y^\mu_a}\Big\}\frac{\partial^2}{\partial y^\epsilon_a\partial y^\lambda_a}+{\cal O}(c^{-4}).
\label{eq:D'Alemb_x}
\end{eqnarray}
The terms of order $c^{-2}$ on the right-hand side of this expression can be further simplified with the help of harmonic gauge conditions. Indeed, using Eqs.~(\ref{eq:DeDoGa-Q}), (\ref{eq:form-inv_0a}) and (\ref{eq:form-inv_ab}) the expression above takes the following form:
{}
\begin{eqnarray}
\frac{\partial^2}{\partial {x^0}^2}+\gamma^{\epsilon\lambda}
\frac{\partial^2}{\partial x^\epsilon \partial x^\lambda}
&=&
\frac{\partial^2}{\partial {y^0_a}^2}+\gamma^{\epsilon\lambda}
\frac{\partial^2}{\partial y^\epsilon_a \partial y^\lambda_a}+
\frac{2}{c^2}\Big(\frac{\partial {\cal K}_a}{\partial y^0_a}+
{\textstyle\frac{1}{2}}v_{a_0}{}_\epsilon v_{a_0}^\epsilon\Big)
\gamma^{\epsilon\lambda}
\frac{\partial^2}{\partial y^\epsilon_a \partial y^\lambda_a}+{\cal O}(c^{-4}).
\label{eq:D'Alemb_x+}
\end{eqnarray}
The last equation may be given in the equivalent form:
{}
\begin{eqnarray}
\Box_x &=& \Big\{1+
\frac{2}{c^2}\Big(\frac{\partial {\cal K}_a}{\partial y^0_a}+
{\textstyle\frac{1}{2}}v^{}_{a_0\epsilon} v_{a_0}^\epsilon\Big)
+{\cal O}(c^{-4})\Big\} \Box_{y_a}.
\label{eq:D'Alemb_y+=+}
\end{eqnarray}
Therefore, taking into account that, for any given $y_b$, the following relation holds
\begin{eqnarray}
 \Box_{y_b} w_b(y_b(x))
&=&
{4\pi G}\sigma_b(y_b(x))+{\cal O}(c^{-4}),
\label{eq:(1R00)-N1+00}
\end{eqnarray}
hence the following result is valid
for any body $b$:
\begin{eqnarray}
 \Box_x w_b(y_b(x))
&=&
{4\pi G}\sigma_b(y_b(x))\Big\{1+\frac{2}{c^2}\Big(
\frac{\partial {\cal K}_b}{\partial y^0_b}+
{\textstyle\frac{1}{2}}v^{}_{b_0\epsilon}v^\epsilon_{b_0}\Big)+{\cal O}(c^{-4})\Big\}.
\label{eq:(1R00)-N1+}
\end{eqnarray}

With the help of Eq.~(\ref{eq:(1R00)-N1+}) we can now rewrite Eq.~(\ref{eq:(1R00)-N1}) as
{}
\begin{eqnarray}
 \Delta_x \Big\{\sum_b\Big[\frac{2}{c^2}
\Big(\frac{\partial {\cal K}_b}{\partial y^0_b}+
{\textstyle\frac{1}{2}}v^{}_{b_0\epsilon}v^\epsilon_{b_0}\Big)w_b(y_b(x))+
\frac{1}{c^2}w^2_b(y_b(x))\Big]-
\frac{1}{c^2}\Big(\sum_bw_b(y_b(x))\Big)^2+
\frac{1}{2c^2}h^{[4]00}_{\rm int}(x)\Big\}={\cal O}(c^{-4}).\label{eq:(1R00)-N2}
\end{eqnarray}

This equation can be solved for $h^{[4]00}_{\rm int}(x)$. The solution to this partial differential equation is composed of two families: one is a particular solution that can be found by equating the entire expression under the D'Alembert operator to zero, $h^{[4]00}_{\rm int (0)}(x)$, while the other is a family of solutions involving symmteric and trace free (STF) tensors that are divergent at spatial infinity:
{}
\begin{eqnarray}
h^{[4]00}_{\rm int}(x)=h^{[4]00}_{\rm int (0)}(x)+
\sum_{k\ge0}\frac{1}{k!}\delta h^{00}_{\rm int \,\mu_1...\mu_k}(x^0)x^{\mu_1}{}^{...}x^{\mu_k}+{\cal O}(|x^{\mu}|^K)+{\cal O}(c^{-2}),
\label{eq:h-int}
\end{eqnarray}
where $\delta h^{00}_{\rm int \,\mu_1...\mu_k}(x^0)$ are arbitrary  STF tensors that depend only on the time-like coordinate $x^0$.
Requiring that the solution to this equation be asymptotically flat and that there be no gravitational radiation coming from outside the system Eq.~(\ref{eq:bound-cond}), namely
\begin{equation}
\lim_{\substack{r\rightarrow\infty\\t+r/c={\rm const}}}h^{[4]00}_{\rm int}(x) = 0 ~~~ {\rm and} ~~~\lim_{\substack{r\rightarrow\infty\\t+r/c={\rm const}}}
\big[(rh^{[4]00}_{\rm int})_{,r}+(rh^{[4]00}_{\rm int})_{,0}\big]=0,
\label{eq:assympt-flat}
\end{equation}
implies that $\delta h^{00}_{\rm int \,\mu_1...\mu_k}(x^0)=0,$ for all $k$. Therefore, Eq.~(\ref{eq:(1R00)-N2}) yields the following solution for the interaction term:
\begin{eqnarray}
h^{[4]00}_{\rm int}(x)=2\Big\{\Big(\sum_bw_b(y_b(x))\Big)^2-
\sum_b w^2_b(y_b(x))\Big\}-
4\sum_bw_b(y_b(x))\Big(\frac{\partial {\cal K}_b}{\partial y^0_b}+
{\textstyle\frac{1}{2}}v^{}_{b_0\epsilon}v^\epsilon_{b_0}\Big)+{\cal O}(c^{-2}).
\label{eq:int-term}
\end{eqnarray}

The first two terms on the right-hand side of the equation above represent the usual composition of the gravitational interaction term for $N$-body system. The last term is the coupling of the local inertia field to the local gravity potential transformed to the global coordinates. The presence of this coupling reduces the total energy in the system, acting as a binding energy for the combined system of gravity and inertia.

Substituting the interaction term Eq.~(\ref{eq:int-term}) into the $g^{00}(x)$ component (\ref{eq:g00-trans-brf}) of the metric tensor we obtain the following expression for the metric tensor in the global frame:
\begin{eqnarray}
g^{00}(x)&=& 1+\frac{2}{c^2}\sum_b {\hat w}_b(x)+
\frac{2}{c^4}\Big(\sum_b {\hat w}_b(x)\Big)^2+{\cal O}(c^{-6}),
\label{eq:g00-bar_ctv_tot*}\\
g^{0\alpha}(x)&=& \frac{4}{c^3}\sum_b
{\hat w}^\alpha_b(x)+{\cal O}(c^{-5}),
\label{eq:g0a-bar_ctv_tot*}\\
g^{\alpha\beta}(x)&=& \gamma^{\alpha\beta}-\gamma^{\alpha\beta}\frac{2}{c^2} \sum_b {\hat w}_b(x)+{\cal O}(c^{-4}),\label{eq:gab-bar_ctv_tot*}
\end{eqnarray}
where the gravitational scalar, $\hat w_b$, and vector $\hat w_b^\alpha$ potentials of body $b$, expressed in the global coordinates $\{x^m\}$, have the following form (see also \cite{DSX-I}):
\begin{eqnarray}
{\hat w}_b(x)&=& \big(1 -
\frac{2}{c^2}v^{}_{b_0\epsilon}v^\epsilon_{b_0}\big)w_b(y_b(x))
-\frac{4}{c^2}v^{}_{b_0\epsilon}w^\epsilon_b(y_b(x))+{\cal O}(c^{-4}),\label{eq:w_0_ctv_bar}\\
{\hat w}^\alpha_b(x)&=& w^\alpha_b(y_b(x))+v^\alpha_{b_0}w_b(y_b(x))
+{\cal O}(c^{-2}). \label{eq:w_a_ctv_bar}
\end{eqnarray}
It is convenient to express these potentials in terms of integrals over the volumes of the bodies in the global coordinates.  To do this, we use Eqs.~(\ref{eq:(1R0a)-N}), (\ref{eq:(1Rab)-N}) and (\ref{eq:(1R00)-N1+}) to determine the following solutions for the scalar $w_b(y_b(x))$ and vector $w^\alpha_b(y_b(x))$ potentials that satisfy the boundary conditions given by Eq.~(\ref{eq:bound-cond}):
{}
\begin{eqnarray}
w_b(y_b(x))&=& G\int d^3x'
\frac{\sigma_b(y_b(x'))}{|\vec{x}-\vec{x}'|}+
\frac{1}{2c^2}G\frac{c^2\partial^2}{\partial {x^0}^2}\int d^3x'
{\sigma_b(y_b(x'))}{|\vec{x}-\vec{x}'|}+\nonumber\\
&&\hskip 86pt +~
\frac{2G}{c^2}\int d^3x'\frac{\sigma_b(y_b(x'))}{|\vec{x}-\vec{x'}|}
\Big(\frac{\partial {\cal K}'_b}{\partial y^0_b}+
{\textstyle\frac{1}{2}}
v^{}_{b_0\epsilon}v^\epsilon_{b_0}\Big)+{\cal O}(c^{-3}),
\label{eq:w_0-ctv_bar++}\\
w^\alpha_b(y_b(x))&=& G\int d^3x'\frac{\sigma^\alpha_b(y_b(x'))}{|\vec{x}-\vec{x}'|}+{\cal O}(c^{-2}).
\label{eq:w_a-ctv_bar+++}
\end{eqnarray}

Substituting these results into Eqs.~(\ref{eq:w_0_ctv_bar})--(\ref{eq:w_a_ctv_bar}) in the form of the integrals over the body's volume leads to the following expression for gravitational potentials of the body $b$ in the global coordinates of inertial reference frame:
{}
\begin{eqnarray}
{\hat w}_b(x)&=& \big(1 -
\frac{2}{c^2}v^{}_{b_0\epsilon}v^\epsilon_{b_0}\big)
G\int d^3x'\frac{\sigma_b(y_b(x'))}{|\vec{x}-\vec{x}'|}
-\frac{4}{c^2}v^{}_{b_0\epsilon}
G\int d^3x'\frac{\sigma^\epsilon_b
(y_b(x'))}{|\vec{x}-\vec{x}'|}+ \nonumber\\
&&\hskip 15pt +~
\frac{1}{2c^2}G\frac{c^2\partial^2}{\partial {x^0}^2}\int d^3x'
{\sigma_b(y_b(x'))}{|\vec{x}-\vec{x}'|}+
\frac{2G}{c^2}\int d^3x'\frac{\sigma_b(y_b(x'))}{|\vec{x}-\vec{x'}|}
\Big(\frac{\partial {\cal K}'_b}{\partial y^0_b}+
{\textstyle\frac{1}{2}}
v^{}_{b_0\epsilon}v^\epsilon_{b_0}\Big)+{\cal O}(c^{-3}),
\label{eq:w_0_ctv_bar+fin*}\\
{\hat w}^\alpha_b(x)&=& G\int d^3x'\frac{\sigma^\alpha_b(y_b(x'))}{|\vec{x}-\vec{x}'|}+
v^\alpha_{b_0}G\int d^3x'\frac{\sigma_b(y_b(x'))}{|\vec{x}-\vec{x}'|}
+{\cal O}(c^{-2}),
\label{eq:w_a_ctv_bar+fin*}
\end{eqnarray}
which are consistent with transformation rules for relativistic mass and current densities (\ref{eq:sig0_tranf})--(\ref{eq:siga_tranf}), correspondingly.

We have obtained a solution for the gravitational field equations for the $N$-body problem in the post-Newtonian approximation.  In the global (barycentric) reference frame this solution is represented by the metric tensor given by Eqs.~(\ref{eq:g00-bar_ctv_tot*})--(\ref{eq:gab-bar_ctv_tot*}) together with the scalar and vector potentials given by Eqs.~(\ref{eq:w_0_ctv_bar+fin*}) and (\ref{eq:w_a_ctv_bar+fin*}). Note that the form of the metric (\ref{eq:g00-bar_ctv_tot*})--(\ref{eq:gab-bar_ctv_tot*}) is identical to that of the isolated one-body problem (\ref{eq:g^mn1B}), except that these two structurally-identical metric tensors have different  potentials. Such form-invariance of the metric tensor is due to harmonic gauge conditions and the covariant structure of the total metric sought in the form of Eq.~(\ref{eq:ansatz}) together with the covariantly-superimposed form of the $N$-body source term Eq.~(\ref{eq:N-source^mn}).

In the following section we transform this solution to the local frame, verify that this solution satisfies the equations of general relativity in that frame, and improve the form of this solution using the harmonic gauge.

\section{Transformation to the local frame}
\label{sec:xformlocal}

We begin this section by investigating the transformation properties of the covariant form of the metric tensor from the global frame to the local frame. The use of the covariant form makes it possible to find the transformation functions ${\cal K}$, ${\cal L}$ and ${\cal Q}$ in Eqs.~(\ref{eq:trans-0}) and (\ref{eq:trans-a}) more expediently.

\subsection{Gravitational field equations in the local frame}
\label{sec:grav-field-eqs}

The metric tensor in the local frame can be obtained by applying the standard rules of tensor transformations to the metric tensor in the global frame given by Eqs.~(\ref{eq:g00-bar_ctv_tot*})--(\ref{eq:gab-bar_ctv_tot*}):
\begin{equation}
g_{mn}(y_a)=\frac{\partial x^k}{\partial y^m_a}\frac{\partial x^l}{\partial y^n_a}~g_{kl}(x(y_a)).
\label{eq:ansatz-loc_cov}
\end{equation}

Using the coordinate transformations given by Eqs.~(\ref{eq:trans-0})--(\ref{eq:trans-a}) together with Eqs.~(\ref{eq:(C1a)})--(\ref{eq:(C1d)}), we determine the form of the metric tensor in the local reference frame associated with a body $a$:
{}
\begin{eqnarray}
g_{00}(y_a)&=& 1+\frac{2}{c^2}\Big\{\frac{\partial {\cal K}_a}{\partial y^0_a}+{\textstyle\frac{1}{2}}v^{}_{a_0\epsilon} v_{a_0}^\epsilon\Big\}
+\frac{2}{c^4}\Big\{ \frac{\partial {\cal L}_a}{\partial y^0_a}+
{\textstyle\frac{1}{2}}\Big(\frac{\partial {\cal K}_a}{\partial y^0_a}\Big)^2+cv^{}_{a_0\epsilon}\frac{\partial {\cal Q}^\epsilon_a}{\partial y^0_a}-\Big(\frac{\partial {\cal K}_a}{\partial y^0_a}+
{\textstyle\frac{1}{2}}v^{}_{a_0\epsilon} v_{a_0}^\epsilon\Big)^2\Big\}-\nonumber\\[2pt]
{}
&&\hskip 8pt-~\frac{2}{c^2}\sum_b\Big\{\Big(1-\frac{2}{c^2}v^{}_{a_0\epsilon} v_{a_0}^\epsilon\Big)\hat w_b(x(y_a))+
\frac{4}{c^2}v^{}_{a_0\epsilon} \hat w^\epsilon_b(x(y_a))\Big\}+\nonumber\\[-5pt]
{}
&&\hskip 8pt+~
\frac{2}{c^4}\Big(\sum_b \hat w_b(x(y_a))-
\frac{\partial {\cal K}_a}{\partial y^0_a}- {\textstyle\frac{1}{2}}v_{a_0}{}_\epsilon v_{a_0}^\epsilon\Big)^{\!2}+{\cal O}(c^{-6}),\label{eq:g00-loc_cov}\\
{}
g_{0\alpha}(y_a)&=& \frac{1}{c}\Big(
\frac{1}{c}{\partial {\cal K}_a\over\partial y^\alpha_a}+
v^{}_{a_0\alpha}\Big) -\frac{4}{c^3}\gamma_{\alpha\epsilon}\sum_b \Big(\hat w^\epsilon_b(x(y_a))-v_{a_0}^\epsilon \hat w_b(x(y_a))\Big)+
\nonumber\\
&&\hskip 8pt+~
\frac{1}{c^3}\Big\{\frac{1}{c}\frac{\partial {\cal L}_a}{\partial y^\alpha_a}+c\gamma_{\alpha\epsilon}\frac{\partial {\cal Q}^\epsilon_a}{\partial y^0_a}+
v^{}_{a_0\lambda}\frac{\partial {\cal Q}^\lambda_a}{\partial y^\alpha_a}+\frac{1}{c}{\partial {\cal K}_a\over\partial y^\alpha_a}\frac{\partial {\cal K}_a}{\partial y^0_a}\Big\}+{\cal O}(c^{-5}),
\label{eq:g0a-loc_cov}\\[3pt]
{}
g_{\alpha\beta}(y_a)&=& \gamma_{\alpha\beta}+\frac{1}{c^2}\Big\{\frac{1}{c}{\partial {\cal K}_a\over\partial y^\alpha_a}
\frac{1}{c}{\partial {\cal K}_a\over\partial y^\beta_a}
+\gamma_{\alpha\lambda}\frac{\partial Q_a^\lambda}{\partial y^\beta_a}+\gamma_{\beta\lambda}\frac{\partial Q_a^\lambda}{\partial y^\alpha_a}\Big\}+\gamma_{\alpha\beta}\frac{2}{c^2} \sum_b \hat w_b(x(y_a))+{\cal O}(c^{-4}).
\label{eq:gab-loc_cov}
\end{eqnarray}

The form of the metric in the local reference frame given by Eqs.~(\ref{eq:g00-loc_cov})--(\ref{eq:gab-loc_cov}) still has a significant number of degrees of freedom reflecting the presence of the yet arbitrary transformation functions. Some of these degrees of freedom can be eliminated by imposing the harmonic gauge conditions. Eqs.~(\ref{eq:(DeDgGa)OKs01}) and (\ref{eq:(DeDgGa)OKsa}) can provide valuable constraints on the form of the metric tensor in the local frame. Indeed, using the actual structure of the metric expressed by Eqs.~(\ref{eq:g00-loc_cov})--(\ref{eq:gab-loc_cov}) in combination with the gauge conditions (\ref{eq:(DeDgGa)OKs01})--(\ref{eq:(DeDgGa)OKsa}), and along with Eqs.~(\ref{eq:form-inv_0a})--(\ref{eq:form-inv_ab}), we can now use these sets of constraints on the transformation functions and improve the form of the metric tensor in the local frame. As a result, the metric tensor in the harmonic coordinates of the local frame has the following form:
{}
\begin{eqnarray}
g_{00}(y_a)&=& 1+\frac{2}{c^2}\Big\{\frac{\partial {\cal K}_a}{\partial y^0_a}+{\textstyle\frac{1}{2}}v^{}_{a_0\epsilon} v_{a_0}^\epsilon\Big\}
+\frac{2}{c^4}\Big\{ \frac{\partial {\cal L}_a}{\partial y^0_a}+
{\textstyle\frac{1}{2}}\Big(\frac{\partial {\cal K}_a}{\partial y^0_a}\Big)^2+cv^{}_{a_0\epsilon}\frac{\partial {\cal Q}^\epsilon_a}{\partial y^0_a}-\Big(\frac{\partial {\cal K}_a}{\partial y^0_a}+
{\textstyle\frac{1}{2}}v^{}_{a_0\epsilon} v_{a_0}^\epsilon\Big)^2\Big\}-\nonumber\\[2pt]
{}
&&\hskip 8pt-~\frac{2}{c^2}\sum_b\Big\{\Big(1-\frac{2}{c^2}v^{}_{a_0\epsilon} v_{a_0}^\epsilon\Big)\hat w_b(x(y_a))+
\frac{4}{c^2}{v^{}_{a_0}}_{\!\epsilon} \hat w^\epsilon_b(x(y_a))\Big\}+\nonumber\\[-5pt]
{}
&&\hskip 8pt+~
\frac{2}{c^4}\Big(\sum_b \hat w_b(x(y_a))-
\frac{\partial {\cal K}_a}{\partial y^0_a}- {\textstyle\frac{1}{2}}v^{}_{a_0\epsilon} v_{a_0}^\epsilon\Big)^{\!2}+{\cal O}(c^{-6}),\label{eq:g00-loc+}\\
{}
g_{0\alpha}(y_a)&=& -\frac{4}{c^3}\gamma_{\alpha\epsilon}\sum_b \Big(\hat w^\epsilon_b(x(y_a))-v_{a_0}^\epsilon \hat w_b(x(y_a))\Big)
+
\nonumber\\
&&\hskip 8pt+~
\frac{1}{c^3}\Big\{\frac{1}{c}\frac{\partial {\cal L}_a}{\partial y^\alpha_a}+c\gamma_{\alpha\epsilon}\frac{\partial {\cal Q}^\epsilon_a}{\partial y^0_a}+
v^{}_{a_0\lambda}\frac{\partial {\cal Q}^\lambda_a}{\partial y^\alpha_a}-
v^{}_{a_0\alpha}\frac{\partial {\cal K}_a}{\partial y^0_a}\Big\}+{\cal O}(c^{-5}),~~~~~
\label{eq:g0a-loc+}\\[3pt]
{}
g_{\alpha\beta}(y_a)&=& \gamma_{\alpha\beta}+
\gamma_{\alpha\beta}\frac{2}{c^2} \Big\{\sum_b \hat w_b(x(y_a))-
\frac{\partial {\cal K}_a}{\partial y^0_a}- {\textstyle\frac{1}{2}}v^{}_{a_0\epsilon} v_{a_0}^\epsilon\Big\}+{\cal O}(c^{-4}).
\label{eq:gab-loc+}
\end{eqnarray}

The covariant components of the source term in the local reference frame, $S_{mn}(y_a)$, are related to the same in the global frame, $S_{mn}(x)$, via the usual tensor transformation:
\begin{equation}
S_{mn}(y_a)=\frac{\partial x^k}{\partial y^m_a}\frac{\partial x^l}{\partial y^n_a}~S_{kl}(x(y_a)).
\label{eq:N-source_low}
\end{equation}
On the other hand, given the definition for the scalar and vector densities (which are given by Eqs.~(\ref{eq:(sig)})--(\ref{eq:(sig_a)})), for which we need covariant components of the source term $S^{mn}$, we can determine $S_{mn}(y_a)$ in the local frame via the covariant components $S^{mn}$ in the following form
{}
\begin{equation}
S_{mn}(y_a)= g_{mk}(y_a)g_{nl}(y_a)S^{kl}(y_a)=g_{mk}(y_a)g_{nl}(y_a)\frac{\partial y^k_a}{\partial x^p}\frac{\partial y^l_a}{\partial x^q}~S^{pq}(x(y_a)),
\end{equation}
where ${\partial y^k_a}/{\partial x^p}$ are given by Eqs.~(\ref{eq:(C1a)_inv})--(\ref{eq:(C1d)_inv}), which lead to the following expression for the components of the source term $S_{mn}(y_a)$:
{}
\begin{eqnarray}
S_{00}(y_a)&=& \Big(1+\frac{2}{c^2}g^{[2]}_{00}(y_a)+{\cal O}(c^{-4})\Big)\times
\nonumber\\
&&\hskip -30pt \times~
\frac{1}{2}c^2\sum_b\Big\{\Big(1 -
\frac{2}{c^2}v_{a_0\epsilon}v^\epsilon_{a_0}\Big)\hat \sigma_b(x(y_a))+
\frac{4}{c^2}v_{a_0\epsilon} \hat \sigma^\epsilon_b(x(y_a))
+\frac{2}{c^2}
\big(\frac{\partial \hat{\cal K}_a}{\partial x^0}+
{\textstyle\frac{1}{2}}v_{a_0\epsilon}v^\epsilon_{a_0}\big)
\hat \sigma_b(x(y_a))+{\cal O}(c^{-4})\Big\},~~~
\label{eq:S00_tranf-loc-ctv-2}\\[3pt]
{}
S_{0\alpha}(y_a)&=& \gamma^{}_{\alpha\lambda}c\sum_b\Big\{\hat \sigma^{\lambda}_b(x(y_a))-
v^\lambda_{a_0}\hat \sigma_b(x(y_a))+{\cal O}(c^{-2})\Big\},
\label{eq:S0a_tranf-loc-ctv-2}\\[3pt]
{}
S_{\alpha\beta}(y_a)&=& -\gamma_{\alpha\beta}\frac{1}{2}c^2\Big\{\sum_b
\hat \sigma_b(x(y_a))+{\cal O}(c^{-2})\Big\},
\label{eq:Sab_tranf-loc-ctv-2}
\end{eqnarray}
where $\hat{\cal K}_a$ comes as a part of the rules Eqs.~(\ref{eq:(C1a)_inv})--(\ref{eq:(C1d)_inv}) that are developed for the inverse transformations $y^m=f^m(x^k)$.

To put the source term in the form (\ref{eq:Sab_tranf-glob}), we introduce transformation rules for the scalar and vector densities of a body $b$ under coordinate transformation from global, $\{x\}$, to local coordinates, $\{y_a\}$ of a particular body $a$:
{}
\begin{eqnarray}
\sigma_b(y_a)&=& \Big(1 -
\frac{2}{c^2}v_{a_0\epsilon}v^\epsilon_{a_0}\Big)\hat \sigma_b(x(y_a))+
\frac{4}{c^2}v_{a_0\epsilon} \hat \sigma^\epsilon_b(x(y_a))
+\frac{2}{c^2}
\big(\frac{\partial \hat{\cal K}_a}{\partial x^0}+
{\textstyle\frac{1}{2}}v_{a_0\epsilon}v^\epsilon_{a_0}\big)
\hat \sigma_b(x(y_a))+{\cal O}(c^{-4}),~~~
\label{eq:sig_0_tranf-loc}\\[3pt]
{}
\sigma^\alpha_b(y_a)&=& \hat \sigma^{\alpha}_b(x(y_a))-
v^\alpha_{a_0}\hat \sigma_b(x(y_a))+{\cal O}(c^{-2}),
\label{eq:sig_a_tranf-loc}
\end{eqnarray}
where $\hat \sigma_b(x)$ and $\hat \sigma^\alpha_b(x)$ are given by (\ref{eq:sig0_tranf})--(\ref{eq:siga_tranf}), correspondingly.

Similarly, to compute the Ricci tensor in the global frame, we use the following expression:
\begin{equation}
R_{mn}(y_a)= g_{mk}(y_a)g_{nl}(y_a)R^{kl}(y_a),
\end{equation}
with $R^{kl}(y_a)$ being expressed in terms of the covariant metric $g_{mn}$, $R^{kl}[g_{mn}(y_a)]$, namely:
{}
\begin{eqnarray}
R_{00}(y_a) &=& \Big(1+\frac{2}{c^2}g_{00}^{[2]}(x)+{\cal O}(c^{-4})\Big)\nonumber\\
&&\hskip2pt \times~
\Big(-{1\over2} \Box_{y_a}
\Big(c^{-2}g^{[2]}_{00}+ c^{-4}\Big\{g^{[4]}_{00}-
{\textstyle\frac{1}{2}}\big(g^{[2]}_{00}\big)^2\Big\}\Big) +
c^{-4}\frac{1}{2}\Big(g^{[2]}_{\epsilon\lambda}+\gamma_{\epsilon\lambda}g^{[2]}_{00}\Big)\frac{\partial^2 g^{[2]}_{00}}{\partial y_{a\epsilon}\partial y_{a\lambda}} +{\cal O}(c^{-6})\Big),
 \label{eq:(1R00)-loc_cov}\\
{}
R_{0\alpha}(y_a) &=& -c^{-3}{1\over2} \Delta_{y_a}
g^{[3]}_{0\alpha} + {\cal O}(c^{-5}),
\label{eq:(1R0a)-loc_cov}\\
{}
R_{\alpha\beta}(y_a) &=& -c^{-2}{1\over2}
\Delta_{y_a}
g^{[2]}_{\alpha\beta}  + {\cal O}(c^{-4}),
\label{eq:(1Rab)-loc_cov}
\end{eqnarray}
where $\Box_{y_a}=\partial^2/(\partial y^0_a)^2+\Delta_{y_a}$ and $\Delta_{y_a} =\gamma^{\epsilon\lambda}\partial^2/\partial y_a^\epsilon\partial y_a^\lambda$ are the d'Alembert and Laplace operators with respect to the local coordinates $\{y^k_a\}$, correspondingly.

The metric tensor given by Eqs.~(\ref{eq:g00-loc+})--(\ref{eq:gab-loc+}) and the condition given by Eq.~(\ref{eq:(DeDgGa)OKsa}) allow us to present the gravitational field equations (\ref{eq:GR-eq}) of the general theory of relativity in local coordinates in the following form:
{}
\begin{eqnarray}
\Box_{y_a}\Big[\sum_b\Big\{\Big(1-\frac{2}{c^2}v^{}_{a_0\epsilon} v_{a_0}^\epsilon\Big)\hat w_b(x(y_a))&+&
\frac{4}{c^2}{v^{}_{a_0}}_{\!\epsilon} \hat w^\epsilon_b(x(y_a))\Big\}- \nonumber\\-
\frac{\partial {\cal K}_a}{\partial y^0_a}-
{\textstyle\frac{1}{2}}v^{}_{a_0\epsilon} v_{a_0}^\epsilon
-\frac{1}{c^2}\Big\{ \frac{\partial {\cal L}_a}{\partial y^0_a}&+&
{\textstyle\frac{1}{2}}\Big(\frac{\partial {\cal K}_a}{\partial y^0_a}\Big)^2+cv^{}_{a_0\epsilon}\frac{\partial {\cal Q}^\epsilon_a}{\partial y^0_a}-\Big(\frac{\partial {\cal K}_a}{\partial y^0_a}+
{\textstyle\frac{1}{2}}v^{}_{a_0\epsilon} v_{a_0}^\epsilon\Big)^2\Big\}+{\cal O}(c^{-4})\Big]=\nonumber\\[2pt]
= {4\pi G}\sum_b\Big\{\Big(1 -
\frac{2}{c^2}v^{}_{a_0\epsilon}v^\epsilon_{a_0}\Big)\hat \sigma_b(x(y_a))&+&
\frac{4}{c^2}v^{}_{a_0\epsilon} \hat \sigma^\epsilon_b(x(y_a))
+\frac{2}{c^2}
\big(\frac{\partial \hat{\cal K}_a}{\partial x^0}+
\textstyle{\frac{1}{2}}v^{}_{a_0\epsilon}v^\epsilon_{a_0}\big)
\hat \sigma_b(x(y_a))+{\cal O}(c^{-4})\Big\},~~~~
\label{eq:GR00_loc-cov}\\[3pt]
{}
\Delta_{y_a}  \Big[\sum_b \Big(\hat w^\alpha_b(x(y_a))-v_{a_0}^\alpha \hat w_b(x(y_a))\Big)&-&
{\textstyle\frac{1}{4}}\Big\{\frac{1}{c}\gamma^{\alpha\epsilon}\frac{\partial {\cal L}_a}{\partial y^\epsilon_a}+c\frac{\partial {\cal Q}^\alpha_a}{\partial y^0_a}+
\gamma^{\alpha\epsilon}v^{}_{a_0\lambda}\frac{\partial {\cal Q}^\lambda_a}{\partial y^\epsilon_a}-
v^{\alpha}_{a_0}\frac{\partial {\cal K}_a}{\partial y^0_a}\Big\}+{\cal O}(c^{-5})\Big] =\nonumber\\
&=& {4\pi G}\sum_b\Big\{\hat \sigma^{\alpha}_b(x(y_a))-
v^\alpha_{a_0}\hat \sigma_b(x(y_a))\Big\}+ {\cal O}(c^{-2}),
\label{eq:GR0a_loc-cov}\\
{}
\Delta_{y_a} \Big[\sum_b \hat w_b(x(y_a))-
\frac{\partial {\cal K}_a}{\partial y^0_a} &-& {\textstyle\frac{1}{2}}v^{}_{a_0\epsilon} v_{a_0}^\epsilon+{\cal O}(c^{-2})\Big]
= 8\pi G \gamma_{\alpha\beta} \sum_b
\hat \sigma_b(x(y_a))+{\cal O}(c^{-2}). \label{eq:GRab_loc-cov}
\end{eqnarray}

Using the fact that $\Delta_{y_a}=\Delta_{x}+{\cal O}(c^{-2})$ and the gauge equations (\ref{eq:DeDoGa-K})--(\ref{eq:DeDoGa-Q}) we can show that Eqs.~(\ref{eq:GR0a_loc-cov}) and (\ref{eq:GRab_loc-cov}) are identically satisfied. Furthermore, Eq.~(\ref{eq:GR00_loc-cov}) reduces to
{}
\begin{eqnarray}
\Box_{y_a}\sum_b \hat w_b(x(y_a))
= {4\pi G}\sum_b\Big\{\hat \sigma_b(x(y_a))+\frac{2}{c^2}
\big(\frac{\partial \hat{\cal K}_a}{\partial x^0}+
\textstyle{\frac{1}{2}}v^{}_{a_0\epsilon}v^\epsilon_{a_0}\big)
\hat \sigma_b(x(y_a))+{\cal O}(c^{-4})\Big\}.
\label{eq:GR00_loc-cov+0}
\end{eqnarray}

We can verify the validity of this equation in a manner similar to the derivation shown by Eqs.~(\ref{eq:D'Alemb_x}) and (\ref{eq:D'Alemb_x+}): We can express the d'Alembertian in the local coordinates $\Box_{y_a}$ via global coordinates $\{x^m\}$ as below:
{}
\begin{eqnarray}
\frac{\partial^2}{\partial {y^0_a}^2}+\gamma^{\epsilon\lambda}
\frac{\partial^2}{\partial y^\epsilon_a \partial y^\lambda_a}=
\frac{\partial^2}{\partial {x^0}^2}+\gamma^{\epsilon\lambda}
\frac{\partial^2}{\partial x^\epsilon \partial x^\lambda}
&+&
\frac{1}{c^2}\Big\{a_{a_0}^\mu-\gamma^{\epsilon\lambda}
\frac{\partial^2 \hat{\cal Q}^\mu_a}{\partial x^\epsilon \partial x^\lambda}\Big\}\frac{\partial}{\partial x^\mu}-\nonumber\\
&-&
\frac{1}{c^2}\Big\{v_{a_0}^\epsilon v_{a_0}^\lambda+\gamma^{\epsilon\mu}\frac{\partial \hat{\cal Q}^\lambda_a}{\partial x^\mu}+\gamma^{\lambda\mu}\frac{\partial \hat{\cal Q}^\epsilon_a}{\partial x^\mu}\Big\}\frac{\partial^2}{\partial x^\epsilon\partial x^\lambda}+{\cal O}(c^{-4}),
\label{eq:D'Alemb_y}
\end{eqnarray}
where ${\partial y^k_a}/{\partial x^p}$ needed to derive Eq.~(\ref{eq:D'Alemb_y}) are given by Eqs.~(\ref{eq:(C1a)_inv})--(\ref{eq:(C1d)_inv}). Terms of order $c^{-2}$ that are present on the right-hand side of this equation can be simplified using the harmonic gauge conditions (\ref{eq:DeDoGa-Q-hat}) and (\ref{eq:form-inv_ab+}), so that Eq.~(\ref{eq:D'Alemb_y}) takes the form:
{}
\begin{eqnarray}
\frac{\partial^2}{\partial {y^0_a}^2}+\gamma^{\epsilon\lambda}
\frac{\partial^2}{\partial y^\epsilon_a \partial y^\lambda_a}
&=&
\frac{\partial^2}{\partial {x^0}^2}+\gamma^{\epsilon\lambda}
\frac{\partial^2}{\partial x^\epsilon \partial x^\lambda}
+
\frac{2}{c^2}\Big(\frac{\partial \hat{\cal K}_a}{\partial x^0}+
{\textstyle\frac{1}{2}}v^{}_{a_0\epsilon} v_{a_0}^\epsilon\Big)
\gamma^{\epsilon\lambda}
\frac{\partial^2}{\partial x^\epsilon \partial x^\lambda}+{\cal O}(c^{-4}).
\label{eq:D'Alemb_y+}
\end{eqnarray}
The last equation may be given in the equivalent form:
{}
\begin{eqnarray}
\Box_{y_a} &=&
\Big\{1+
\frac{2}{c^2}\Big(\frac{\partial \hat{\cal K}_a}{\partial x^0}+
{\textstyle\frac{1}{2}}v^{}_{a_0\epsilon} v_{a_0}^\epsilon\Big)
+{\cal O}(c^{-4})\Big\} \Box_x.
\label{eq:D'Alemb_y+==}
\end{eqnarray}
As a side result, comparing this expression with Eq.~(\ref{eq:D'Alemb_y+=+}), we see that the following approximate relation holds:
{}
\begin{equation}
\Big(\frac{\partial \hat{\cal K}_a}{\partial x^0}+
{\textstyle\frac{1}{2}}v^{}_{a_0\epsilon} v_{a_0}^\epsilon\Big)
= -\Big(\frac{\partial {\cal K}_a}{\partial y^0_a}+
{\textstyle\frac{1}{2}}v^{}_{a_0\epsilon} v_{a_0}^\epsilon\Big)
+{\cal O}(c^{-2}).
\label{eq:K-K-hat-rel}
\end{equation}
Using Eq.~(\ref{eq:eq-w^a*}) we see that the following relation is satisfied for any $\{y_a\}$:
{}
\begin{eqnarray}
\Box_x \hat w_b(x(y_a)) &=& {4\pi G}\hat \sigma_b(x(y_a)) +{\cal O}(c^{-4}),\label{eq:Box_y-w-00}
\end{eqnarray}
and relying on (\ref{eq:D'Alemb_y+==}), we observe that the following relation also holds:
\begin{eqnarray}
\Box_{y_a} \hat w_b(x(y_a))&=&
{4\pi G}\hat \sigma_b(x(y_a))\Big\{1+\frac{2}{c^2}\Big(
\frac{\partial \hat{\cal K}_a}{\partial x^0}+
{\textstyle\frac{1}{2}}v^{}_{a_0\epsilon}v^\epsilon_{a_0}\Big)+{\cal O}(c^{-4})\Big\},
\label{eq:Box_y-w}
\end{eqnarray}
which proves Eq.~(\ref{eq:GR00_loc-cov+0}). As a result, with the help of Eqs.~(\ref{eq:DeDoGa-K})--(\ref{eq:DeDoGa-Q}), and also (\ref{eq:Box_y-w}), we verified that Eqs.~(\ref{eq:GR00_loc-cov})--(\ref{eq:GRab_loc-cov}) are satisfied identically.

In the next subsection, we use the harmonic gauge conditions to constrain this remaining freedom which would lead to a particular structure of both the metric tensor and the  coordinate transformations.

\subsection{The form of the metric tensor in the local frame}
\label{sec:harm-metric-loc}

Eqs.~(\ref{eq:g00-loc+})--(\ref{eq:gab-loc+}) represent the metric tensor $g_{mn}$ given by Eqs.~(\ref{eq:(DeDgGa)OKs01})--(\ref{eq:(DeDgGa)OKsa}) in the coordinates of a local reference frame that satisfies the harmonic gauge. This set of gauge conditions forms the foundation of our method of constructing a proper reference frame of a gravitationally accelerated observer. As a result, the metric representing spacetime in the proper frame may be  presented in the following elegant isotropic form that depends only on two harmonic potentials:
\begin{eqnarray}
g_{00}(y_a)&=& 1-\frac{2}{c^2}w(y_a)+\frac{2}{c^4}w^2(y_a)+O(c^{-6}),
\label{eq:g00-cov*}\\
g_{0\alpha}(y_a)&=& -\gamma_{\alpha\lambda}\frac{4}{c^3}w^\lambda(y_a)+O(c^{-5}),
\label{eq:g0a-cov*}\\
g_{\alpha\beta}(y_a)&=& \gamma_{\alpha\beta}+\gamma_{\alpha\beta}\frac{2}{c^2} w(y_a)+O(c^{-4}),
\label{eq:gab-cov*}
\end{eqnarray}
where the scalar $w$ and vector $w^\alpha$ potentials, expressed as functions of the local coordinates $\{y^k_a\}$ of body $a$, read as follows:
{}
\begin{eqnarray}
w(y_a)&=&\sum_bw_b(y_a)-\frac{\partial {\cal K}_a}{\partial y^0_a}-
{\textstyle\frac{1}{2}}v^{}_{a_0\epsilon} v_{a_0}^\epsilon -
\frac{1}{c^2}\Big\{ \frac{\partial {\cal L}_a}{\partial y^0_a}+cv^{}_{a_0\epsilon}\frac{\partial {\cal Q}^\epsilon_a}{\partial y^0_a}+
{\textstyle\frac{1}{2}}\Big(\frac{\partial {\cal K}_a}{\partial y^0_a}\Big)^2-\Big(\frac{\partial {\cal K}_a}{\partial y^0_a}+
{\textstyle\frac{1}{2}}v^{}_{a_0\epsilon} v_{a_0}^\epsilon\Big)^2\Big\}+O(c^{-4}), \hskip 20pt
\label{eq:pot_loc-w_0-cov*}\\
w^\alpha(y_a)&=& \sum_bw^\alpha_b(y_a)-
{\textstyle\frac{1}{4}}\Big\{ \gamma^{\alpha\epsilon}\frac{1}{c}\frac{\partial {\cal L}_a}{\partial y^\epsilon_a}+c\frac{\partial {\cal Q}^\alpha_a}{\partial y^0_a}+\gamma^{\alpha\epsilon}v^{}_{a_0\lambda}\frac{\partial {\cal Q}^\lambda_a}{\partial y^\epsilon_a}-v_{a_0}^\alpha\frac{\partial {\cal K}_a}{\partial y^0_a}\Big\}+O(c^{-2}),
\label{eq:pot_loc-w_a-cov*}
\end{eqnarray}
with gravitational  potentials $w_b(y_a)$ and $w^\alpha_b(y_a)$ transforming between the global and local frames as below
\begin{eqnarray}
w_b(y_a)&=& \Big(1-\frac{2}{c^2}v^{}_{a_0\epsilon} v_{a_0}^\epsilon\Big)\hat w_b(x(y_a))+\frac{4}{c^2}v^{}_{a_0\epsilon}
\hat w^\epsilon_b(x(y_a))+{\cal O}(c^{-4}),
\label{eq:pot_loc_grav-w_0-cov+}\\[3pt]
w^\alpha_b(y_a)&=& \hat w^\alpha_b(x(y_a))-v_{a_0}^\alpha
\hat w_b(x(y_a))+{\cal O}(c^{-2}).
\label{eq:pot_loc_grav-w_a-cov+}
\end{eqnarray}

It follows from Eq.~(\ref{eq:(DeDgGa)OKs0}) that potentials $w_b(y_a)$ and $w^\alpha_b(y_a)$ satisfy the continuity equation:
\begin{equation}
c\frac{\partial w}{\partial y^0_a}+
\frac{\partial w^\epsilon}{\partial y^\epsilon_a}=
{\cal O}(c^{-2}).  \label{eq:(DeDgGa)cont-eq}
\end{equation}

With the help of Eqs.~(\ref{eq:DeDoGa-K})--(\ref{eq:DeDoGa-Q}) and (\ref{eq:form-inv_0a})--(\ref{eq:form-inv_ab}) (as captured in the field equations (\ref{eq:GR00_loc-cov})-(\ref{eq:GRab_loc-cov})), we can verify that the new potentials $w$ and $w^\alpha$ satisfy the following post-Newtonian harmonic equations:
{}
\begin{eqnarray}
\Box_{y_a} {w}(y_a)&=&{4\pi G}\sum_b\Big\{\Big(1 -
\frac{2}{c^2}v_{a_0\epsilon}v^\epsilon_{a_0}\Big)\hat \sigma_b(x(y_a))+
\frac{4}{c^2}v^{}_{a_0\epsilon} \hat \sigma^\epsilon_b(x(y_a))
+\nonumber\\[-3pt]
&&\hskip 100pt +\,\frac{2}{c^2}
\big(\frac{\partial \hat{\cal K}_a}{\partial x^0}+
\textstyle{\frac{1}{2}}v_{a_0\epsilon}v^\epsilon_{a_0}\big)
\hat \sigma_b(x(y_a))+{\cal O}(c^{-4})\Big\},
\label{eq:eq-w-}\\[3pt]
{}
\Delta_{y_a} w^\alpha(y_a)&=&4\pi G\sum_b\Big\{\hat \sigma^{\alpha}_b(x(y_a))-v^\alpha_{a_0}\hat \sigma_b(x(y_a))+{\cal O}(c^{-2})\Big\}.
\label{eq:eq-w^a-}
\end{eqnarray}

One may verify that according to Eqs.~(\ref{eq:Box_y-w}) and (\ref{eq:GR0a_loc-cov}), the scalar and vector potentials can be presented as functions of the local coordinates only:
{}
\begin{eqnarray}
\hat w_b(x(y_a))&=& G\int d^3y'_a\frac{\hat \sigma_b(x(y'_a))}{|\vec{y}_a-\vec{y}'_a|}+\frac{1}{2c^2}G\frac{c^2\partial^2}{\partial {y^0_a}^2}\int d^3y'_a
{\hat \sigma_b(x(y'_a))}{|\vec{y}_a-\vec{y}'_a|}+\nonumber\\
&&\hskip 86pt+~
\frac{2G}{c^2}\int d^3y'_a\frac{\hat \sigma_b(x(y'_a))}{|\vec{y}_a-\vec{y}'_a|}
\Big(
\frac{\partial \hat{\cal K}_a}{\partial x^0}+
{\textstyle\frac{1}{2}}v^{}_{a_0\epsilon}v^\epsilon_{a_0}\Big)+{\cal O}(c^{-4}),
\label{eq:w_0-ctv_bar+=-*}\\
\hat w^\alpha_b(x(y_a))&=& G\int d^3y'_a\frac{\hat \sigma^\alpha_b(x(y'_a))}{|\vec{y}_a-\vec{y}'_a|}+{\cal O}(c^{-2}).
\label{eq:w_a-ctv_bar+=-*}
\end{eqnarray}
These expressions allow us to present the scalar and vector potentials, given by Eqs.~(\ref{eq:pot_loc_grav-w_0-cov+})--(\ref{eq:pot_loc_grav-w_a-cov+}), via barycentric densities given as functions of the local coordinates:
{}
\begin{eqnarray}
{w}_b(y_a)&=& \big(1 -
\frac{2}{c^2}v^{}_{a_0\epsilon}v^\epsilon_{a_0}\big)
G\int d^3y'_a\frac{\hat \sigma_b(x(y'_a))}{|\vec{y}_a-\vec{y}'_a|}
+\frac{4}{c^2}v^{}_{a_0\epsilon}
G\int d^3y'_a\frac{\hat \sigma^\epsilon_b(x(y'_a))}{|\vec{y}_a-\vec{y}'_a|}+ \nonumber\\
&&\hskip 6pt +~
\frac{1}{2c^2}G\frac{c^2\partial^2}{\partial {y^0_a}^2}\int d^3y'_a
{\hat \sigma_b(x(y'_a))}{|\vec{y}_a-\vec{y}'_a|}+
\frac{2G}{c^2}\int d^3y'_a\frac{\hat \sigma_b(x(y'_a))}{|\vec{y}_a-\vec{y}'_a|}
\Big(
\frac{\partial \hat{\cal K}_a}{\partial x^0}+
{\textstyle\frac{1}{2}}v^{}_{a_0\epsilon}v^\epsilon_{a_0}\Big)+{\cal O}(c^{-3}),
\label{eq:w_0_ctv_bar+fin0}\\
{w}^\alpha_b(y_a)&=& G\int d^3y'_a\frac{\hat \sigma^\alpha_b(x(y'_a))}{|\vec{y}_a-\vec{y}'_a|}-
v^\alpha_{a_0}G\int d^3y'_a\frac{\hat \sigma_b(x(y'_a))}{|\vec{y}_a-\vec{y}'_a|}
+{\cal O}(c^{-2}),
\label{eq:w_a_ctv_bar+fin0}
\end{eqnarray}
which are consistent with transformation rules for relativistic mass and current densities (\ref{eq:sig_0_tranf-loc})--(\ref{eq:sig_a_tranf-loc}), correspondingly.

One can also present the transformation rules for the relativistic gravitational potentials from the proper coordinates $\{y^m_b\}$ associated with body $b$ to the coordinates $\{y^m_a\}$ associated with body $a$.  With the help of Eqs.~(\ref{eq:w_0_ctv_bar}) and (\ref{eq:w_a_ctv_bar}), we can express the potentials $w_b(x)$ and $w^\alpha_b(x)$ in global coordinate $\{x^m\}$ and as functions of $\{x(y_a)\}$, and substitute the results into Eqs.~(\ref{eq:pot_loc_grav-w_0-cov+}) and (\ref{eq:pot_loc_grav-w_a-cov+}), which yeilds\footnote{To avoid a proliferation of cumbersome notations, we do not use hats or other typography at this point to distinguish $w_b(y_a)$, meaning the scalar potential of body $b$, in the coordinates of body $a$, expressed as a function of the coordinates $\{y_a^m\}$, from $w_b(y_b)$, meaning the scalar potential of body $b$, in the coordinates of body $b$, expressed as a function of the coordinates $\{y_b^m\}$. The same applies to $w_b^\alpha(y_a)$ vs. $w_b^\alpha(y_b)$.}:
{}
\begin{eqnarray}
w_b(y_a)&=& \Big(1-\frac{2}{c^2}v^{}_{ba_0\epsilon} v_{ba_0}^\epsilon\Big)
w_b(y_b(y_a))+\frac{4}{c^2}v^{}_{ba_0\epsilon} w^\epsilon_b(y_b(y_a))+{\cal O}(c^{-4}),
\label{eq:pot_loc_grav-w_0-cov-1}\\[3pt]
w^\alpha_b(y_a)&=& w^\alpha_b(y_b(y_a))-v_{ba_0}^\alpha w_b(y_b(y_a))+{\cal O}(c^{-2}),
\label{eq:pot_loc_grav-w_a-cov-1}
\end{eqnarray}
with $v_{ba_0}^\alpha=v_{a_0}^\alpha-v_{b_0}^\alpha$ being the relative velocity between the two bodies. Assuming the existence of the coordinate transformations between the two bodies, $a$ and $b$, in the form
\begin{eqnarray}
y^0_b&=& y^0_a+c^{-2}{\cal K}_{ba}(y^0_a,y^\epsilon_a)+{\cal O}(c^{-4})y^0_a,
\label{eq:trans-0-ab}\\[3pt]
y^\alpha_b&=& y^\alpha_a+z^\alpha_{ba_0}(y^0_a)+{\cal O}(c^{-2}),
\label{eq:trans-a-ab}
\end{eqnarray}
with function ${\cal K}_{ba}$ given \cite{Turyshev-96} as ${\cal K}_{ba}(y^0_a,y^\epsilon_a)={\cal K}_{a}(y^0_a,y^\epsilon_a)-{\cal K}_{b}(y^0_a,y^\epsilon_a+z^\epsilon_{ba_0})+{\cal O}(c^{-2}),$
we note that the scalar and vector potentials, $w_b(y_b(y_a))$ and $w_b^\alpha(y_b(y_a))$ correspondingly, satisfy the following equations:
\begin{eqnarray}
\Box_{y_a} w_b(y_b(y_a))&=&
{4\pi G}\sigma_b(y_b(y_a))\Big\{1-\frac{2}{c^2}\Big(
\frac{\partial {\cal K}_{ba}}{\partial y^0_a}+
{\textstyle\frac{1}{2}}v^{}_{ba_0\epsilon}v^\epsilon_{ba_0}\Big)+{\cal O}(c^{-4})\Big\},
\label{eq:Box_y-w-0-ba}\\
\Delta_{y_a} w^\alpha_b(y_b(y_a))&=&
{4\pi G}\sigma^\alpha_b(y_b(y_a)) +{\cal O}(c^{-2}).
\label{eq:Box_y-w-a-ba}
\end{eqnarray}
In Eq.~(\ref{eq:Box_y-w-0-ba}), similarly to Eq.~(\ref{eq:K-K-hat-rel}) we used the following identity:
{}
\begin{equation}
\Big(\frac{\partial {\cal K}_{ab}}{\partial y^0_b}+
{\textstyle\frac{1}{2}}v^{}_{ba_0\epsilon} v_{ba_0}^\epsilon\Big)
= -\Big(\frac{\partial {\cal K}_{ba}}{\partial y^0_a}+
{\textstyle\frac{1}{2}}v^{}_{ba_0\epsilon} v_{ba_0}^\epsilon\Big)
+{\cal O}(c^{-2}).
\label{eq:K-K-hat-rel-ba}
\end{equation}

Eqs.~(\ref{eq:Box_y-w-0-ba}) and (\ref{eq:Box_y-w-a-ba}) have the following solutions
{}
\begin{eqnarray}
w_b(y_b(y_a))&=& G\int d^3y'_a\frac{\sigma_b(y_b(y'_a))}{|\vec{y}_a-\vec{y}'_a|}+\frac{1}{2c^2}G\frac{c^2\partial^2}{\partial {y^0_a}^2}\int d^3y'_a
{\sigma_b(y_b(y'_a))}{|\vec{y}_a-\vec{y}'_a|}-\nonumber\\
&&\hskip 86pt-~
\frac{2G}{c^2}\int d^3y'_a\frac{\sigma_b(y_b(y'_a))}{|\vec{y}_a-\vec{y}'_a|}
\Big(
\frac{\partial {\cal K}_{ba}}{\partial y^0_a}+
{\textstyle\frac{1}{2}}v^{}_{ba_0\epsilon}v^\epsilon_{ba_0}\Big)+{\cal O}(c^{-4}),
\label{eq:w_0-ctv_bar-ba}\\
w^\alpha_b(y_b(y_a))&=& G\int d^3y'_a\frac{\sigma^\alpha_b(y_b(y'_a))}{|\vec{y}_a-\vec{y}'_a|}+{\cal O}(c^{-2}).
\label{eq:w_a-ctv_bar-ba}
\end{eqnarray}

Therefore, we finally obtain the following expressions for the scalar and vector gravitational potentials, expressed as integrals over the compact volumes of the bodies given in the local coordinates:
{}
\begin{eqnarray}
w_b(y_a)&=& \big(1 -
\frac{2}{c^2}v^{}_{ba_0\epsilon}v^\epsilon_{ba_0}\big)
G\int d^3y'_a\frac{\sigma_b(y_b(y'_a))}{|\vec{y}_a-\vec{y}'_a|}
+\frac{4}{c^2}v^{}_{ba_0\epsilon}
G\int d^3y'_a\frac{\sigma^\epsilon_b
(y_b(y'_a))}{|\vec{y}_a-\vec{y}'_a|}+ \nonumber\\
&&\hskip 1pt +~
\frac{1}{2c^2}G\frac{c^2\partial^2}{\partial {y^0_a}^2}\int d^3y'_a
{\sigma_b(y_b(y'_a))}{|\vec{y}_a-\vec{y}'_a|}-
\frac{2G}{c^2}\int d^3y'_a
\frac{\sigma_b(y_b(y'_a))}{|\vec{y}_a-\vec{y}'_a|}
\Big(\frac{\partial {\cal K}_{ba}}{\partial y^0_a}+
{\textstyle\frac{1}{2}}
v^{}_{ba_0\epsilon}v^\epsilon_{ba_0}\Big)+{\cal O}(c^{-3}),~~~
\label{eq:w_0_ctv_bar+fin000}\\
w^\alpha_b(y_a)&=& G\int d^3y'_a
\frac{\sigma^\alpha_b(y_b(y'_a))}{|\vec{y}_a-\vec{y}'_a|}-
v^\alpha_{ba_0}G\int d^3y'_a
\frac{\sigma_b(y_b(y'_a))}{|\vec{y}_a-\vec{y}'_a|}
+{\cal O}(c^{-2}).
\label{eq:w_a_ctv_bar+fin000}
\end{eqnarray}

We use the phrase {\em harmonic metric tensor} to describe the metric tensor (\ref{eq:g00-cov*})--(\ref{eq:gab-cov*}), defined in terms of the harmonic potentials $w$ and $w^\alpha$, which in turn are given by Eqs.~(\ref{eq:pot_loc-w_0-cov*})--(\ref{eq:pot_loc-w_a-cov*}) and satisfy the harmonic equations (\ref{eq:eq-w-})--(\ref{eq:eq-w^a-}).

\subsection{Applying the harmonic gauge conditions to reconstruct the transformation functions}
\label{sec:harm-fun_struct}

The harmonic conditions in our derivation play a second role, which is to put further constraints on coordinate transformations, taking us closer to determine in full the form of the functions ${\cal K}_a$, ${\cal Q}_a^\alpha$ and ${\cal L}_a$. As we discussed in Sec.~\ref{seq:harmform}, these functions must satisfy Eqs.~(\ref{eq:DeDoGa-K})--(\ref{eq:DeDoGa-Q}).
The general solution to these elliptic-type equations can be written in the form of a Taylor series expansion in terms of irreducible Cartesian symmetric trace-free (STF) tensors.  Furthermore, the solutions for the functions ${\cal K}_a, {\cal L}_a$ and ${\cal Q}^\alpha_a$ in Eqs.~(\ref{eq:DeDoGa-K})--(\ref{eq:DeDoGa-Q}) consist of two parts: a fundamental solution of the homogeneous Laplace equation and a particular solution of the inhomogeneous Poisson equation (except for Eq.~(\ref{eq:DeDoGa-K}), which is homogeneous). We discard the part of the fundamental solution that has a singularity at the origin of the local coordinates, where $y^\alpha=0$. Below we derive solutions for each of these transformation functions.

\subsubsection{Determining the structure of ${\cal K}_a$}

The general solution to Eq.~(\ref{eq:DeDoGa-K}) with regular behavior on the world-line (i.e., omitting terms divergent when $|\vec y_a|\rightarrow 0$ or solutions not differentiable at $|\vec{y}_a|=0$) can be given in the following form:
\begin{equation}
{\cal K}_a (y_a) = \kappa_{a_0}+ \kappa_{a_0\mu} y^\mu_a + \delta \kappa_a(y_a)+ {\cal O}(c^{-4}), \qquad {\rm where} \qquad
\delta \kappa_a(y_a)=\sum_{k=2}\frac{1}{k!}\kappa_{a_0\mu_1...\mu_k} (y^0_a)y^{\mu_1...}_ay^{\mu_k}_a + {\cal O}(c^{-4}),
\label{eq:K-gen}
\end{equation}
with $\kappa_{a_0\mu_1...\mu_k} (y^0_a)$ being STF tensors \cite{Thorne-1980}, which depend only on the time-like coordinate $y^0_a$. Substituting this form of the function ${\cal K}_a$ into Eq.~(\ref{eq:form-inv_0a}), we then find solutions for $\kappa_{a_0\mu}$ and $\kappa_{a\mu_1...\mu_k}$:
{}
\begin{equation}
\kappa^{}_{a_0\mu} =- c v^{}_{a_0\mu}  + {\cal O}(c^{-4}),
\qquad \kappa_{a_0\mu_1...\mu_k} ={\cal O}(c^{-4}), \qquad k\ge 2.
\label{eq:Ka-cov0}
\end{equation}

As a result, the function ${\cal K}_a$ that satisfies the harmonic gauge conditions is determined to be
{}
\begin{equation}
{\cal K}_a (y_a) = \kappa_{a_0} - c (v^{}_{a_0\mu}y^\mu_a)  + {\cal O}(c^{-4}).
\label{eq:Ka-cov}
\end{equation}
This expression completely fixes the spatial dependence of the function ${\cal K}_a$, but still has an arbitrary dependence on the time-like coordinate via the function $\kappa_{a_0} (y^0_a)$.

\subsubsection{Determining the structure of ${\cal Q}^\alpha_a$}

The general solution for the function ${\cal Q}^\alpha$ that satisfies Eq.~(\ref{eq:DeDoGa-Q}) may be presented as a sum of a solution of the inhomogeneous Poisson equation and a solution of the homogeneous Laplace equation. In particular, solutions with regular behavior in the vicinity of the world-line may be given in the following form:
\begin{equation}
{\cal Q}^\alpha_a (y_a) = q^\alpha_{a_0}+ q^\alpha_{a_0\mu} y^\mu_a +{\textstyle\frac{1}{2}}q^\alpha_{a_0\mu\nu} y^\mu_a y^\nu_a + \delta\xi^\alpha_a(y_a)+ {\cal O}(c^{-2}),
\label{eq:K-gen+}
\end{equation}
where $q^\alpha_{a_0\mu\nu}$ can be determined directly from Eq.~(\ref{eq:DeDoGa-Q}) and the function $\delta\xi^\alpha_a$ satisfies Laplace's equation
\begin{eqnarray}
\gamma^{\epsilon\lambda}\frac{\partial^2}{\partial y^\epsilon_a \partial y^\lambda_a}
\delta\xi^\alpha_a &=& {\cal O}(c^{-2}).
\label{eq:DeDoGa-a_xi+}
\end{eqnarray}

We can see that Eq.~(\ref{eq:DeDoGa-Q}) can be used to determine $q^\alpha_{a_0\mu\nu}$, but would leave the other terms in the equation unspecified. To determine these terms, we use Eq.~(\ref{eq:form-inv_ab}) together with Eq.~(\ref{eq:form-inv_0a}), and get:
{}
\begin{eqnarray}
v^{}_{a_0\alpha} v^{}_{a_0\beta}+\gamma_{\alpha\lambda}\frac{\partial {\cal Q}^\lambda_a}{\partial y^\beta_a}+\gamma_{\beta\lambda}\frac{\partial {\cal Q}^\lambda_a}{\partial y^\alpha_a}+2\gamma_{\alpha\beta}\Big(\frac{\partial {\cal K}_a}{\partial y^0_a}+
{\textstyle\frac{1}{2}}v^{}_{a_0\epsilon} v_{a_0}^\epsilon\Big)&=&{\cal O}(c^{-2}).
\label{eq:Q-form-inv-cov}
\end{eqnarray}

Using the intermediate solution (\ref{eq:Ka-cov}) for the function ${\cal K}_a$ in Eq.~(\ref{eq:Q-form-inv-cov}), we obtain the following equation for ${\cal Q}^\alpha_a$:
{}
\begin{eqnarray}
v_{a_0\alpha} v_{a_0\beta}+\gamma_{\alpha\lambda}\frac{\partial {\cal Q}^\lambda_a}{\partial y^\beta_a}+\gamma_{\beta\lambda}\frac{\partial {\cal Q}^\lambda_a}{\partial y^\alpha_a}+2\gamma_{\alpha\beta}\Big(\frac{\partial \kappa_{a_0}}{\partial y^0_a}+
{\textstyle\frac{1}{2}}v^{}_{a_0\epsilon} v_{a_0}^\epsilon- a^{}_{a_0\epsilon} y^\epsilon_a\Big)&=&{\cal O}(c^{-2}).
\label{eq:form-inv_exp-cov+}
\end{eqnarray}

A trial solution to Eq.~(\ref{eq:form-inv_exp-cov+}) may be given in the following general from:
{}
\begin{eqnarray}
{\cal Q}^\alpha_a &=& q^{\alpha}_{a_0}+
c_1 v_{a_0}^\alpha v^{}_{a_0\epsilon} y^\epsilon_a +
c_2 v^{}_{a_0\epsilon} v_{a_0}^\epsilon y^\alpha_a +
c_3 a^{\alpha}_{a_0} y^{}_{a\epsilon} y^\epsilon_a+
c_4 a^{}_{a_0\epsilon} y^\epsilon_a y^\alpha_a +
c_5\big(\frac{\partial \kappa_{a_0}}{\partial y^0_a}+
{\textstyle\frac{1}{2}}
v^{}_{a_0\epsilon} v_{a_0}^\epsilon\big)y^\alpha_a+\nonumber\\
&&+~y^{}_{a\epsilon}\omega_{a_0}^{\epsilon\alpha}+\delta\xi^\alpha_a(y_a)
+{\cal O}(c^{-2}),
\label{eq:delta-Q-c+}
\end{eqnarray}
where $q^{\alpha}_{a_0}$ and the antisymmetric matrix $\omega_{a_0}^{\alpha\epsilon}=-\omega_{a_0}^{\epsilon\alpha}$ are functions of the time-like coordinate $y^0_a$; $c_1,..., c_5$ are constant coefficients; and $\delta\xi^\mu_a(y_a)$, given by Eq.~(\ref{eq:DeDoGa-a_xi+}), is at least of the third order in spatial coordinates $y^\mu_a$, namely $\delta\xi^\mu_a(y_a)\propto{\cal O}(|y^\mu_a|^3)$. Direct substitution of Eq.~(\ref{eq:delta-Q-c+}) into Eq.~(\ref{eq:form-inv_exp-cov+}) results in the following unique solution for the constant coefficients:
{}
\begin{equation}
c_1 = -\frac{1}{2},\quad
c_2 = 0,\quad
c_3 = -\frac{1}{2}, \quad
c_4 = 1,\quad
c_5 = -1.
\label{eq:delta-Q-c-sol}
\end{equation}
As a result, the function ${\cal Q}^\alpha_a$ has the following structure:
\begin{eqnarray}
{\cal Q}^\alpha_a(y_a)&=& q^{\alpha}_{a_0} -
\Big({\textstyle\frac{1}{2}}v^\alpha_{a_0} v^\epsilon_{a_0}+
\omega_{a_0}^{\alpha\epsilon}+\gamma^{\alpha\epsilon}\big(\frac{\partial \kappa_{a_0}}{\partial y^0_a}+
{\textstyle\frac{1}{2}}v^{}_{a_0\lambda} v_{a_0}^\lambda\big)\Big)y_{a\epsilon} +
a^{}_{a_0\epsilon}\Big(y^\alpha_a y^\epsilon_a-
{\textstyle\frac{1}{2}}\gamma^{\alpha\epsilon}y^{}_{a\lambda} y^\lambda_a\Big)
+\delta\xi^\alpha_a(y_a)+{\cal O}(c^{-2}),
\hskip 25pt
\label{eq:d-Q+}
\end{eqnarray}
where $q^{\alpha}_{a_0}$ and $\omega_{a_0}^{\alpha\epsilon}$ are yet to be determined.

By substituting Eq.~(\ref{eq:d-Q+}) into Eq.~(\ref{eq:form-inv_exp-cov+}), we see that the function $\delta\xi^\alpha_a(y_a)$ in Eq.~(\ref{eq:d-Q+}) must satisfy the equation:
{}
\begin{equation}
\frac{\partial}{\partial y_{a\alpha}}\delta\xi^\beta_a + \frac{\partial}{\partial y_{a\beta}} \delta\xi^\alpha_a ={\cal O}(c^{-2}).
\label{eq:form-inv_xi-cov}
\end{equation}
We keep in mind that the function $\delta\xi^\alpha_a(y_a)$ must also satisfy Eq.~(\ref{eq:DeDoGa-a_xi+}). The solution to the partial differential equation (\ref{eq:DeDoGa-a_xi+}) with regular behavior on the world-line (i.e., when $|\vec y_a|\rightarrow 0$) can be given in powers of $y^\mu_a$ as
{}
\begin{equation}
\delta\xi^\alpha_a(y_a)=\sum_{k\ge3}^K\frac{1}{k!}\delta\xi^\alpha_{a_0\,\mu_1...\mu_k}(y^0_a)y^{\mu_1}_a{}^{...}y^{\mu_k}_a+{\cal O}(|y^{\mu}_a|^{K+1}) + {\cal O}(c^{-2}),
\label{eq:delxi}
\end{equation}
where $\delta\xi^\alpha_{a_0\,\mu_1...\mu_k} (y^0_a)$ are STF tensors that depend only on time-like coordinate. Using the solution (\ref{eq:delxi}) in Eq.~(\ref{eq:form-inv_xi-cov}), we can see that $\delta\xi^\alpha_{0\,\mu_1...\mu_k}$ is also antisymmetric with respect to the index $\alpha$ and any of the spatial indices $\mu_1...\mu_k$. Combination of these two conditions suggests that $\delta\xi^\alpha_{0\,\mu_1...\mu_k}=0$ for all $k\ge 3$, thus
{}
\begin{equation}
\delta\xi^\alpha_a(y_a)={\cal O}(c^{-2}).
\end{equation}

Therefore, application of the harmonic gauge conditions leads to the following structure of the function ${\cal Q}^\alpha_a$:
\begin{eqnarray}
{\cal Q}^\alpha_a(y_a)&=& q^{\alpha}_{a_0} -
\Big({\textstyle\frac{1}{2}}v^\alpha_{a_0} v^\epsilon_{a_0}+
\omega_{a_0}^{\alpha\epsilon}+\gamma^{\alpha\epsilon}\big(\frac{\partial \kappa_{a_0}}{\partial y^0_a}+
{\textstyle\frac{1}{2}}v^{}_{a_0\lambda} v_{a_0}^\lambda\big)\Big)y_{a\epsilon} +
a^{}_{a_0\epsilon}\Big(y^\alpha_a y^\epsilon_a-
{\textstyle\frac{1}{2}}\gamma^{\alpha\epsilon}y^{}_{a\lambda} y^\lambda_a\Big)+ {\cal O}(c^{-2}),
\label{eq:d-Q}
\end{eqnarray}
where $q^{\alpha}_{a_0},\omega_{a_0}^{\alpha\epsilon}$ and $\kappa_{a_0}$ are yet to be determined.

\subsubsection{Determining the structure of ${\cal L}_a$}

We now turn our attention to the second gauge condition on the temporal coordinate transformation, Eq.~(\ref{eq:DeDoGa-L}). Using the intermediate solution (\ref{eq:Ka-cov}) of the function ${\cal K}_a$, we obtain the following equation for ${\cal L}_a$:
{}
\begin{eqnarray}
\gamma^{\epsilon\lambda}\frac{\partial^2 {\cal L}_a}{\partial y^\epsilon_a \partial y^\lambda_a}
&=& -c^2\frac{\partial^2 {\cal K}_a}{\partial {y^0_a}^2}+
{\cal O}(c^{-2})=c\big(v^{}_{a_0\epsilon} a_{a_0}^\epsilon+
\dot a^{}_{a_0\epsilon} y^\epsilon_a\big)-c^2\frac{\partial}{\partial y^0_a}\Big(\frac{\partial \kappa_{a_0}}{\partial y^0_a}+
{\textstyle\frac{1}{2}}v^{}_{a_0\epsilon} v_{a_0}^\epsilon\Big)+{\cal O}(c^{-2}).
\label{eq:DeDoGa-0_LK+}
\end{eqnarray}

The general solution of Eq.~(\ref{eq:DeDoGa-0_LK+}) for ${\cal L}_a$  may be presented as a sum of a solution $\delta {\cal L}_a$ for the inhomogeneous Poisson equation and a solution $\delta{\cal L}_{a0}$ of the homogeneous Laplace equation. A trial solution of the  inhomogeneous equation to this equation, $\delta{\cal L}_a$, is sought in the following form:
{}
\begin{equation}
\delta{\cal L}_a =
ck_1 (v^{}_{a_0\epsilon} a_{a_0}^\epsilon) (y^{}_{a\mu} y^\mu_a)+
ck_2 (\dot{a}^{}_{a_0\epsilon}y^\epsilon_a)(y^{}_{a\nu} y^\nu_a)-
k_3c^2\frac{\partial}{\partial y^0_a}\Big(\frac{\partial \kappa_{a_0}}{\partial y^0_a}+
{\textstyle\frac{1}{2}}v^{}_{a_0\epsilon} v_{a_0}^\epsilon\Big)(y_{a\nu} y^\nu_a) +{\cal O}(c^{-2}),
\label{eq:delta-L=}
\end{equation}
where $k_1, k_2, k_3$ are some constants. Direct substitution of Eq.~(\ref{eq:delta-L=}) into Eq.~(\ref{eq:DeDoGa-0_LK+}) results in the following values for these constants:
{}
\begin{equation}
k_1 = \frac{1}{6},\quad
k_2 = \frac{1}{10}, \quad
k_3 = \frac{1}{6}.
\label{eq:delta-Q-k-sol}
\end{equation}
As a result, the solution for $\delta{\cal L}_a$ that satisfies the harmonic gauge conditions has the following form:
{}
\begin{eqnarray}
\delta{\cal L}_a(y_a)&=&
{\textstyle\frac{1}{6}}c
\big(v^{}_{a_0\epsilon} a_{a_0}^\epsilon\big)(y^{}_{a\nu} y^\nu_a)+
{\textstyle\frac{1}{10}}c(\dot{a}^{}_{a_0\epsilon}y^\epsilon_a)
(y^{}_{a\nu} y^\nu_a) -
{\textstyle\frac{1}{6}}c^2\frac{\partial}{\partial y^0_a}\Big(\frac{\partial \kappa_{a_0}}{\partial y^0_a}+
{\textstyle\frac{1}{2}}v^{}_{a_0\epsilon} v_{a_0}^\epsilon\Big)(y^{}_{a\nu} y^\nu_a) +{\cal O}(c^{-2}).
\end{eqnarray}

The solution for the homogeneous equation with regular behavior on the world-line (i.e., when $|y^\mu_a|\rightarrow0$) may be presented as follows:
{}
\begin{eqnarray}
{\cal L}_{a0}(y_a)&=&\ell_{a_0}(y^0)+\ell_{a_0\lambda}(y^0_a)\,y^\lambda_a+
{\textstyle\frac{1}{2}}\ell_{a_0\lambda\mu}(y^0_a)\,y^\lambda_a y^\mu_a +\delta\ell_a(y_a)+{\cal O}(c^{-2}),
\end{eqnarray}
where $\ell_{a_0\lambda\mu}$ is an STF tensor of second rank and $\delta\ell_a$ is a function formed from similar STF tensors of higher order:
\begin{equation}
\delta\ell_a(y_a)=\sum_{k\ge3}^K\frac{1}{k!}\delta\ell_{a_0\,\mu_1...\mu_k}(y^0_a)y^{\mu_1}_a{}^{...}y^{\mu_k}_a+{\cal O}(|y^{\mu}_a|^{K+1})+{\cal O}(c^{-2}).
\end{equation}

Finally, the general solution of the Eq.~(\ref{eq:DeDoGa-0_LK+}) may be presented as a sum of the special solution $\delta{\cal L}_a$ of the inhomogeneous equation and the solution ${\cal L}_{a0}$ of the homogeneous equation $\Delta {\cal L}_a=0$. Therefore, the general solution for the harmonic gauge equations for the function ${\cal L}_a(y_a)= {\cal L}_{a0}+\delta{\cal L}_a$ has the following form:
{}
\begin{eqnarray}
{\cal L}_a(y_a)&=&\ell_{a_0}+\ell_{a_0\lambda}\,y^\lambda_a+
{\textstyle\frac{1}{2}}\ell_{a_0\lambda\mu}\,y^\lambda_a y^\mu_a
+\delta\ell_a(y_a)+\nonumber\\[3pt]
&+&{\textstyle\frac{1}{6}}c
\big(v^{}_{a_0\epsilon} a_{a_0}^\epsilon\big)(y^{}_{a\nu} y^\nu_a)+
{\textstyle\frac{1}{10}}c(\dot{a}^{}_{a_0\epsilon}y^\epsilon_a)
(y^{}_{a\nu} y^\nu_a) -
{\textstyle\frac{1}{6}}c^2\frac{\partial}{\partial y^0_a}\Big(\frac{\partial \kappa_{a_0}}{\partial y^0_a}+
{\textstyle\frac{1}{2}}v^{}_{a_0\epsilon} v_{a_0}^\epsilon\Big)(y^{}_{a\nu} y^\nu_a) +{\cal O}(c^{-2}).
\label{eq:L-gen-cov}
\end{eqnarray}

\subsection{Harmonic functions of the coordinate transformation to a local frame}

As a result of imposing the harmonic gauge conditions, we were able to determine the structure of the transformation functions ${\cal K}_a, {\cal Q}^\alpha_a$ and ${\cal L}_a$ that satisfy the harmonic gauge:
{}
\begin{eqnarray}
{\cal K}_a (y_a) &=& \kappa_{a_0} - c (v^{}_{a_0\mu}y^\mu_a)  + {\cal O}(c^{-4}),
\label{eq:K-sum}\\
{\cal Q}^\alpha_a(y_a)&=& q^{\alpha}_{a_0} -
\Big({\textstyle\frac{1}{2}}v^\alpha_{a_0} v^\epsilon_{a_0}+
\omega_{a_0}^{\alpha\epsilon}+\gamma^{\alpha\epsilon}\big(\frac{\partial \kappa_{a_0}}{\partial y^0_a}+
{\textstyle\frac{1}{2}}v^{}_{a_0\lambda} v_{a_0}^\lambda\big)\Big)y^{}_{a\epsilon} +
a^{}_{a_0\epsilon}\Big(y^\alpha_a y^\epsilon_a-
{\textstyle\frac{1}{2}}\gamma^{\alpha\epsilon}y^{}_{a\lambda} y^\lambda_a\Big)+ {\cal O}(c^{-2}),
\label{eq:Q-sum}\\
{\cal L}_a(y_a)&=&\ell_{a_0}+\ell_{a_0\lambda}\,y^\lambda_a+
{\textstyle\frac{1}{2}}\ell_{a_0\lambda\mu}\,y^\lambda_a y^\mu_a
+\delta\ell_a(y_a)+\nonumber\\[3pt]
&&+~{\textstyle\frac{1}{6}}c\Big(
\big(v^{}_{a_0\epsilon} a_{a_0}^\epsilon\big)  -
c\frac{\partial}{\partial y^0_a}\big(\frac{\partial \kappa_{a_0}}{\partial y^0_a}+
{\textstyle\frac{1}{2}}v^{}_{a_0\epsilon} v_{a_0}^\epsilon\big)\Big)(y^{}_{a\nu} y^\nu_a)+
{\textstyle\frac{1}{10}}c(\dot{a}^{}_{a_0\epsilon}y^\epsilon_a)
(y^{}_{a\nu} y^\nu_a) +{\cal O}(c^{-2}).
\label{eq:L-sum}
\end{eqnarray}

Note that the harmonic gauge conditions allow us to reconstruct the structure of these functions with respect to spatial coordinate $y^\mu_a$. However, the time-dependent functions $\kappa_{a_0}, q^{\alpha}_{a_0},\omega_{a_0}^{\alpha\epsilon}$, $\ell_{a_0}, \ell_{a_0\lambda}, \ell_{a_0\lambda\mu}$, and $\delta\ell_{a_0\,\mu_1...\mu_k}$ cannot be determined from the gauge conditions alone. We need to apply another set of conditions that would dynamically define the proper reference frame of a moving observer, thereby fixing these time-dependent functions. This procedure will be discussed in the following section.

\section{Dynamical conditions for a proper reference frame}
\label{sec:dyn}

As we saw in the preceding section, introduction of the harmonic gauge allows one to determine the harmonic structure of the transformation functions ${\cal K}_a$, ${\cal L}_a$ and ${\cal Q}^\alpha_a$, but it was insufficient to completely determine these functions. To proceed further, we investigate the dynamics of a co-moving observer in the accelerated reference frame.

To define a proper reference frame for a body $a$, we consider the metric tensor $g_{mn}(y_a)$, given by Eqs.~(\ref{eq:g00-cov*})--(\ref{eq:gab-cov*}). Note that the scalar, $w$, and vector, $w^\alpha$, potentials given by Eqs.~(\ref{eq:pot_loc-w_0-cov*}) and (\ref{eq:pot_loc-w_a-cov*}) could be interpreted as a linear superposition of three different types of potentials, representing gravity and inertia, namely:
{}
\begin{eqnarray}
w(y_a)&=&w_a(y_a)+
\sum_{b\not=a}w_b(y_a)+u_a(y_a)+O(c^{-4}),
\label{eq:loc_pot_0}\\
w^\alpha(y_a)&=& w^\alpha_a(y_a)+\sum_{b\not=a}w^\alpha_b(y_a)
+u^\alpha_a(y_a)+O(c^{-2}),
\label{eq:loc_pot_a}
\end{eqnarray}
where $w_a(y_a)$ and $w^\alpha_a(y_a)$, are the scalar and vector gravity potentials produced by the body $a$ itself. Potentials $w_{\rm ext}(y_a)=\sum_{b\not=a}w_b(y_a)$ and $w^\alpha_{\rm ext}(y_a)=\sum_{b\not=a}w^\alpha_b(y_a)$ represent external gravitational field produced in the vicinity of the body $a$ by other bodies ($b\not=a$) of the $N$-body system. Finally, the quantities $u_a(y_a)$ and $u^\alpha_a(y_a)$ are the scalar and vector potentials of field of inertia in the frame associated with the of the body $a$, which are given as below
{}
\begin{eqnarray}
u_a(y_a)&=&
-\frac{\partial {\cal K}_a}{\partial y^0_a}-
{\textstyle\frac{1}{2}}v^{}_{a_0\epsilon} v_{a_0}^\epsilon -
\frac{1}{c^2}\Big\{ \frac{\partial {\cal L}_a}{\partial y^0_a}+cv^{}_{a_0\epsilon}\frac{\partial {\cal Q}^\epsilon_a}{\partial y^0_a}+
{\textstyle\frac{1}{2}}\Big(\frac{\partial {\cal K}_a}{\partial y^0_a}\Big)^2-\Big(\frac{\partial {\cal K}_a}{\partial y^0_a}+
{\textstyle\frac{1}{2}}v^{}_{a_0\epsilon} v_{a_0}^\epsilon\Big)^2\Big\}+O(c^{-4}), \hskip 20pt
\label{eq:in_pot_loc_0}\\
u^\alpha_a(y_a)&=&
-{\textstyle\frac{1}{4}}\Big\{ \gamma^{\alpha\epsilon}\frac{1}{c}\frac{\partial {\cal L}_a}{\partial y^\epsilon_a}+c\frac{\partial {\cal Q}^\alpha_a}{\partial y^0_a}+\gamma^{\alpha\epsilon}v^{}_{a_0\lambda}\frac{\partial {\cal Q}^\lambda_a}{\partial y^\epsilon_a}-v_{a_0}^\alpha\frac{\partial {\cal K}_a}{\partial y^0_a}\Big\}+O(c^{-2}).
\label{eq:in_pot_loc_a}
\end{eqnarray}
As both inertial potentials depend on the coordinate transformation functions, imposing conditions on $u_a(y_a)$ and $u^\alpha_a(y_a)$ will allow us to fix the time-dependent functions still present in Eqs.~(\ref{eq:K-sum})--(\ref{eq:L-sum}). Conversely, a particular choice of these potentials will correspond to a particular choice of the reference frame associated with the body $a$.

Physically, we choose the coordinate system $\{y^m_a\}$ in the proper reference frame of the body $a$ such that along the world-line of the body\footnote{As a first step, we treat the body as a test particle that follows a geodesic in the external gravity field. In the 1PNA, this allows one to determine the structure and the largest contributions to all the quantities of interest. However, as we discussed in Sec.~\ref{sec:introduction}, interaction of the body's intrinsic multipole moments with the external gravity causes the body to deviate from a geodesic trajectory. In Sec.~\ref{sec:ext-body}, we present a procedure to determine the small corrections to the body's equations of motion and relevant terms in the coordinate transformations resulting from such a deviation. For this purpose, we treat the body as an extended object and integrate equations of the energy-momentum conservation over the body's volume, which yields the desired corrections.} the external gravitational potentials $w_b$ and $w^\alpha_b$, $b\not=a$ and the first spatial derivatives $\partial_\beta w_b$ and $\partial_\beta w^\alpha_b$ are absent, so that the spacetime of the $N$-body system in the local coordinates along the world-line $g_{mn}(y_a)$ reduces to the spacetime produced by the gravity of body $a$ only, $g_{mn}^a(y_a)$.  To put it another way, on the world-line of the body $a$, the presence of the external gravity will be completely compensated by the frame-reaction potentials. Consequently, in accordance with the Equivalence Principle (EP), the external gravitational field in the body-centric metric will be manifested only via relativistic tidal effects in the vicinity of the world-line of body $a$. These requirements constitute a generalized set of Fermi-conditions in the case of the $N$-body system.

The approach outlined above was originally developed for an accelerated observer in \cite{TMT:2011}; below, we shall implement this approach for the gravitational $N$-body problem.

\subsection{`Good' proper reference frame and local metric conditions}
\label{sec:good-frame}

An observer that remains at rest with respect to the accelerating frame does so because of the balance between an external force and the fictitious frame-reaction force \cite{TMT:2011}, which is exerted by the local inertia represented by the harmonic potentials (\ref{eq:in_pot_loc_0})--(\ref{eq:in_pot_loc_a}). If the external force is universal (in the sense of the weak EP) and in the case of complete balance of external and this fictitious frame-reaction force, there will be no net force acting on the observer allowing him to be at rest in what we shall call its proper reference frame. The motion of the observer in this frame, then, will resemble a free-fall that follows a geodesic with respect to the local metric $g_{mn}$. It would, thus, be natural to require that in order for the observer to be at rest in his proper reference frame associated with the local metric, the observer's ordinary relativistic three-momentum in that frame should be conserved. We can explore these conditions by writing down the Lagrangian of a test particle that represents the observer and find that it is possible to eliminate all the remaining unknown components of ${\cal K}_a$, ${\cal L}_a$ and ${\cal Q}^\alpha_a$.

The metric tensor $g_{mn}(y_a)$, given by Eqs.~(\ref{eq:g00-cov*})--(\ref{eq:gab-cov*}) allows one to study the dynamics of the reference frame that moves in response to the presence of the external force. The Lagrangian $L$ of a test particle that moves in this system can be obtained directly from the metric $g_{mn}$ and written as below \cite{Landau-Lifshitz-1988,TMT:2011}:
{}
\begin{eqnarray}
L_a&=&-m_ac^2\frac{ds}{dy^0_a}=-m_ac^2\Big({g_{mn}\frac{dy^m_a}{dy^0_a}\frac{dy^n_a}{dy^0_a}}\Big)^{1/2}=\nonumber\\
&=&-m_ac^2\Big\{1+c^{-2}\Big({\textstyle\frac{1}{2}}v{}_\epsilon v^\epsilon-w\Big)+c^{-4}\Big({\textstyle\frac{1}{2}}w^2-{\textstyle\frac{1}{8}}(v{}_\epsilon v^\epsilon)^2-4v_\epsilon w^\epsilon+{\textstyle\frac{3}{2}}v{}_\epsilon v^\epsilon w+{\cal O}(c^{-6})\Big\}. \label{eq:lagr}
\end{eqnarray}

The Lagrangian given by Eq.~(\ref{eq:lagr}) leads to the following equation of motion:
{}
\begin{eqnarray}
c\frac{d}{dy^0_a}\Big\{v^\alpha\Big(1+c^{-2}\big(3w-{\textstyle\frac{1}{2}}v{}_\epsilon v^\epsilon\big)-4c^{-2}w^\alpha+{\cal O}(c^{-4})\Big)\Big\}=-\partial^\alpha_a w\Big\{1-c^{-2}\big({\textstyle\frac{3}{2}}v{}_\epsilon v^\epsilon+w\big)\Big\}
&-&
\frac{4}{c^2}v_\epsilon \partial^\alpha_a w^\epsilon+{\cal O}(c^{-4}),~~~~~~
 \label{eq:eq-mot}
\end{eqnarray}
where $\partial_{a\alpha}= \partial/\partial y^\alpha_a$.
It can be shown that Eq.~(\ref{eq:eq-mot}) is equivalent to the geodesic equation for the local reference frame that is freely falling in the external gravitational field. However, Eq.~(\ref{eq:eq-mot}) has the advantage that it allows to separate relativistic quantities and to study the motion of the system in a more straightforward way.

The condition that the particle, representing an observer, is to remain at rest with respect to the accelerating frame, then, amounts to demanding conservation of its three-dimensional momentum $p^\alpha=\partial L_a/\partial v^\alpha$:
\begin{equation}
p^\alpha=v^\alpha\Big(1+c^{-2}\big(3w-{\textstyle\frac{1}{2}}v{}_\epsilon v^\epsilon\big)\Big)-4c^{-2}w^\alpha+{\cal O}(c^{-4}).\label{eq:canmon}
\end{equation}
In other words, the total time derivative of $p^\alpha$ must vanish. We assume that the observer is located at the spatial origin of the local coordinates, $y_a\equiv|y^\alpha_a|=0$. Then this condition amounts to requiring that the first spatial derivatives $\partial_{a\beta} w_b$ and $\partial_{a\beta} w^\alpha_b$ are equal to that produced by the gravity of body $a$ only. Thus, we require the following two conditions to hold:
{}
\begin{eqnarray}
\lim_{y_a\rightarrow 0} \frac{\partial w(y_a)}{\partial y_a^\beta}  &=& \frac{\partial w_a}{\partial y_a^\beta}\Big|_{y_a=0}+
{\cal O}(c^{-4}),
\label{eq:fermi-cov_part-0pot-0}\\
\lim_{y_a\rightarrow 0} \frac{\partial w^\alpha(y_a)}{\partial y_a^\beta}  &=&  \frac{\partial w^\alpha_a}{\partial y_a^\beta} \Big|_{y_a=0}+{\cal O}(c^{-2}).
\label{eq:fermi-cov_part-0pot-a}
\end{eqnarray}
We also require that on the world-line of the body its momentum (\ref{eq:canmon}) is to depend only on the potentials produced by the body itself and is independent on external gravitational fields, namely:
\begin{equation}
p^\alpha|_{y_a}=v^\alpha\Big(1+c^{-2}\big(3w_a-{\textstyle\frac{1}{2}}v{}_\epsilon v^\epsilon\big)\Big)-4c^{-2}w^\alpha_a+{\cal O}(c^{-4}).\label{eq:canmon+}
\end{equation}
Thus, we require that another pair of conditions hold:
{}
\begin{eqnarray}
 \lim_{y_a\rightarrow 0} w(y_a)     &=& w_a\big|_{y_a=0} +
{\cal O}(c^{-4}),
\label{eq:fermi-cov_0pot-0}\\
\lim_{y_a\rightarrow 0} w^\alpha(y_a)     &=&  w^\alpha_a\big|_{y_a=0} + {\cal O}(c^{-2}).
\label{eq:fermi-cov_0pot-a}
\end{eqnarray}

Therefore, summarizing Eqs.~(\ref{eq:fermi-cov_part-0pot-0})--(\ref{eq:fermi-cov_part-0pot-a}) and (\ref{eq:fermi-cov_0pot-0})--(\ref{eq:fermi-cov_0pot-a}), we require that the following relations involving scalar and vector potentials $w$ and $w^\alpha$ hold along the world-line:
{}
\begin{eqnarray}
 \lim_{|y_a|\rightarrow 0} w(y_a)     &=& w_a|_{y_a=0} +
{\cal O}(c^{-4}), \qquad\qquad  ~
\lim_{|y_a|\rightarrow 0} \frac{\partial w(y_a)}{\partial y_a^\beta} = \frac{\partial w_a}{\partial y_a^\beta} \Big|_{y_a=0}+
{\cal O}(c^{-4})
 \label{eq:fermi-cov_pot-0}\\
 \lim_{|y_a|\rightarrow 0} w^\alpha(y_a)     &=&  w^\alpha_a|_{y_a=0} +
{\cal O}(c^{-2}),  \qquad\qquad
\lim_{|y_a|\rightarrow 0} \frac{\partial w^\alpha(y_a)}{\partial y_a^\beta}  =  \frac{\partial w^\alpha_a}{\partial y_a^\beta} \Big|_{y_a=0}+
{\cal O}(c^{-2}).
 \label{eq:fermi-cov_pot-a}
\end{eqnarray}

In practical terms, the procedure proposed here can be formulated by proceeding with the following three steps:
\begin{inparaenum}[i)]
\item replace body $a$ with a test particle that moves on a geodesic in the background of the other bodies; in this case, the quantities $w$ and $w^\alpha$ entering the Lagrangian (\ref{eq:lagr}) are given by Eqs.~(\ref{eq:loc_pot_0}) and (\ref{eq:loc_pot_a}), representing external gravity and inertia: $w\sim w_{\rm ext}+u_a,$ $w^\alpha \sim w^\alpha_{\rm ext}+u^\alpha_a$;
\item choose a Fermi-type coordinate system where the combination of the external and inertial metric components together with their gradients are zero (to satisfy the weak EP);
the 3-momentum of the test particle $p^\alpha$ satisfies the geodesic equation (\ref{eq:eq-mot}); and
\item switch on the self-gravitational field of body $a$, which is still assumed to be a point-like particle; this yields the conditions (\ref{eq:fermi-cov_pot-0})--(\ref{eq:fermi-cov_pot-a}).
\end{inparaenum}

Taking the decomposition of the total harmonic potentials in the local frame and using this decomposition into the proper and external gravity potentials, and inertial potentials (\ref{eq:loc_pot_0})--(\ref{eq:loc_pot_a}), conditions (\ref{eq:fermi-cov_pot-0})--(\ref{eq:fermi-cov_pot-a}) may equivalently be presented as
{}
\begin{eqnarray}
\Big(\sum_{b\not=a}w_b(y_a)+ u_a(y_a)\Big)\Big|_{y_a=0}&=&{\cal O}(c^{-4}), \qquad\qquad  ~
\Big(\sum_{b\not=a}\frac{\partial w_b(y_a)}{\partial y_a^\beta}
+ \frac{\partial u_a(y_a)}{\partial y_a^\beta}\Big)\Big|_{y_a=0}=\,
{\cal O}(c^{-4}),
 \label{eq:fermi-cov_pot-0*}\\
\Big(\sum_{b\not=a}w^\alpha_b(y_a)+ u^\alpha_a(y_a)\Big)\Big|_{y_a=0} &=&
{\cal O}(c^{-2}), \qquad\qquad
\Big(\sum_{b\not=a}\frac{\partial w^\alpha_b(y_a)}{\partial y_a^\beta}+ \frac{\partial u^\alpha_a(y_a)}{\partial y_a^\beta}\Big)\Big|_{y_a=0}=\,
{\cal O}(c^{-2}).
 \label{eq:fermi-cov_pot-a*}
\end{eqnarray}
These conditions yield additional equations needed to
determine the explicit form of the coordinate transformation functions ${\cal K}_a, {\cal L}_a$ and ${\cal Q}^\alpha_a$, allowing us to fix the time in the local coordinates. As it was demonstrated in \cite{TMT:2011}, the conditions (\ref{eq:fermi-cov_pot-0*})--(\ref{eq:fermi-cov_pot-a*}) can be used to determine frame-reaction potentials. These potentials are the combination of the terms on the right-hand sides of Eqs.~(\ref{eq:pot_loc-w_0-cov*})--(\ref{eq:pot_loc-w_a-cov*}) that depend only on the coordinate transformation functions ${\cal K}_a, {\cal L}_a$ and ${\cal Q}^\alpha_a$ and their derivatives. In addition, this procedure also enables us to determine the equations of motion of an observer with respect to the inertial reference system (logical, but somewhat unexpected result), written in his own coordinates $\{y^k_a\}$. As a result, the dynamical conditions discussed above are necessary to fix the remaining freedom in the harmonic coordinate transformation functions ${\cal K}_a, {\cal L}_a$ and ${\cal Q}^\alpha_a$ and uniquely determine their structure.

\subsection{Application of the dynamical conditions to fix the proper reference frame}

Imposing the generalized Fermi conditions (\ref{eq:fermi-cov_pot-0})--(\ref{eq:fermi-cov_pot-a}) on the potentials $w$ and $w^\alpha$, which are given by Eqs.~(\ref{eq:pot_loc-w_0-cov*})--(\ref{eq:pot_loc-w_a-cov*}), and using $\overline\rho$ to denote the value of any function $\rho(y_a)$ on the world-line of body $a$ via
\begin{equation}
\lim_{|y_a|\rightarrow 0} \rho(y_a)  =\rho|_{y_a=0}= \overline\rho,
\label{eq:u-bar}
\end{equation}
results in the following set of partial differential equations on the world-line of the body $a$:
{}
\begin{eqnarray}
\sum_{b\not=a} \overline{w}_b-\frac{\partial \kappa_{a_0}}{\partial y^0_a}-
{\textstyle\frac{1}{2}}v^{}_{a_0\epsilon} v_{a_0}^\epsilon -
\frac{1}{c^2}\Big\{\frac{\partial {\cal L}_a}{\partial y^0_a}+cv^{}_{a_0\epsilon}\frac{\partial {\cal Q}^\epsilon_a}{\partial y^0_a}+
{\textstyle\frac{1}{2}}\Big(\frac{\partial \kappa_{a_0}}{\partial y^0_a}\Big)^2-\Big(\frac{\partial \kappa_{a_0}}{\partial y^0_a}+
{\textstyle\frac{1}{2}}v^{}_{a_0\epsilon} v_{a_0}^\epsilon\Big)^2\Big\}\Big|_{y_a=0}&=&
{\cal O}(c^{-4}),\hskip24pt
\label{eq:match-w_0-cov}\\
\sum_{b\not=a} \overline{\frac{\partial {w}_b}{\partial y_a^\beta}}+a^{}_{a_0\beta} -
\frac{1}{c^2}\Big\{ \frac{\partial^2 {\cal L}_a}{\partial y^\beta_a\partial y^0_a}+cv_{a_0\epsilon}\frac{\partial^2 {\cal Q}^\epsilon_a}{\partial y^\beta_a\partial y^0_a}+
a^{}_{a_0\beta}\Big(\frac{\partial \kappa_{a_0}}{\partial y^0_a}+
v^{}_{a_0\epsilon} v_{a_0}^\epsilon\Big)\Big\}\Big|_{y_a=0}&=&{\cal O}(c^{-4}),\hskip24pt
\label{eq:match-w_0dir-cov}\\
\sum_{b\not=a} \overline{w^\alpha_b}-
{\textstyle\frac{1}{4}}\Big\{ \gamma^{\alpha\epsilon}\frac{1}{c}\frac{\partial {\cal L}_a}{\partial y^\epsilon_a}+c\frac{\partial {\cal Q}^\alpha_a}{\partial y^0_a}+\gamma^{\alpha\epsilon}v^{}_{a_0\lambda}\frac{\partial {\cal Q}^\lambda_a}{\partial y^\epsilon_a}-v_{a_0}^\alpha\frac{\partial \kappa_{a_0}}{\partial y^0_a}\Big\}\Big|_{y_a=0}&=&{\cal O}(c^{-2}),
\label{eq:match-w_a-cov}\\
\sum_{b\not=a} \overline{\frac{\partial {w}_\alpha^b}{\partial y_a^\beta}}-
{\textstyle\frac{1}{4}}\Big\{ \frac{1}{c}\frac{\partial^2 {\cal L}_a}{\partial y^\alpha_a\partial y^\beta_a}+
c\gamma_{\alpha\lambda}\frac{\partial^2 {\cal Q}^\lambda_a}{\partial y^0_a\partial y^\beta_a}+v^{}_{a_0\lambda}\frac{\partial^2 {\cal Q}^\lambda_a}{\partial y^\alpha_a\partial y^\beta_a}+v^{}_{a_0\alpha}
a^{}_{a_0\beta}\Big\}\Big|_{y_a=0}&=&{\cal O}(c^{-2}).
\label{eq:match-w_adir-cov}
\end{eqnarray}

Equations~(\ref{eq:match-w_0-cov})--(\ref{eq:match-w_adir-cov}) may be used to determine uniquely the transformation functions ${\cal K}_a, {\cal L}_a$ and ${\cal Q}^\alpha_a$. Indeed, from the first two equations above, (\ref{eq:match-w_0-cov}) and (\ref{eq:match-w_0dir-cov}), we immediately have:
{}
\begin{eqnarray}
\sum_{b\not=a} \overline{{U}_b}-\frac{\partial \kappa_{a_0}}{\partial y^0_a}-
{\textstyle\frac{1}{2}}v^{}_{a_0\epsilon} v_{a_0}^\epsilon&=& {\cal O}(c^{-4}), \label{eq:dir0-K+}\\
\sum_{b\not=a} \overline{\frac{\partial {U}_b}{\partial y_a^\beta}}+a^{}_{a_0\beta}&=&
{\cal O}(c^{-4}),
\label{eq:a-Newton}
\end{eqnarray}
where we, for convenience, have split the scalar potential $w_b$ into Newtonian ($U_b$) and post-Newtonian ($\delta w_b$) parts:
{}
\begin{equation}
w^{}_b=U_b+\frac{1}{c^2}\delta w^{}_b+{\cal O}(c^{-4}).
\label{eq:w-U-cov}
\end{equation}
With such a decomposition, the gravitational scalar and vector potentials in the proper reference frame of the body $a$, defined by Eqs.~(\ref{eq:pot_loc_grav-w_0-cov+})--(\ref{eq:pot_loc_grav-w_a-cov+}), have the following form as functions of local coordinates:
\begin{eqnarray}
{w}^{}_b(y_a)&=& U_b(x(y_a)) +\frac{1}{c^2}\delta w^{}_b(y_a)
+{\cal O}(c^{-4}),
\label{eq:pot_loc_grav-w_0-cov*}\\[3pt]
w^\alpha_b(y_a)&=& \hat w^\alpha_b(x(y_a))-v_{a_0}^\alpha U_b(x(y_a))+{\cal O}(c^{-2}),
\label{eq:pot_loc_grav-w_a-cov*}
\end{eqnarray}
with $\delta w^{}_b(y_a)$ given by the following expression:
{}
\begin{equation}
\delta w^{}_b(y_a)=\delta \hat w^{}_b(x(y_a))- 2(v^{}_{a_0\epsilon} v_{a_0}^\epsilon)U_b(x(y_a))+4v^{}_{a_0\epsilon} \hat w^\epsilon_b(x(y_a)) +{\cal O}(c^{-2}).
\label{eq:delta_w(y)}
\end{equation}

Eqs.~(\ref{eq:K-sum}) and (\ref{eq:dir0-K+}) allow us to determine $\kappa_{a_0}$ and present the solution for the function ${\cal K}_a$ as:
{}
\begin{equation}
{\cal K}_a(y_a)= \int_{y^0_{a_0}}^{y^0_a}\!\!\!
\Big(\sum_{b\not=a} \overline{{U}_b}-
{\textstyle\frac{1}{2}}v^{}_{a_0\epsilon} v_{a_0}^\epsilon\Big)dy'^0_a-
c(v^{}_{a_0\epsilon} y_a^\epsilon) + {\cal O}(c^{-4}),
\label{eq:Ka-sol-cov}
\end{equation}
where the Newtonian potential $\overline{U_b}$ is given evaluated on the world-line of the body $a$, namely $\overline{U_b}= U_b(x(y_a))|_{y_a=0}$. From Eq.~(\ref{eq:a-Newton}) we also have the well-known Newtonian equation of motion of the body $a$:
{}
\begin{eqnarray}
a_{a_0}^\alpha&=&-\gamma^{\alpha\epsilon}\sum_{b\not=a} \overline{\frac{\partial {U}_b}{\partial y_a^\epsilon}} +{\cal O}(c^{-4}),
\label{eq:a-Newton+}
\end{eqnarray}
where the gradient of the Newtonian potential $\partial U_b/\partial y^\alpha_a $ is taken on the world-line of the body $a$.

Substituting the expression (\ref{eq:dir0-K+}) into Eq.~(\ref{eq:Q-sum}), we can determine the function ${\cal Q}^\alpha_a$:
{}
\begin{eqnarray}
{\cal Q}^\alpha_a(y_a)&=& q^{\alpha}_{a_0} -
\Big({\textstyle\frac{1}{2}}v^\alpha_{a_0} v^\epsilon_{a_0}+
\gamma^{\alpha\epsilon}\sum_{b\not=a} \overline{{U}_b}+\omega_{a_0}^{\alpha\epsilon}\Big)y_{a\epsilon} +a^{}_{a_0\epsilon}
\Big(y^\alpha_a y^\epsilon_a-
{\textstyle\frac{1}{2}}\gamma^{\alpha\epsilon}y^{}_{a\lambda} y_a^\lambda\Big)+ {\cal O}(c^{-2}).
\label{eq:d-Q=}
\end{eqnarray}

Finally, the general solution for the function ${\cal L}_a$, given by Eq.~(\ref{eq:L-sum}), now takes the following form:
{}
\begin{equation}
{\cal L}_a(y_a)=\ell_{a_0}+\ell_{a_0\lambda}\,y_a^\lambda+
{\textstyle\frac{1}{2}}\ell_{a_0\lambda\mu}\,y_a^\lambda y_a^\mu
-{\textstyle\frac{1}{6}}c
\Big(\sum_{b\not=a}c\overline{\frac{\partial U_b}{\partial y^0_a}} -
v^{}_{a_0\epsilon} a_{a_0}^\epsilon\Big)(y^{}_{a\nu}y^\nu_a)+
{\textstyle\frac{1}{10}}c(\dot{a}^{}_{a_0\epsilon}y^\epsilon_a)
(y^{}_{a\nu}y^\nu_a)+
\delta\ell_a(y_a)+{\cal O}(c^{-2}).
\label{eq:L-gen-cov+}
\end{equation}

The next task is to find the remaining undetermined time-dependent functions present in ${\cal Q}^\alpha_a$ and ${\cal L}_a$, as given by Eqs.~(\ref{eq:d-Q=}) and (\ref{eq:L-gen-cov+}). To do this, we rewrite the remaining parts of Eqs.~(\ref{eq:match-w_0-cov})--(\ref{eq:match-w_adir-cov}) as a system of partial differential equations, again set on the world-line of the body $a$:
{}
\begin{eqnarray}
\sum_{b\not=a} \overline{\delta {w}_b} -
\Big\{\frac{\partial {\cal L}_a}{\partial y^0_a}+cv^{}_{a_0\epsilon}\frac{\partial {\cal Q}^\epsilon_a}{\partial y^0_a}+
{\textstyle\frac{1}{2}}\Big(\frac{\partial \kappa_{a_0}}{\partial y^0_a}\Big)^2-\Big(\frac{\partial \kappa_{a_0}}{\partial y^0_a}+
{\textstyle\frac{1}{2}}v^{}_{a_0\epsilon} v_{a_0}^\epsilon\Big)^2\Big\}\Big|_{y_a=0}&=&
{\cal O}(c^{-2}),\hskip24pt
\label{eq:match-w_0-cov2}\\
\sum_{b\not=a} \overline{\frac{\partial \delta {w}_b}{\partial y_a^\beta}}  -
\Big\{ \frac{\partial^2 {\cal L}_a}{\partial y^\beta_a\partial y^0_a}+cv^{}_{a_0\epsilon}\frac{\partial^2 {\cal Q}^\epsilon_a}
{\partial y^\beta_a\partial y^0_a}+
a^{}_{a_0\beta}\Big(\frac{\partial \kappa_{a_0}}{\partial y^0_a}+
v^{}_{a_0\epsilon} v_{a_0}^\epsilon\Big)\Big\}\Big|_{y_a=0}&=&{\cal O}(c^{-2}),\hskip24pt
\label{eq:match-w_0dir-cov2}\\
\sum_{b\not=a}\overline{{w}^\alpha_b}-
{\textstyle\frac{1}{4}}\Big\{ \gamma^{\alpha\epsilon}\frac{1}{c}\frac{\partial {\cal L}_a}{\partial y^\epsilon_a}+c\frac{\partial {\cal Q}^\alpha_a}{\partial y^0_a}+\gamma^{\alpha\epsilon}v^{}_{a_0\lambda}\frac{\partial {\cal Q}^\lambda_a}{\partial y^\epsilon_a}-v_{a_0}^\alpha\frac{\partial \kappa_{a_0}}{\partial y^0_a}\Big\}\Big|_{y_a=0}&=&{\cal O}(c^{-2}),
\label{eq:match-w_a-cov2}\\
\sum_{b\not=a} \overline{\frac{\partial {w}_{b\alpha}}{\partial y_a^\beta}}-
{\textstyle\frac{1}{4}}\Big\{\frac{1}{c}\frac{\partial^2 {\cal L}_a}{\partial y^\alpha_a\partial y^\beta_a}+
c\gamma_{\alpha\lambda}\frac{\partial^2 {\cal Q}^\lambda_a}{\partial y^0_a\partial y^\beta_a}+v^{}_{a_0\lambda}\frac{\partial^2 {\cal Q}^\lambda_a}{\partial y^\alpha_a\partial y^\beta_a}+v^{}_{a_0\alpha}
a^{}_{a_0\beta}\Big\}\Big|_{y_a=0}&=&{\cal O}(c^{-2}).
\label{eq:match-w_adir-cov2}
\end{eqnarray}

These equations may be used to determine the remaining unknown time-dependent functions $\ell_{a_0}$, $\ell_{a_0\lambda}$,  $\ell_{a_0\lambda\mu}$, and also $q^{\alpha}_{a_0}$ and $\omega_{a_0}^{\alpha\epsilon}$, which are still present in the transformation functions. Thus, substituting the solutions for ${\cal K}_a$ and ${\cal Q}^\alpha_a$, given respectively by Eqs.~(\ref{eq:delta_w(y)}), (\ref{eq:Ka-sol-cov}) and (\ref{eq:d-Q=}), into Eq.~(\ref{eq:match-w_0-cov2}) leads to the following solution for ${\dot \ell}_{a_0}$:
{}
\begin{eqnarray}
{\textstyle\frac{1}{c}}\dot\ell_{a_0}&=&-v^{}_{a_0\epsilon}{\dot q}^\epsilon_{a_0}-
{\textstyle\frac{1}{8}}(v^{}_{a_0\epsilon}v^\epsilon_{a_0})^2-
{\textstyle\frac{3}{2}}(v^{}_{a_0\epsilon}v^\epsilon_{a_0})
\sum_{b\not=a} \overline{{U}_b}+
4v^{}_{a_0\epsilon}\sum_{b\not=a} \overline{{\hat w}^\epsilon_b}+
{\textstyle\frac{1}{2}}(\sum_{b\not=a} \overline{{U}_b})^2+
\sum_{b\not=a} \overline{\delta {\hat w}_b}+{\cal O}(c^{-2}).
\label{eq:ell_0}
\end{eqnarray}

Next, Eq.~(\ref{eq:match-w_0dir-cov2}) results in the following equation for $\dot\ell^\alpha_{a_0}$:
{}
\begin{equation}
{\textstyle\frac{1}{c}}\dot\ell^\alpha_{a_0}=
\sum_{b\not=a} \Big(
\overline{\frac{\partial \delta{\hat w}_b}{\partial y_{a\alpha}}}+
4v^{}_{a_0\epsilon}\overline{\frac{\partial {\hat w}^\epsilon_b}{\partial y_{a\alpha}}}\Big)-
a^\alpha_{a_0}\!\sum_{b\not=a} \overline{{U}_b}+
2(v^{}_{a_0\epsilon}v^\epsilon_{a_0})a^\alpha_{a_0}+
{\textstyle\frac{1}{2}}v^\alpha_{a_0}\!(v^{}_{a_0\epsilon}a^\epsilon_{a_0})+
v^\alpha_{a_0}\!\sum_{b\not=a} c\overline{\frac{\partial {U}_b}{\partial y^0_a}}-v^{}_{a_0\epsilon}{\dot\omega}_{a_0}^{\alpha\epsilon}
+{\cal O}(c^{-2}).
\label{eq:ell_a-dot}
\end{equation}

From Eq.~(\ref{eq:match-w_a-cov2}) we can determine $\ell^\alpha_{a_0}$:
{}
\begin{eqnarray}
{\textstyle\frac{1}{c}}\ell^\alpha_{a_0}&=&-{\dot q}^\alpha_{a_0}+4\sum_{b\not=a} \overline{\hat w^\alpha_b}-2v^\alpha_{a_0}\sum_{b\not=a} \overline{{U}_b}-
v^{}_{a_0\epsilon}\,\omega_{a_0}^{\alpha\epsilon}+{\cal O}(c^{-2}).
\label{eq:ell_a}
\end{eqnarray}

Eq.~(\ref{eq:match-w_adir-cov2}) leads to the following solution for $\ell^{\alpha\beta}_{a_0}$:
{}
\begin{eqnarray}
{\textstyle\frac{1}{c}}\ell^{\alpha\beta}_{a_0}&=&4\sum_{b\not=a}\overline{\frac{\partial {\hat w}^{\alpha}_b}{\partial y_{a\beta}}}+{\textstyle\frac{4}{3}}\gamma^{\alpha\beta}\sum_{b\not=a}
c\overline{\frac{\partial U_b}{\partial y^0_a}}+
{\textstyle\frac{2}{3}}\gamma^{\alpha\beta}v^{}_{a_0\epsilon}a_{a_0}^\epsilon+
{\textstyle\frac{5}{2}}v^\alpha_{a_0}a^\beta_{a_0}-
{\textstyle\frac{1}{2}}v^\beta_{a_0}a^\alpha_{a_0}+
{\dot\omega}_{a_0}^{\alpha\beta}+{\cal O}(c^{-2}).
\label{eq:ell_ab}
\end{eqnarray}

The quantity $\ell^{\alpha\beta}_{a_0}$ is an STF tensor. The expression on the right-hand side must, therefore, be also symmetric. This can be achieved by choosing the anti-symmetric tensor ${\dot\omega}_{a_0}^{\alpha\beta}$ appropriately. With the help of (\ref{eq:a-Newton+}), this can be done uniquely, resulting in
\begin{eqnarray}
\dot\omega_{a_0}^{\alpha\beta}&=&
{\textstyle\frac{1}{2}}(v^{\alpha}_{a_0}a^{\beta}_{a_0}-v^{\beta}_{a_0}a^{\alpha}_{a_0})+2\sum_{b\not=a}\Big(
\overline{\frac{\partial ({\hat w}^{\beta}_b-v_{a_0}^\beta U_b)}{\partial y_{a\alpha}}}-
\overline{\frac{\partial ({\hat w}^{\alpha}_b-v_{a_0}^\alpha U_b)}{\partial y_{a\beta}}}\Big)+{\cal O}(c^{-2}),
\label{eq:omega+}
\end{eqnarray}
which leads to the following solution for $\ell^{\alpha\beta}_{a_0}$:
{}
\begin{equation}
{\textstyle\frac{1}{c}}\ell^{\alpha\beta}_{a_0}=2\sum_{b\not=a}\Big(
\overline{\frac{\partial ({\hat w}^{\beta}_b-v_{a_0}^\beta U_b)}{\partial y_{a\alpha}}}+
\overline{\frac{\partial ({\hat w}^{\alpha}_b-v_{a_0}^\alpha U_b)}{\partial y_{a\beta}}}\Big)+{\textstyle\frac{4}{3}}\gamma^{\alpha\beta}\sum_{b\not=a}
c\overline{\frac{\partial U_b}{\partial y^0_a}}+
\Big({\textstyle\frac{2}{3}}\gamma^{\alpha\beta}v^{}_{a_0\epsilon}a_{a_0}^\epsilon-
v^\alpha_{a_0}a^\beta_{a_0}-v^\beta_{a_0}a^\alpha_{a_0}\Big)+{\cal O}(c^{-2}).
\label{eq:ellb-dd}
\end{equation}
Using the rules (\ref{eq:pot_loc_grav-w_0-cov+})--(\ref{eq:pot_loc_grav-w_a-cov+}) for transforming the potentials, together with the continuity equation (\ref{eq:(DeDgGa)cont-eq}), one can verify that (\ref{eq:ellb-dd}) is trace-free.

Also, using Eq.~(\ref{eq:omega+}), one can present a solution for $\dot\ell^\alpha_{a_0}$, Eq.~(\ref{eq:ell_a-dot}), in the form
{}
\begin{equation}
{\textstyle\frac{1}{c}}\dot\ell^\alpha_{a_0}=
\sum_{b\not=a} \overline{\frac{\partial\delta{\hat w}_b}{\partial y_{a\alpha}}}+
2v^{}_{a_0\epsilon}\!\sum_{b\not=a} \Big(
\overline{\frac{\partial \hat w^\epsilon_b}{\partial y_{a\alpha}}}+
\overline{\frac{\partial \hat w^\alpha_b}{\partial y_{a\epsilon}}}\Big)+
\Big({\textstyle\frac{1}{2}}(v^{}_{a_0\epsilon}v^\epsilon_{a_0})-
\sum_{b\not=a} \overline{{U}_b}\Big)a^\alpha_{a_0}+
2v^\alpha_{a_0}\!(a^{}_{a_0\epsilon}v^\epsilon_{a_0})+
v^\alpha_{a_0}\!\sum_{b\not=a} c
\overline{\frac{\partial{U}_b}{\partial y^0_a}}
+{\cal O}(c^{-2}).~~
\label{eq:ell_a-dot+}
\end{equation}

Finally, Eqs.~(\ref{eq:ell_a-dot}) and (\ref{eq:ell_a}) allow us to determine the equation for $q_{a_0}^{\alpha}$. Indeed, differentiating Eq.~(\ref{eq:ell_a}) with respect to time and subtracting the result from Eq.~(\ref{eq:ell_a-dot}), we obtain the following equation for $\ddot q_{a_0}^{\alpha}$:
{}
\begin{eqnarray}
\ddot q_{a_0}^{\alpha}&=& -\gamma^{\alpha\epsilon}\sum_{b\not=a} \overline{\frac{\partial\delta {\hat w}_b}{\partial y^\epsilon_a}}+
4\sum_{b\not=a} c\overline{\frac{\partial{\hat w}^\alpha_b}{\partial y^0_a}}-
4v^{}_{a_0\epsilon}\sum_{b\not=a}
\overline{\frac{\partial \hat w^\epsilon_b}{\partial y_{a\alpha}}}-
a^\alpha_{a_0}\sum_{b\not=a} \overline{{U}_b}-
3v^\alpha_{a_0}\sum_{b\not=a} c
\overline{\frac{\partial {U}_b}{\partial y^0_a}}-\nonumber\\
&&{}-
\Big({\textstyle\frac{1}{2}}v^\alpha_{a_0}v^\epsilon_{a_0}+\omega_{a_0}^{\alpha\epsilon}\Big)a^{}_{a_0\epsilon}-2a^\alpha_{a_0}(v^{}_{a_0\epsilon}v^\epsilon_{a_0})+
{\cal O}(c^{-2}).
\label{eq:q-ddot}
\end{eqnarray}
Therefore, we can use Eq.~(\ref{eq:q-ddot}) to completely determine the function $q_{0}^{\alpha}$.

The true position of a body $a$ includes terms to all orders, not just the first-order (Galilean) term. This led us to introduce the post-Galilean vector, $x^\alpha_{a_0}(y^0_a)$, defined by Eq.~(\ref{eq:totvec}). Combining this definition with Eqs.~(\ref{eq:a-Newton}) and (\ref{eq:q-ddot}), we can now relate the magnitude of the frame-reaction force (acting on a unit mass), written in the local reference frame of the body $a$, to the acceleration produced by the external gravitational field $\ddot x^\alpha_{a_0}=  \ddot z^\alpha_{a_0}+c^{-2}\ddot q_{a_0}^{\alpha} + {\cal O}(c^{-4})$:
{}
\begin{eqnarray}
\ddot x^\alpha_{a_0}&=&-\sum_{b\not=a}
\overline{\frac{\partial {\hat w}_b}{\partial y^\epsilon_a}}
\Big\{\gamma^{\alpha\epsilon}-\frac{1}{c^2}\Big(
\gamma^{\alpha\epsilon}\sum_{b\not=a} \overline{{U}_b}+
{\textstyle\frac{1}{2}}v^\alpha_{a_0}v^\epsilon_{a_0}+\omega_{a_0}^{\alpha\epsilon}+2\gamma^{\alpha\epsilon}(v^{}_{a_0\epsilon}v^\epsilon_{a_0})\Big)\Big\}+\nonumber\\
&&\hskip 70pt+~
\frac{1}{c^2}\Big(4\sum_{b\not=a} c
\overline{\frac{\partial {\hat w}^\alpha_b}{\partial y^0_a}}-
4v^{}_{a_0\epsilon}\sum_{b\not=a}
\overline{\frac{\partial \hat w^\epsilon_b}{\partial y_{a\alpha}}}-
3v^\alpha_{a_0}\sum_{b\not=a} c
\overline{\frac{\partial {U}_b}{\partial y^0_a}}\Big)+ {\cal O}(c^{-4}).~~
\label{eq:geod_eq-local-cov+}
\end{eqnarray}

The equation of motion (\ref{eq:geod_eq-local-cov+}) establishes the correspondence between the external gravitational field and the fictitious frame-reaction acceleration $\ddot x^\alpha_{a_0}(y^0)$ that is needed to keep the body at rest in its proper reference frame. This frame-reaction force balances the effect of the external gravity acting on the observer co-moving with the body.

\subsection{Summary of results for the direct transformations}
\label{sec:sum-direct}

In the preceding sections, we sought to write the general post-Galilean transformation from a global, inertial frame of reference to a local, accelerating frame in the form
{}
\begin{eqnarray}
x^0&=& y^0_a+c^{-2}{\cal K}_a(y^k_a)+c^{-4}{\cal L}_a(y^k_a)+{\cal O}(c^{-6})y^0_a,\\[3pt]
x^\alpha&=& y^\alpha_a+z^\alpha_{a_0}(y^0_a)+c^{-2}{\cal Q}^\alpha_a(y^k_a)+{\cal O}(c^{-4}).
\end{eqnarray}

To determine the unknown functions ${\cal K}_a$, ${\cal Q}^\alpha_a$, and ${\cal L}_a$, we used the following approach:
\begin{inparaenum}[i)]
\item we imposed the harmonic gauge conditions on the metric tensors in the global and local reference frames;
\item we ensured that the non-inertial local frame is non-rotating;
\item we introduced a co-moving accelerating observer at rest with respect to the proper accelerating frame of a body $a$;
\item we required that a co-moving observers's ordinary three-dimensional three-momentum be conserved on the world-line occupied by the accelerating frame.
\end{inparaenum}

Together, these conditions were sufficient to determine ${\cal K}_a$, ${\cal Q}^\alpha_a$, and ${\cal L}_a$ unambiguously. ${\cal K}_a$ and ${\cal Q}^\alpha_a$ are given by:
{}
\begin{eqnarray}
{\cal K}_a(y_a)&=& \int_{y^0_{a_0}}^{y^0_a}\!\!\!
\Big(\sum_{b\not=a} \overline{{U}_b}-
{\textstyle\frac{1}{2}}v^{}_{a_0\epsilon} v_{a_0}^\epsilon\Big)dy'^0_a-
c(v^{}_{a_0\epsilon} y_a^\epsilon) +{\cal O}(c^{-4}), \\
{}
{\cal Q}^\alpha_a(y_a)&=& q^{\alpha}_{a_0} -
\Big({\textstyle\frac{1}{2}}v^\alpha_{a_0} v^\epsilon_{a_0}+
\gamma^{\alpha\epsilon}\sum_{b\not=a} \overline{{U}_b}+\omega_{a_0}^{\alpha\epsilon}\Big)y^{}_{a\epsilon} +
a^{}_{a_0\epsilon}\Big(y^\alpha_a y^\epsilon_a-
{\textstyle\frac{1}{2}}\gamma^{\alpha\epsilon}y^{}_{a\lambda} y_a^\lambda\Big) +{\cal O}(c^{-2}),
\end{eqnarray}
with the anti-symmetric relativistic precession matrix $\omega_{a_0}^{\alpha\beta}$ given by Eq.~(\ref{eq:omega+})
and the post-Newtonian component of the spatial coordinate in the local frame, $q_{a_0}^{\alpha}$, given by Eq.~(\ref{eq:q-ddot}).

To present the last transformation function, ${\cal L}_a(y_a)$,
we use the explicit dependence of the Newtonian and post-Newtonian components of scalar and vector potentials on the body-centric potentials, as given by Eqs.~(\ref{eq:pot_loc_grav-w_0-cov+}) and  (\ref{eq:pot_loc_grav-w_a-cov+}). As a result, ${\cal L}_a(y_a)$ is given by:

{}
\begin{eqnarray}
{\cal L}_a(y_a)&=&\ell_{a_0}+\ell_{a_0\lambda}\,y_a^\lambda+
{\textstyle\frac{1}{2}} \Big(\ell_{a_0\lambda\mu}+
{\textstyle\frac{1}{3}}c\gamma_{\lambda\mu}
\big(v^{}_{a_0\epsilon} a_{a_0}^\epsilon-
\sum_{b\not=a}\dot{\overline{U_b}}\big)\Big)y_a^\lambda y_a^\mu+
{\textstyle\frac{1}{10}}c(\dot{a}^{}_{a_0\epsilon}y^\epsilon_a)
(y_{a\nu}y^\nu_a)+
\delta\ell_a(y_a)+{\cal O}(c^{-2}),~~~~~~
\label{eq:L-gen-cov+=*}
\end{eqnarray}
where the functions $\ell_{a_0}$, $\ell^\alpha_{a_0}$ and $\ell^{\alpha\beta}_{a_0}$ are defined using the following expressions:
{}
\begin{eqnarray}
{\textstyle\frac{1}{c}}\dot\ell_{a_0}&=&-v^{}_{a_0\epsilon}{\dot q}^\epsilon_{a_0}-
{\textstyle\frac{1}{8}}(v^{}_{a_0\epsilon}v^\epsilon_{a_0})^2-
{\textstyle\frac{3}{2}}(v^{}_{a_0\epsilon}v^\epsilon_{a_0})
\sum_{b\not=a} \overline{{U}_b}+
4v^{}_{a_0\epsilon}\sum_{b\not=a} \overline{{\hat w}^\epsilon_b}+
{\textstyle\frac{1}{2}}(\sum_{b\not=a} \overline{{U}_b})^2+
\sum_{b\not=a} \overline{\delta {\hat w}_b}+{\cal O}(c^{-2}),
\label{eq:ell_0-fin*}\\
{\textstyle\frac{1}{c}}\ell^\alpha_{a_0}&=&-{\dot q}^\alpha_{a_0}+4\sum_{b\not=a} \overline{{\hat w}^\alpha_b}-2v^\alpha_{a_0}\sum_{b\not=a} \overline{{U}_b}-
v^{}_{a_0\epsilon}\,\omega_{a_0}^{\alpha\epsilon}+{\cal O}(c^{-2}),
\label{eq:ell_a-fin*}\\
{\textstyle\frac{1}{c}}\ell^{\alpha\beta}_{a_0}&=&2\sum_{b\not=a}\Big(
\overline{\frac{\partial {w}^{\beta}_b}{\partial y_{a\alpha}}}+
\overline{\frac{\partial {w}^{\alpha}_b}{\partial y_{a\beta}}}\Big)+{\textstyle\frac{4}{3}}\gamma^{\alpha\beta}\sum_{b\not=a}
c\overline{\frac{\partial U_b}{\partial y^0_a}}+
\Big({\textstyle\frac{2}{3}}\gamma^{\alpha\beta}v^{}_{a_0\epsilon}a_{a_0}^\epsilon+
v^\alpha_{a_0}a^\beta_{a_0}+v^\beta_{a_0}a^\alpha_{a_0}\Big)+{\cal O}(c^{-2}).
\label{eq:ellb-dd-fin*}
\end{eqnarray}

Substituting these solutions for the functions ${\cal K}_a, {\cal Q}^\alpha_a$ and ${\cal L}_a$ into the expressions for the potentials $w$ and $w^\alpha$ given by Eqs.~(\ref{eq:pot_loc-w_0-cov*})--(\ref{eq:pot_loc-w_a-cov*}), we find the following form for these potentials:
{}
\begin{eqnarray}
w(y_a)&=&\sum_b w_b(y_a)-\sum_{b\not=a}\Big(\overline{w_b} +y_a^\epsilon \overline{\frac{\partial {w}_b}{\partial y^\epsilon_a}} \Big)-\nonumber\\
&&\hskip 40pt ~-
\frac{1}{c^2}\Big\{{\textstyle\frac{1}{2}}
y^{\epsilon}_{a}y^{\lambda}_{a}\Big[\gamma_{\epsilon\lambda} a^{}_{a_0\mu} a_{a_0}^\mu+a^{}_{a_0\epsilon} a^{}_{a_0\lambda}+
2\dot{a}^{}_{a_0\epsilon} v^{}_{a_0\lambda}+2v^{}_{a_0\epsilon} \dot{a}^{}_{a_0\lambda}
+\gamma_{\epsilon\lambda}\sum_{b\not=a}
\ddot{\overline{U_b}}+
\nonumber\\
&&\hskip 70pt ~+
 2\sum_{b\not=a}\Big(
\dot{\overline{\frac{\partial {w}_{b\lambda}}{\partial y_a^\epsilon}}}+
\dot{\overline{\frac{\partial {w}_{b\epsilon}}{\partial y_a^\lambda}}}\Big)\Big]+{\textstyle\frac{1}{10}}
(\ddot a^{}_{a_0\epsilon} y^\epsilon_a)(y^{}_{a\mu} y_a^\mu)+\partial_0 \delta\ell_a\Big\}+
{\cal O}(c^{-4}),~~~
\label{eq:pot_loc-w_0-cov+}\\[3pt]
w^\alpha(y_a)&=&\sum_b w^\alpha_b(y_a)-\sum_{b\not=a}\Big(\overline{w^\alpha_b} +y_a^\epsilon \overline{\frac{\partial {w}^\alpha_b}{\partial y^\epsilon_a}} \Big) -
{\textstyle\frac{1}{10}}\big\{3y^\alpha_a y^\epsilon_a-
\gamma^{\alpha\epsilon}y^{}_{a\mu} y^\mu_a\big\}{\dot a}^{}_{a_0\epsilon}-{\textstyle\frac{1}{4}}{\textstyle\frac{1}{c}}\frac{\partial \delta\ell_a}{\partial y_{a\alpha}} +{\cal O}(c^{-2}),
\label{eq:pot_loc-w_a-cov+}
\end{eqnarray}
where $w_b(y_a)$ and $w^\alpha_b(y_a)$ are given by Eqs.~(\ref{eq:w_0_ctv_bar+fin000})--(\ref{eq:w_a_ctv_bar+fin000}), correspondingly.

Substitution of these solutions for the potentials into the expressions for the metric tensor given by Eqs.~(\ref{eq:g00-cov*})--(\ref{eq:gab-cov*}) leads to the following form of the local metric tensor:
{}
\begin{eqnarray}
g_{00}(y_a)&=& 1-\frac{2}{c^2}\Big\{\sum_b w_b(y_a)-\sum_{b\not=a}\Big(\overline{w_b} +y_a^\epsilon \overline{\frac{\partial {w}_b}{\partial y^\epsilon_a}} \Big)\Big\}+\frac{2}{c^4}\Big\{
\Big[\sum_b U_b(y_a)-\sum_{b\not=a}\Big(\overline{{U}_b} +y_a^\epsilon \overline{\frac{\partial{U}_b}{\partial y^\epsilon_a}} \Big)\Big]^2+\nonumber\\
&&\hskip 75pt ~+
{\textstyle\frac{1}{2}}
y^{\epsilon}_{a}y^{\lambda}_{a}\Big[\gamma_{\epsilon\lambda} a^{}_{a_0\mu} a_{a_0}^\mu+a^{}_{a_0\epsilon} a^{}_{a_0\lambda}+
2\dot{a}^{}_{a_0\epsilon} v^{}_{a_0\lambda}+2v^{}_{a_0\epsilon} \dot{a}^{}_{a_0\lambda}
+\gamma_{\epsilon\lambda}\sum_{b\not=a}
\ddot{\overline{U_b}}+\nonumber\\
&&\hskip 115pt ~+
 2\sum_{b\not=a}\Big(
\dot{\overline{\frac{\partial{w}_{b\lambda}}{\partial y_a^\epsilon}}}+
\dot{\overline{\frac{\partial {w}_{b\epsilon}}{\partial y_a^\lambda}}}\Big)\Big]+
{\textstyle\frac{1}{10}}
(\ddot a^{}_{a_0\epsilon} y^\epsilon_a)(y^{}_{a\mu} y_a^\mu)+ \partial_0 \delta\ell_a\Big\}+{\cal O}(c^{-6}),
\label{eq:eta_00-cov-fin}\\
g_{0\alpha}(y_a)&=& -\gamma_{\alpha\lambda}\frac{4}{c^3}\Big\{\sum_b w^\lambda_b(y_a)-\sum_{b\not=a}\Big(\overline{w^\lambda_b}
+y_a^\epsilon \overline{\frac{\partial {w}^\lambda_b}{\partial y^\epsilon_a}} \Big)-
{\textstyle\frac{1}{10}}\big\{3y^\lambda_a y^\epsilon_a-
\gamma^{\lambda\epsilon}y^{}_{a\mu} y^\mu_a\big\}{\dot a}^{}_{a_0\epsilon}\Big\}
+\frac{1}{c^4}\frac{\partial\delta\ell_a}{\partial y_a^\alpha} +{\cal O}(c^{-5}),
~~
\label{eq:eta_0a-cov-fin}\\
g_{\alpha\beta}(y_a)&=& \gamma_{\alpha\beta}+\gamma_{\alpha\beta}\frac{2}{c^2} \Big\{\sum_b w_b(y_a)-\sum_{b\not=a}\Big(\overline{w_b} +y_a^\epsilon \overline{\frac{\partial {w}_b}{\partial y^\epsilon_a}} \Big)\Big\}+{\cal O}(c^{-4}).
\label{eq:eta_ab-cov-fin}
\end{eqnarray}

All terms in this metric are determined except for the function $\delta\ell_a$, which remains unknown. We note that the potentials Eqs.~(\ref{eq:pot_loc-w_0-cov+})--(\ref{eq:pot_loc-w_a-cov+}) depend on partial derivatives of $\delta \ell_a$. The same partial derivatives appear in the temporal and mixed components of the metric (\ref{eq:eta_00-cov-fin})--(\ref{eq:eta_0a-cov-fin}). The presence of these terms in the metric amounts to adding a full time derivative to the Lagrangian that describes the system of the moving observer. Indeed, separating  the terms that depend on $\delta\ell_a$ from the Lagrangian constructed with Eqs.~(\ref{eq:eta_00-cov-fin})--(\ref{eq:eta_ab-cov-fin}), we get:
{}
\begin{equation}
\delta L_{\delta\ell_a}=-m c^2\Big\{\frac{2}{c^4}\Big(\frac{\partial \delta\ell_a}{\partial {y^0_a}} +\frac{v^\epsilon}{c}
\frac{\partial \delta\ell_a}{\partial {y^\epsilon_a}}\Big)+{\cal O}(c^{-6})\Big\}=-m c^2\Big\{\frac{2}{c^4}\frac{d\delta\ell_a}{dy^0_a}+{\cal O}(c^{-6})\Big\}.
\end{equation}

As a result, the remainder of the gauge transformation depending on $\delta\ell_a$ will not change the dynamics in the system and, thus, it can be omitted, so that the potentials $w$ and $w^\alpha$ take the  form:
{}
\begin{eqnarray}
w(y_a)&=&\sum_b w_b(y_a)-\sum_{b\not=a}\Big(\overline{w_b} +y_a^\epsilon \overline{\frac{\partial{w}_b}{\partial y^\epsilon_a}} \Big)-\nonumber\\
&&\hskip 40pt ~-
\frac{1}{c^2}\Big\{{\textstyle\frac{1}{2}}
y^{\epsilon}_{a}y^{\lambda}_{a}\Big[\gamma_{\epsilon\lambda} a^{}_{a_0\mu} a_{a_0}^\mu+a^{}_{a_0\epsilon} a^{}_{a_0\lambda}+
2\dot{a}^{}_{a_0\epsilon} v^{}_{a_0\lambda}+2v^{}_{a_0\epsilon} \dot{a}^{}_{a_0\lambda}
+\gamma_{\epsilon\lambda}\sum_{b\not=a}
\ddot{\overline{U_b}}+
\nonumber\\
&&\hskip 70pt ~+
 2\sum_{b\not=a}\Big(
\dot{\overline{\frac{\partial {w}_{b\lambda}}{\partial y_a^\epsilon}}}+
\dot{\overline{\frac{\partial {w}_{b\epsilon}}{\partial y_a^\lambda}}}\Big)\Big]+{\textstyle\frac{1}{10}}
(\ddot a^{}_{a_0\epsilon} y^\epsilon_a)(y^{}_{a\mu} y_a^\mu)\Big\}+
{\cal O}(c^{-4}),~~~
\label{eq:pot_loc-w_0-cov+fin*}\\[3pt]
w^\alpha(y_a)&=&\sum_b w^\alpha_b(y_a)-\sum_{b\not=a}\Big(\overline{w^\alpha_b} +y_a^\epsilon \overline{\frac{\partial{w}^\alpha_b}{\partial y^\epsilon_a}} \Big) -
{\textstyle\frac{1}{10}}\big\{3y^\alpha_a y^\epsilon_a-
\gamma^{\alpha\epsilon}y^{}_{a\mu} y^\mu_a\big\}{\dot a}^{}_{a_0\epsilon}+{\cal O}(c^{-2}).
\label{eq:pot_loc-w_a-cov+fin}
\end{eqnarray}

We can now present the local metric in the following final form:
{}
\begin{eqnarray}
g_{00}(y_a)&=& 1-\frac{2}{c^2}\Big\{\sum_b w_b(y_a)-\sum_{b\not=a}\Big(\overline{w_b} +y_a^\epsilon \overline{\frac{\partial{w}_b}{\partial y^\epsilon_a}} \Big)\Big\}+\frac{2}{c^4}\Big\{
\Big[\sum_b U_b(y_a)-\sum_{b\not=a}\Big(\overline{U_b} +y_a^\epsilon \overline{\frac{\partial {U}_b}{\partial y^\epsilon_a}} \Big)\Big]^2+\nonumber\\
&&\hskip 2pt ~+
{\textstyle\frac{1}{2}}
y^{\epsilon}_{a}y^{\lambda}_{a}\Big[\gamma_{\epsilon\lambda} a^{}_{a_0\mu} a_{a_0}^\mu+a^{}_{a_0\epsilon} a^{}_{a_0\lambda}+
2\dot{a}^{}_{a_0\epsilon} v^{}_{a_0\lambda}+2v^{}_{a_0\epsilon} \dot{a}^{}_{a_0\lambda}
+\gamma_{\epsilon\lambda}\sum_{b\not=a}
\ddot{\overline{U_b}}+ 2\sum_{b\not=a}\Big(
\dot{\overline{\frac{\partial {w}_{b\lambda}}{\partial y_a^\epsilon}}}+
\dot{\overline{\frac{\partial {w}_{b\epsilon}}{\partial y_a^\lambda}}}\Big)\Big]+\nonumber\\
&&\hskip 115pt ~+
{\textstyle\frac{1}{10}}
(\ddot a^{}_{a_0\epsilon} y^\epsilon_a)(y^{}_{a\mu} y_a^\mu)\Big\}+{\cal O}(c^{-6}),
\label{eq:g00-cov-fin++}\\
g_{0\alpha}(y_a)&=& -\gamma_{\alpha\lambda}\frac{4}{c^3}\Big\{\sum_b w^\lambda_b(y_a)-\sum_{b\not=a}\Big(\overline{w^\lambda_b}
+y_a^\epsilon \overline{\frac{\partial{w}^\lambda_b}{\partial y^\epsilon_a}} \Big)-
{\textstyle\frac{1}{10}}\big\{3y^\lambda_a y^\epsilon_a-
\gamma^{\lambda\epsilon}y^{}_{a\mu} y^\mu_a\big\}{\dot a}^{}_{a_0\epsilon}\Big\}+{\cal O}(c^{-5}),~~
\label{eq:g0a-cov-fin++}\\
g_{\alpha\beta}(y_a)&=& \gamma_{\alpha\beta}+\gamma_{\alpha\beta}\frac{2}{c^2} \Big\{\sum_b w_b(y_a)-\sum_{b\not=a}\Big(\overline{w_b} +y_a^\epsilon \overline{\frac{\partial{w}_b}{\partial y^\epsilon_a}} \Big)\Big\}+{\cal O}(c^{-4}),
\label{eq:gab-cov-fin++}
\end{eqnarray}
with the equation of motion of the body $a$, $\ddot x^\alpha_{a_0}=  \ddot z^\alpha_{a_0}+c^{-2}\ddot q_{a_0}^{\alpha} + {\cal O}(c^{-4})$ given by Eq.~(\ref{eq:geod_eq-local-cov+}).

The coordinate transformations that place an observer into this reference frame are given below:
{}
\begin{eqnarray}
x^0&=& y^0_a+
c^{-2}\Big\{\int_{y^0_{a_0}}^{y^0_a}\!\!\!
\Big(\sum_{b\not=a} \overline{{\hat w}_b}-
{\textstyle\frac{1}{2}}(v^{}_{a_0\epsilon} +c^{-2}{\dot q}_{a_0\epsilon})(v_{a_0}^\epsilon+c^{-2}{\dot q}^\epsilon_{a_0})\Big)dy'^0_a-
c(v^{}_{a_0\epsilon} +c^{-2}{\dot q}_{a_0\epsilon}) y_a^\epsilon\Big\}+\nonumber\\
&&~~~+
c^{-4}\Big\{\int_{y^0_{a_0}}^{y^0_a}\!\!\!
\Big(-{\textstyle\frac{1}{8}}(v^{}_{a_0\epsilon}v^\epsilon_{a_0})^2-
{\textstyle\frac{3}{2}}(v^{}_{a_0\epsilon}v^\epsilon_{a_0})\sum_{b\not=a} \overline{U_b}+
4v^{}_{a_0\epsilon}\sum_{b\not=a} \overline{{\hat w}^\epsilon_b}+
{\textstyle\frac{1}{2}}\big(\sum_{b\not=a} \overline{U_b}\big)^2
\Big)dy'^0_a+\nonumber\\
&&\hskip 40pt +~
c\Big(4\sum_{b\not=a} \overline{{\hat w}^{}_{b\epsilon}}-
2v^{}_{a_0\epsilon}\sum_{b\not=a} \overline{U_b}-
v^{\lambda}_{a_0}\,\omega^{}_{a_0\epsilon\lambda}\Big)y^\epsilon_a+
{\textstyle\frac{1}{2}} c\Big[
\gamma_{\lambda\mu}v^{}_{a_0\epsilon}a_{a_0}^\epsilon+
v^{}_{a_0\lambda}a^{}_{a_0\mu}+v^{}_{a_0\mu}a^{}_{a_0\lambda}+
\nonumber\\
&&\hskip 40pt +~
\gamma_{\lambda\mu}\sum_{b\not=a}\dot{\overline{U_b}}+
2\sum_{b\not=a}\Big(
\overline{\frac{\partial {w}^{}_{b\mu}}{\partial y_a^\lambda}}+
\overline{\frac{\partial {w}^{}_{b\lambda}}{\partial y_a^\mu}}\Big)\Big]y_a^\lambda y_a^\mu+
{\textstyle\frac{1}{10}}c(\dot{a}^{}_{a_0\epsilon}y^\epsilon_a)
(y_{a\nu}y^\nu_a)\Big\}
+{\cal O}(c^{-6})y^0_a,\label{eq:transfrom-fin-0+}\\[3pt]
x^\alpha&=& y^\alpha_a+z^\alpha_{a_0}+
c^{-2}\Big\{q^{\alpha}_{a_0} -
\Big({\textstyle\frac{1}{2}}v^\alpha_{a_0} v^\epsilon_{a_0}+
\gamma^{\alpha\epsilon}\sum_{b\not=a} \overline{U_b}+\omega_{a_0}^{\alpha\epsilon}\Big){y_a}_\epsilon +
a^{}_{a_0\epsilon}\Big(y^\alpha_a y^\epsilon_a-
{\textstyle\frac{1}{2}}\gamma^{\alpha\epsilon}{y_a}^{}_\lambda y_a^\lambda\Big)\Big\}+{\cal O}(c^{-4}).
\label{eq:transfrom-fin-a+}
\end{eqnarray}

In practice, one can solve (\ref{eq:transfrom-fin-0+}) by expressing it in a differential form $d x^0/dy^0_a$, rather then using integrals over time. This is a typical approach for solving equations of an analogous nature for time coordinate transformations; see for instance  \cite{Soffel:2003cr,Brumberg-book-1991}.

The expressions (\ref{eq:g00-cov-fin++})--(\ref{eq:gab-cov-fin++}) represent the harmonic metric tensor in the local coordinates of the accelerating reference frame. The form of the metric (\ref{eq:g00-cov-fin++})--(\ref{eq:gab-cov-fin++}) and the coordinate transformations (\ref{eq:transfrom-fin-0+})--(\ref{eq:transfrom-fin-a+}) are new and extend previous formulations (discussed, for instance, in \cite{DSX-I}) obtained with different methods. These results can be used to develop models for high precision navigation and gravitational experiments \cite{Turyshev-2008ufn}. However, for a complete description of these experiments we would need to establish the inverse coordinate transformations -- this task will be performed in the next section.

\subsection{Generalization to the case of extended self-gravitating bodies}
\label{sec:ext-body}

The results that we obtained thus far can be generalized to extended bodies, in a manner similar to \cite{DenisovTuryshev:1990}. To do this, we remember the equation for the conservation of the energy-momentum tensor of matter Eq.~(\ref{eq:tem-conserv}). It is known that general relativity, in the weak field and slow motion approximation, possesses integral conservation laws for the energy-momentum tensor of matter and the gravitational field taken jointly \cite{Will_book93,Turyshev-96,DenisovTuryshev:1990}. It means that equation (\ref{eq:tem-conserv}) can be identically represented as the conservation law of the sum of symmetric energy-momentum tensors of the gravitational field $t^{mn}_g$ and matter $t^{mn}_m$ in
spacetime:
{}
\begin{equation}
\nabla_k\Big(\sqrt{-g}{T}^{mk}\Big)=\partial_k\Big(t^{mk}_g+t^{mk}_m\Big)=0.
\label{eqno(6.26)}
\end{equation}
In the proper reference frame of the body $a$ this equation takes the form of the conservation law of the total energy-momentum tensor $\tilde t^{mk}=t^{mk}_i+t^{mk}_g+t^{mk}_m$ of the fields of inertia, gravity, and matter taken jointly, namely:
{}
\begin{equation}
\nabla_k\Big(\sqrt{-g}{T}^{mk}\Big)
={\partial\over\partial y^k_a}\Big(t^{mk}_i+t^{mk}_g+t^{mk}_m\Big)=0.
\label{eqno(6.27)}
\end{equation}
Below we shall explore this relation and investigate the conditions under which this condition is satisfied.

Writing (\ref{eq:tem-conserv}) for $m=0$ and using (\ref{eq:(Tab-sig)}), we obtain
{}
\begin{equation}
c{\partial\over \partial y^0_a}\Big(\frac{1}{c^2}
\sqrt{-g}{T}^{00}\Big)+
{\partial\over \partial y^\epsilon_a}\Big(\frac{1}{c}\sqrt{-g}{T}^{0\epsilon}\Big)
-\frac{1}{c^2}\Big(\sigma c{\partial w \over \partial y^0_a} + 2\sigma^\mu {\partial w \over
\partial y^\mu_a}\Big) = {\cal O}(c^{-4}),
\label{eqno(6.30)}
\end{equation}
where, for convenience, we introduced the symbols $\sigma=\sum_b\sigma_b(y_a)$ and $\sigma^\alpha=\sum_b\sigma_b^\alpha(y_a)$.

To bring relation (\ref{eqno(6.30)}) to the form of (\ref{eqno(6.27)}) it is necessary  to transform the last two terms by extracting from them the partial derivatives with respect to time and the three-dimensional divergence. Such a transformation cannot be carried out in a unique manner. Indeed, using the harmonic equations
{}
\begin{equation}
\Delta_a w=4\pi G\sigma+{\cal O}(c^{-2}) \hskip 20pt {\rm and} \hskip 20pt
\Delta_a w^\alpha=4\pi G\sigma^\alpha+{\cal O}(c^{-2}),
\end{equation}
with $\Delta_a$ being the Laplacian in the coordinates $\{y^m_a\}$ and also the Newtonian continuity equation
{}
\begin{equation}
c{\partial \sigma\over \partial y^0_a}+
{\partial \sigma^{\epsilon}\over \partial y^\epsilon_a}
 = {\cal O}(c^{-2}),
\label{eqno(6.30)-New}
\end{equation}
which can be derived directly from (\ref{eqno(6.30)}), we rewrite the given terms in the most general form reflecting this ambiguity:
{}
\begin{eqnarray}
\sigma c{\partial w\over \partial y^0_a}&+&2\sigma^\mu
{\partial w\over \partial y^\mu_a}  =
c{\partial \over \partial y^0_a}\Big(a_1\sigma w+
{2a_1-3\over 8\pi G}
\partial _\mu w\partial^\mu w\Big)+ \nonumber\\
&+&{\partial \over \partial y^\mu_a}\Big({1-a_1\over4\pi G}c{\partial w\over \partial y^0_a}\partial^\mu
w +
{a_2c\over 4\pi G}w \partial^\mu
{\partial w\over \partial y^0_a} +(a_1+a_2)w \sigma^\mu+
{a_1+a_2-2\over 4\pi G}\partial_\nu w \big(\partial^\nu
w^\mu-\partial^\mu w^\nu\big)\Big), \label{eqno(6.31)}
\end{eqnarray}
where $a_1$ and $a_2$ are arbitrary numbers and $\partial^\mu=\partial/\partial y^\mu_a$. With consideration of this relation, and collecting like terms in (\ref{eqno(6.30)}), we get:
{}
\begin{equation}
c{\partial\over \partial y^0_A}\Big(t^{00}_i+t^{00}_g+t^{00}_m\Big)+
{\partial\over \partial
y^\beta_A}\Big(t^{0\beta}_i+t^{0\beta}_g+t^{0\beta}_m\Big)=
{\cal O}(c^{-5}),
\label{eqno(6.32a)}
\end{equation}
\noindent with the following expressions for the $(00)$ and $(0\alpha)$ components of the density of total energy-momentum tensor:
{}
\begin{eqnarray}
t^{00}_i+t^{00}_g+t^{00}_m&=&\frac{1}{c^2}\sqrt{-g}{ T}^{00}-\frac{1}{c^2}\Big(a_1\sigma w+{2a_1-3\over 8\pi G}
\partial_\mu w\partial ^\mu w\Big)+{\cal O}(c^{-4}),
 \label{eqno(6.32b)}\\
t^{0\alpha}_i+t^{0\alpha}_g+t^{0\alpha}_m&=&\frac{1}{c}\sqrt{-g}{ T}^{0\alpha}+ \frac{1}{c^2}\Big({a_1-1\over 4\pi G} c{\partial w\over \partial y^0_a} \partial^\alpha w -
{a_2 c\over 4\pi G}w\partial^\alpha {\partial w\over
\partial y^0_a}-(a_1+a_2)w\sigma^\alpha+\nonumber\\
&&\hskip 128pt+ {2-a_1-a_2\over 4\pi G }\partial_\nu w \big(\partial^\nu
w^\alpha-\partial^\alpha w^\nu\big)\Big)
+{\cal O}(c^{-5}).
\label{eqno(6.32c)}
\end{eqnarray}

Remembering the definition for the scalar density $\sigma$ given by Eq.~(\ref{eq:(sig)}) as $\sigma=c^{-2}\big({T}^{00}-\gamma_{\lambda\mu}{T}^{\lambda\mu}\big)$  and that for the stress-energy tensor (\ref{eq:(Tab-sig)}), we can write Eq.~(\ref{eq:tem-conserv}) for $n=\alpha$ as
{}
\begin{eqnarray}
c{\partial \over \partial y^0_a}\Big(\frac{1}{c}\sqrt{-g}{T}^{\alpha 0}\Big) &+&
{\partial \over \partial y^\lambda_a}\Big(\sqrt{-g}{T}^{\alpha\lambda}\Big)  +
\sqrt{-g}\sigma\partial^\alpha w
+\nonumber\\
&+&\frac{1}{c^2}\Big\{
-2\sigma\partial^\alpha w^2-4\sigma c\partial_0 w^\alpha-4\sigma_\epsilon \partial^\epsilon w^\alpha +
4\sigma_\epsilon \partial^\alpha w^\epsilon  +
2\sigma^\alpha c\partial_0 w+2\partial_\lambda w \sqrt{-g}T^{\alpha\lambda}\Big\}
= {\cal O}(c^{-4}).~~~~~~~
\label{eqno(6.33)}
\end{eqnarray}
To reduce this equation to the form (\ref{eqno(6.27)}) we use the following identities:
{}
\begin{eqnarray}
4\pi G\sigma \partial^\alpha w =
{\partial \over \partial y^0_a}\Big(\partial_0 w\partial^\alpha w\Big)&+& {\partial \over \partial y^\lambda_a}\Big(
\partial^\alpha w \partial^\lambda w-
\frac{1}{2}\gamma^{\alpha\lambda}\partial_\mu w \partial^\mu w-
\frac{1}{2}\gamma^{\alpha\lambda}(\partial_0 w)^2
\Big), \label{eq:ident-1} \\
4\pi G\Big(\sigma c\partial_0 w^\alpha +
\sigma_\epsilon (\partial^\epsilon w^\alpha -\partial^\alpha w^\epsilon)\Big)&=&
c{\partial \over \partial y^0_a}\Big(c\partial_0 w\partial^\alpha w+\partial_\epsilon w (\partial^\alpha w^\epsilon-
\partial^\epsilon w^\alpha)\Big)+ \nonumber\\
&&\hskip -50pt +~
{\partial \over \partial y^\epsilon_a}\Big(
(\partial^\lambda w^\alpha-\partial^\alpha w^\lambda)
(\partial^\epsilon w_\lambda-\partial_\lambda w^\epsilon)+
\partial^\alpha w \, c\partial_0 w^\epsilon +
\partial^\epsilon w \, c\partial_0 w^\alpha +\nonumber\\
&&\hskip -50pt +~
\gamma^{\alpha\epsilon}\big(\partial_\mu w_\nu\partial^\mu w^\nu 
-\partial_\mu w^\nu\partial_\nu w^\mu -
\partial_\lambda w\, c \partial_0 w^\lambda -\frac{1}{2}c^2(\partial_0 w)^2\big)\Big). ~~~
\label{eq:ident-2}
\end{eqnarray}

Substituting (\ref{eq:ident-1}) and (\ref{eq:ident-2}) into Eq.~(\ref{eqno(6.33)}) and collecting the like terms, we obtain
{}
\begin{eqnarray}
c{\partial\over \partial
y^0_a}\Big(t^{\alpha0}_i+t^{\alpha0}_g+t^{\alpha0}_m\Big)+
{\partial\over \partial
y^\beta_a}\Big(t^{\alpha\beta}_i+t^{\alpha\beta}_g+t^{\alpha\beta}_m\Big)={\cal O}(c^{-4}),
\label{eqno(6.34)}
\end{eqnarray}
with the following expressions for the $(\alpha0)$ and
$(\alpha\beta)$ components of the density of the total energy-momentum tensor:
{}
\begin{eqnarray}
t^{\alpha0}_i+t^{\alpha0}_g+t^{\alpha0}_m&=&\frac{1}{c}\sqrt{-g} {T}^{0\alpha}+ \frac{1}{c^2}\Big(2w\sigma^\alpha-\frac{1}{4\pi G}\Big(3c{\partial w\over \partial y^0_a} \partial^\alpha w +4\partial_\nu w \big(\partial^\alpha w^\nu-\partial^\nu w^\alpha\big)\Big)\Big)
+{\cal O}(c^{-5}), \label{eqno(6.35)}\\
4\pi G(t^{\alpha\beta}_i+t^{\alpha\beta}_g+t^{\alpha\beta}_m)&=&
4\pi G\,\sqrt{-g}{T}^{\alpha\beta}\Big(1+\frac{2}{c^2}w\Big)+
\partial^\alpha w \partial^\beta w-
\frac{1}{2}\gamma^{\alpha\beta}\partial_\lambda w \partial^\lambda w+
\frac{3}{2}\gamma^{\alpha\beta}(\partial_0 w)^2+\nonumber\\
&+&
\frac{4}{c^2}
\Big(
(\partial^\lambda w^\alpha-\partial^\alpha w^\lambda)(\partial_\lambda w^\beta-\partial^\beta w_\lambda)-
\partial^\alpha w \,c\partial_0 w^\beta -
\partial^\beta w \,c\partial_0 w^\alpha+\nonumber\\
&&\hskip 20pt
+\,
\gamma^{\alpha\lambda}\big(
\partial_\lambda w\, c\partial_0 w^\lambda+ 
\partial_\lambda w^\epsilon \partial_\epsilon w^\lambda -
\partial_\lambda w_\epsilon \partial^\lambda w^\epsilon\big)\Big)+ {\cal O}(c^{-6}). ~~~
\label{eqno(6.36)}
\end{eqnarray}

Given the conservation laws imposed by general relativity, together with Eq.~(\ref{eqno(6.27)}), the expression (\ref{eqno(6.36)}) must then contain the components of the complete  energy-momentum tensor including matter, inertia, and gravitational field. Below, we are mainly interested in the components $\tilde t^{\alpha0}$ of this tensor. Comparing the expressions for it given by (\ref{eqno(6.32c)}) and (\ref{eqno(6.35)}), we can see that
$t^{0\alpha}_i+t^{0\alpha}_g+t^{0\alpha}_m \not=
t^{\alpha0}_i+t^{\alpha0}_g+t^{\alpha0}_m$.
Therefore, although it allows one to obtain the conservation laws of energy and momentum, it is  not yet sufficient for obtaining the remaining conservation laws for which the components of the complete energy-momentum tensor of the system must be symmetric. In order to ensure the symmetry of the complete energy-momentum tensor of the system we should set:
{}
\begin{equation}
a_1=-2,\qquad a_2=0.\label{eq:a1,a2}
\end{equation}
Thus, a necessary (but not sufficient) condition for the
existence of all conservation laws is that relations (\ref{eq:a1,a2}) should hold. With consideration of these equalities, the component $\tilde t^{00}=t^{00}_i+t^{00}_g+t^{00}_m$ of
(\ref{eqno(6.32b)})
of the complete energy-momentum tensor given will have the form:
{}
\begin{equation}
\tilde t^{00}=
\frac{1}{c^2}\sqrt{-g}{T}^{00}+\frac{1}{c^2}\Big(2\sigma w+{7\over 8\pi G} \partial_\mu w\partial ^\mu w\Big)+{\cal O}(c^{-4}).
\label{eqno(6.39)}
\end{equation}
This expression can be used to describe the density of energy distribution of the system of inertia, gravity and matter in spacetime, while the component
$\tilde t^{\alpha0}=t^{\alpha0}_i+t^{\alpha0}_g+t^{\alpha0}_m$ of (\ref{eqno(6.35)}) can be used to describe the density of its momentum. Integrating  the expression for the energy-momentum tensor over the volume of space occupied by body  $a$ and using the trivial relation:
{}
\begin{equation}
\int_a d^3y'_a   \partial_\mu w \partial ^\mu w
= -4\pi G\int_a  d^3y'_a   \sigma w,
%+\oint_a  dS^{\mu}_a  \hskip 3pt  w \partial_\mu  w,
\label{eqno(6.40)}
\end{equation}
we obtain the following expression for the energy $P^0$ of the system of matter, inertia and gravitational field, which is defined in the vicinity of the body $a$ as:
{}
\begin{equation}
P^0_a\equiv m_a c^2= c^2\int_a d^3y'_a  \Big( t^{00}_i+t^{00}_g+t^{00}_m\Big)=\int_a d^3y'_a \sqrt{-g}{T}^{00}\Big(1- \frac{1}{c^2}{3\over 2} w+{\cal O}(c^{-4})\Big).
\label{eqno(6.41a)}
\end{equation}

The momentum $P_a^\alpha$ of the system of fields in the coordinates of this local reference frame  is determined in an entirely analogous way: that is, by integrating the component $\tilde t^{\alpha 0}$ given by (\ref{eqno(6.35)}) of  the complete energy-momentum tensor over the compact volume of the body $a$:
{}
\begin{equation}
P_a^\alpha =\int_a d^3y'_a  \Big(
t^{\alpha0}_i+t^{\alpha0}_g+t^{\alpha0}_m\Big),
\label{eqno(6.43a)}
\end{equation}
which gives for the momentum $P_a^\alpha$ the following expression:
{}
\begin{equation}
P^\alpha_a= \int_a d^3y'_a  \Big[\frac{1}{c}\sqrt{-g}{T}^{0\alpha}\Big(1+\frac{2}{c^2}w\Big)-\frac{1}{c^2}\frac{1}{4\pi G}\Big(3 c{\partial w\over \partial y^0_a} \partial^\alpha w  +
4\partial_\nu w \big(\partial^\alpha w^\nu-
\partial^\nu w^\alpha\big)\Big)\Big]
+{\cal O}(c^{-5}).
\label{eqno(6.43b)}
\end{equation}

Therefore, we shall henceforth use the following expression
for the density of the total momentum   of matter, inertia and the gravitational field in the volume occupied by the body:
{}
\begin{equation}
{\tilde t}^{0\alpha}= \frac{1}{c}\sqrt{-g}{T}^{0\alpha}\Big(1+\frac{2}{c^2}w\Big)-\frac{1}{c^2}\frac{1}{4\pi G}\Big(3 c{\partial w\over \partial y^0_a} \partial^\alpha w  +
4\partial_\nu w \big(\partial^\alpha w^\nu-
\partial^\nu w^\alpha\big)\Big)
+{\cal O}(c^{-4}), \label{eqno(6.45)}
\end{equation}
and  for the total energy density  we shall use the expression:
{}
\begin{equation}
{\tilde t}^{00}=  \sqrt{-g}{T}^{00}\Big(1- \frac{1}{c^2}{3\over 2} w+{\cal O}(c^{-4})\Big).\label{eqno(6.46)}
\end{equation}

The relationships obtained will be used to define the equations of motion of the extended bodies form with respect to the coordinates of the proper reference frame of a body $a$. Note  that by integrating the expressions (\ref{eqno(6.45)}) and (\ref{eqno(6.46)}) over the compact volumes of the bodies in the system, one may obtain the mass and the momentum of these bodies measured with respect to the proper reference frame of a body $a$.

To formulate the coordinate transformations to the `good' proper reference frame of an extended body $a$, we require that in the 1PNA of the general theory of relativity, the energy of body $a$ in its proper reference frame would not depend on external bodies. We substitute in Eq.~(\ref{eqno(6.41a)}) the expression for the scalar potential $w(y_a)$ from (\ref{eq:pot_loc-w_0-cov*}) and the solution for the harmonic function ${\cal K}_a$ from (\ref{eq:K-sum}). To the required order we have:
{}
\begin{eqnarray}
P^0_a&=&
\int_a d^3y'_a \sqrt{-g}{T}^{00}\Big(1- \frac{1}{c^2}{3\over 2} \Big(\sum_bU_b(y_a)-\frac{\partial \kappa_a}{\partial y^0_a}-
{\textstyle\frac{1}{2}}v^{}_{a_0\epsilon} v_{a_0}^\epsilon+
a^{}_{a_0\epsilon}y^\epsilon_a\Big)+{\cal O}(c^{-4})\Big).
\label{eq:ext-energ}
\end{eqnarray}
The last term, proportional to $y^\epsilon_a$, vanishes because of the definition of a dipole moment $d_a^\epsilon=\int_a d^3y'_a \sigma_a y^\epsilon_a=0$. Now, if we choose $\kappa_a$ to satisfy the equation
{}
\begin{eqnarray}
\int_a d^3y'_a \sigma_a\Big(\sum_{b\not=a}U_b(y_a)-\frac{\partial \kappa_a}{\partial y^0_a}-
{\textstyle\frac{1}{2}}v^{}_{a_0\epsilon} v_{a_0}^\epsilon\Big)={\cal O}(c^{-4}),
\label{eq:ext-energ-2}
\end{eqnarray}
we would reach our objective. Therefore, the solution for the function $\kappa_a$  has the following form:
{}
\begin{eqnarray}
\frac{\partial \kappa_a}{\partial y^0_a}= \sum_{b\not=a}\left<U_b\right>_a-
{\textstyle\frac{1}{2}}v^{}_{a_0\epsilon} v_{a_0}^\epsilon+{\cal O}(c^{-4}), ~~~{\rm where}~~~ \left<U_b\right>_a=m_a^{-1}\int_a d^3y'_a \sigma_a U_b(y_a)~~~{\rm and}~~~
m_a=\int_a d^3y'_a \sigma_a.
\end{eqnarray}
This leads to a solution for the function ${\cal K}_a$ derived for the extended bodies (compare with that derived for test particles by Eq.~(\ref{eq:Ka-sol-cov})) which has the following form:
{}
\begin{equation}
{\cal K}_a(y_a)= \int_{y^0_{a_0}}^{y^0_a}\!\!\!
\Big(\sum_{b\not=a} \left<{U}_b\right>_a-
{\textstyle\frac{1}{2}}v^{}_{a_0\epsilon} v_{a_0}^\epsilon\Big)dy'^0_a-
c(v^{}_{a_0\epsilon} y_a^\epsilon) + {\cal O}(c^{-4}).
\label{eq:Ka-sol-cov!2!}
\end{equation}

Similarly, we need to update the solution for the acceleration $a_{a_0}^\alpha$ of test a particle (\ref{eq:a-Newton+}). This can be done using Eq.~(\ref{eqno(6.34)}) with (\ref{eqno(6.35)}) and (\ref{eqno(6.36)}), which, at the Newtonian level, has the form:
{}
\begin{eqnarray}
c{\partial {\tilde t}^{0\alpha}\over \partial y^0_a} +
{\partial {\tilde t}^{\alpha\lambda}\over \partial y^\lambda_a}  =-
\sigma \partial^\alpha w +{\cal O}(c^{-2}).
\label{eq:Newt}
\end{eqnarray}
Integrating this equation over the volume of the body we have
{}
\begin{eqnarray}
c{\partial {\tilde P}^{\alpha}\over \partial y^0_a} +
\oint_adS^a_\lambda{\tilde t}^{\alpha\lambda} =-\int_ad^3y'_a
\sigma_a \partial^\alpha w +{\cal O}(c^{-2}).
\label{eq:Newt-1}
\end{eqnarray}
The second term on the left hand side of this equation vanishes. Now, if we require that the body is to remain at rest in its proper reference frame, the following equation must be valid:
{}
\begin{eqnarray}
\int_ad^3y'_a
\sigma_a \partial^\alpha w=\int_ad^3y'_a\sigma_a \Big(\sum_b\partial^\alpha w_b +a_{a_0}^\alpha\Big)={\cal O}(c^{-4}).
\label{eq:Newt-2}
\end{eqnarray}
The fact that bodies in the system all have compact support results in $\int_ad^3y'_a \sigma \partial^\alpha w_a=0$. Thus, Eq.~(\ref{eq:Newt-2}) leads to the  well-known Newtonian equation of motion of the extended body $a$:
{}
\begin{eqnarray}
a_{a_0}^\alpha&=&-\sum_{b\not=a} \left<\partial^\alpha U_b\right>_a +{\cal O}(c^{-4}),
~~~{\rm where}~~~ \left<\partial^\alpha U_b\right>_a=m^{-1}_a\int_ad^3y'_a\sigma_a \partial^\alpha U_b.
\label{eq:a-Newton+01}
\end{eqnarray}

At this point, one would need to evaluate Eq.~(\ref{eqno(6.34)}) to the post-Newtonian order $\propto {\cal O}(c^{-2})$. By doing so, one determines the post-Newtonian acceleration of the body $\ddot {q}_{a_0}^{\alpha}$ given by Eq.~(\ref{eq:q-ddot}) in the case of a problem involving $N$ extended bodies \cite{Turyshev-96}. Although the corresponding calculation is straightforward, it is quite lengthy and tedious. Furthermore, any anticipated modifications of  Eq.~(\ref{eq:q-ddot}) will be too small to be detected by measurements in the solar system in the near future. Thus, we will not be presenting this calculation here, leaving it for subsequent publications.

Finally, given the accuracy of the measurements in the solar system, only the following two modifications need to be made to our point particle solution summarized in Sec.~\ref{sec:sum-direct}, namely those involving ${\cal K}_a(y_a)$ and $a_{a_0}^\alpha$:
{}
\begin{eqnarray}
{\cal K}_a(y_a)&=& \int_{y^0_{a_0}}^{y^0_a}\!\!\!
\Big(\sum_{b\not=a} \left<{U}_b\right>_a-
{\textstyle\frac{1}{2}}v^{}_{a_0\epsilon} v_{a_0}^\epsilon\Big)dy'^0_a-
c(v^{}_{a_0\epsilon} y_a^\epsilon) + {\cal O}(c^{-4}),
\label{eq:Ka-sol-cov!!} \\
a_{a_0}^\alpha&=&-\sum_{b\not=a} \left<\partial^\alpha U_b\right>_a +{\cal O}(c^{-4}).
\label{eq:a-Newton+!!}
\end{eqnarray}
The difference between the expressions for ${\cal K}_a(y_a)$ given by Eqs.~(\ref{eq:Ka-sol-cov}) and (\ref{eq:Ka-sol-cov!!}) and also between those derived for $a_{a_0}^\alpha$ and given by Eqs.~(\ref{eq:a-Newton+}) and (\ref{eq:a-Newton+!!}) is that the new expressions account for the interaction of the body's gravitational multipole moments with the background gravitational field (via the procedure of averaging the external gravitational field over the body's volume in the form of $\left<{U}_b\right>_a$ and $\left<\partial^\alpha U_b\right>_a$) as opposed to accounting only for the value of that field along the pointlike body's world-line (via the limiting procedure defined by Eq.~(\ref{eq:u-bar}) taken for $\overline{U_b}$ and $\overline{\partial^\alpha U_b}$).

\section{Inverse transformations}
\label{sec:inverse-trans}

In the preceding sections, we constructed an explicit form of the direct transformation between a global inertial and local accelerating reference frames by applying the harmonic gauge and dynamical conditions on the metric. Given the Jacobian matrix Eqs.~(\ref{eq:(C1a)})--(\ref{eq:(C1d)}), it was most convenient to work with the covariant form of the metric tensor, which could be expressed in terms of the accelerating coordinates by trivial application of the tensorial transformation rules Eq.~(\ref{eq:ansatz-loc_cov}).

The same logic suggests that if we were to work on the inverse transformation: that is, when it is the inverse Jacobian matrix $\{\partial y^m_a/\partial x^n\}$ that is given in explicit form, it is more convenient to work with the contravariant form of the metric tensor, to which this Jacobian can be applied readily. This simple observation leads us to the idea that we can get the inverse transformations---i.e., from the accelerated to the inertial frame---by simply repeating the previous calculations, but with the contravariant form of the metric tensor instead of the covariant form.

In this section, we show that this is indeed feasible, and accomplish something not usually found in the literature: construction of a method that can be applied for both direct and inverse transformations between inertial and accelerating reference frames at the same time, in a self-consistent manner.\footnote{Clearly, the functional form of the inverse transformations, together with the corresponding expressions for the metric tensor and potentials can be determined by simply transforming the expressions that we obtained in Sec.~\ref{sec:xformlocal} to global coordinates. The calculations are tedious but straightforward, leading to known results (e.g., \cite{Soffel:2003cr}). In this section, however, we demonstrate that one can derive the inverse transformations more expeditiously, in a manner that is analogous to the case of the direct transformations.}

\subsection{The contravariant metric tensor in the local frame}

We have obtained the solution of the gravitational field equations for the $N$-body problem, which in the barycentric reference frame has the form Eqs.~(\ref{eq:g00-bar_ctv_tot*})--(\ref{eq:gab-bar_ctv_tot*}) and (\ref{eq:w_0_ctv_bar}), (\ref{eq:w_a_ctv_bar}). The solution in the local frame could be written in the following form:
\begin{equation}
g^{mn}(y_a(x))=\frac{\partial y^m_a}{\partial x^k}\frac{\partial y^n_a}{\partial x^l}~g^{kl}(x).
\label{eq:ansatz-loc_ctv}
\end{equation}
Using the coordinate transformations given by Eqs.~(\ref{eq:trans-0_inv})--(\ref{eq:trans-a_inv}) together with Eqs.~(\ref{eq:(C1a)_inv})--(\ref{eq:(C1d)_inv}), we determine the form of this metric tensor in the local reference frame associated with the body $a$:
{}
\begin{eqnarray}
g^{00}(y_a(x))&=&1+\frac{2}{c^2}\Big\{\frac{\partial \hat {\cal K}_a}{\partial x^0}+
{\textstyle\frac{1}{2}}\gamma^{\epsilon\lambda}
\frac{1}{c}{\partial \hat{\cal K}_a\over\partial x^\epsilon}
\frac{1}{c}{\partial \hat{\cal K}_a\over\partial x^\lambda}\Big\}+\nonumber\\[2pt]
{}
&&\hskip 8pt+~\frac{2}{c^4}\Big\{ \frac{\partial \hat {\cal L}_a}{\partial x^0}+
{\textstyle\frac{1}{2}}
\Big(\frac{\partial \hat {\cal K}_a}{\partial x^0}\Big)^{\!2}
+\gamma^{\epsilon\lambda}
\frac{1}{c}{\partial \hat{\cal K}_a\over\partial x^\epsilon}
\frac{1}{c}\frac{\partial \hat {\cal L}_a}{\partial x^\lambda}
-
\Big(\frac{\partial \hat {\cal K}_a}{\partial x^0}+
{\textstyle\frac{1}{2}}\gamma^{\epsilon\lambda}
\frac{1}{c}{\partial \hat{\cal K}_a\over\partial x^\epsilon}
\frac{1}{c}{\partial \hat{\cal K}_a\over\partial x^\lambda}\Big)^{\!2}\Big\}+\nonumber\\[2pt]
{}
&&\hskip 8pt+~\frac{2}{c^2}\sum_b\Big\{\Big(1-\frac{2}{c^2}v^{}_{a_0\epsilon} v_{a_0}^\epsilon\Big)\hat w_b(x)+
\frac{4}{c^2}{v^{}_{a_0}}_{\!\epsilon} \hat w^\epsilon_b(x)\Big\}+\nonumber\\[-5pt]
{}
&&\hskip 8pt+~
\frac{2}{c^4}\Big(\sum_b \hat w_b(x)+
\frac{\partial \hat {\cal K}_a}{\partial x^0}+
{\textstyle\frac{1}{2}}\gamma^{\epsilon\lambda}
\frac{1}{c}{\partial \hat{\cal K}_a\over\partial x^\epsilon}
\frac{1}{c}{\partial \hat{\cal K}_a\over\partial x^\lambda}\Big)^{\!2}+{\cal O}(c^{-6}),\label{eq:g00-loc_ctv}\\
{}
g^{0\alpha}(y_a(x))&=& \frac{1}{c}\Big(\gamma^{\alpha\epsilon}\frac{1}{c}{\partial \hat{\cal K}_a\over\partial x^\epsilon}-v_{a_0}^\alpha\Big)+\frac{4}{c^3}\sum_b \Big(\hat w^\alpha_b(x)-v_{a_0}^\alpha \hat w_b(x)\Big)+\nonumber\\[-5pt]
{}
&&\hskip 8pt+~
\frac{1}{c^3}\Big\{\gamma^{\alpha\epsilon}\frac{1}{c}\frac{\partial \hat {\cal L}_a}{\partial x^\epsilon}+
c\frac{\partial \hat {\cal Q}^\alpha_a}{\partial x^0}+
\gamma^{\epsilon\lambda}\frac{1}{c}{\partial \hat{\cal K}_a\over\partial x^\epsilon}\frac{\partial \hat {\cal Q}^\alpha_a}{\partial x^\lambda}-
v^\alpha_{a_0}\frac{\partial \hat {\cal K}_a}{\partial x^0}\Big\}+{\cal O}(c^{-5}),
\label{eq:g0a-loc_ctv}\\[3pt]
{}
g^{\alpha\beta}(y_a(x))&=& \gamma^{\alpha\beta}+\frac{1}{c^2}\Big\{
v_{a_0}^\alpha v_{a_0}^\beta+\gamma^{\alpha\lambda}\frac{\partial \hat {\cal Q}_a^\beta}{\partial x^\lambda}+\gamma^{\beta\lambda}\frac{\partial \hat {\cal Q}_a^\alpha}{\partial x^\lambda}\Big\}-\gamma^{\alpha\beta}\frac{2}{c^2} \sum_b \hat w_b(x)+{\cal O}(c^{-4}).
\label{eq:gab-loc_ctv}
\end{eqnarray}

As in the case of the direct transformation discussed in Sec.~\ref{sec:grav-field-eqs}, we impose the harmonic gauge condition on the metric, to help us to constrain the remaining degrees of freedom in the metric and also to establish explicit forms of the transformation functions $\hat{\cal K}_a$, $\hat{\cal L}_a$, and $\hat{\cal Q}^\alpha_a$. Using (\ref{eq:form-inv_0a+})--(\ref{eq:form-inv_ab+})
to constrain the form of the metric tensor, given by Eqs.~(\ref{eq:g00-loc_ctv})--(\ref{eq:gab-loc_ctv}) in the local frame, we obtain:
{}
\begin{eqnarray}
g^{00}(y_a(x))&=&1+\frac{2}{c^2}\Big\{\frac{\partial \hat {\cal K}_a}{\partial x^0}+ {\textstyle\frac{1}{2}}v^{}_{a_0\epsilon} v_{a_0}^\epsilon\Big\}
+\frac{2}{c^4}\Big\{ \frac{\partial \hat {\cal L}_a}{\partial x^0}+
{\textstyle\frac{1}{2}}
\Big(\frac{\partial \hat {\cal K}_a}{\partial x^0}\Big)^{\!2}+
\frac{v_{a_0}^\epsilon}{c}\frac{\partial \hat {\cal L}_a}{\partial x^\epsilon}-
\Big(\frac{\partial \hat {\cal K}_a}{\partial x^0}+
{\textstyle\frac{1}{2}}
v^{}_{a_0\epsilon} v_{a_0}^\epsilon\Big)^{\!2}\Big\}+\nonumber\\[2pt]
{}
&&\hskip 8pt+~\frac{2}{c^2}\sum_b\Big\{\Big(1-\frac{2}{c^2}v^{}_{a_0\epsilon} v_{a_0}^\epsilon\Big)\hat w_b(x)+
\frac{4}{c^2}{v^{}_{a_0}}_{\!\epsilon} \hat w^\epsilon_b(x)\Big\}+\nonumber\\[-5pt]
{}
&&\hskip 8pt+~
\frac{2}{c^4}\Big(\sum_b \hat w_b(x)+
\frac{\partial \hat {\cal K}_a}{\partial x^0}+ {\textstyle\frac{1}{2}}v^{}_{a_0\epsilon} v_{a_0}^\epsilon\Big)^{\!2}+{\cal O}(c^{-6}),\label{eq:g00-loc_ctv2}\\
{}
g^{0\alpha}(y_a(x))&=& \frac{4}{c^3}\sum_b \Big(\hat w^\alpha_b(x)-v_{a_0}^\alpha \hat w_b(x)\Big)+
\frac{1}{c^3}\Big\{\gamma^{\alpha\epsilon}\frac{1}{c}\frac{\partial \hat {\cal L}_a}{\partial x^\epsilon}+
c\frac{\partial \hat {\cal Q}^\alpha_a}{\partial x^0}+v_{a_0}^\epsilon\frac{\partial \hat {\cal Q}^\alpha_a}{\partial x^\epsilon}-
v_{a_0}^\alpha\frac{\partial \hat {\cal K}_a}{\partial x^0}\Big\}+{\cal O}(c^{-5}),~~~
\label{eq:g0a-loc_ctv2}\\[3pt]
{}
g^{\alpha\beta}(y_a(x))&=& \gamma^{\alpha\beta}
-\gamma^{\alpha\beta}\frac{2}{c^2} \Big(\sum_b \hat w_b(x)+
\frac{\partial \hat {\cal K}_a}{\partial x^0}+ {\textstyle\frac{1}{2}}v^{}_{a_0\epsilon} v_{a_0}^\epsilon\Big)+{\cal O}(c^{-4}).
\label{eq:gab-loc_ctv2}
\end{eqnarray}

In analogy with Eq.~(\ref{eq:N-source^mn}), we determine the source term, $S^{mn}(y_a(x))$, for the $N$-body problem in the local reference frame as:
\begin{equation}
S^{mn}(y_a(x))=\sum_b\frac{\partial y^m_a}{\partial x^k}\frac{\partial y^n_a}{\partial x^l}~S^{kl}_{b}(x),
\label{eq:N-source}
\end{equation}
where $S^{kl}_{b}(x(y_a))$ is the source term in the global frame (\ref{eq:Sab_tranf-glob}). Thus, $S^{mn}(y_a(x))$ takes the form:
{}
\begin{eqnarray}
S^{00}(y_a(x))&=& \frac{1}{2}c^2\sum_b\Big\{\Big(1 -
\frac{2}{c^2}v^{}_{a_0\epsilon}v^\epsilon_{a_0}\Big)\hat\sigma_b(x)+\frac{4}{c^2}v^{}_{a_0\epsilon} \hat\sigma^\epsilon_b(x)+
\frac{2}{c^2}
\big(\frac{\partial \hat{\cal K}_a}{\partial x^0}+
{\textstyle\frac{1}{2}}v^{}_{a_0\epsilon}v^\epsilon_{a_0}\big)\hat
\sigma_b(x)+{\cal O}(c^{-4})\Big\},
\label{eq:S00_tranf-loc-ctv}\\[3pt]
{}
S^{0\alpha}(y_a(x))&=& c\sum_b\Big\{\hat\sigma^{\alpha}_b(x)-
v^\alpha_{a_0}\hat\sigma_b(x)+{\cal O}(c^{-2})\Big\},
\label{eq:S0a_tranf-loc-ctv}\\[3pt]
{}
S^{\alpha\beta}(y_a(x))&=& -\gamma^{\alpha\beta}\frac{1}{2}c^2\sum_b
\Big\{\hat\sigma_b(x)+{\cal O}(c^{-2})\Big\}. \label{eq:Sab_tranf-loc-ctv}
\end{eqnarray}
We can the introduce transformation rules for the mass (scalar) and current (vector) densities of a body $b$ under coordinate transformation from the local coordinates $\{y^m_a\}$ of a body $a$ to global coordinates $\{x^m\}$, given by Eqs.~(\ref{eq:trans-0_inv})--(\ref{eq:trans-a_inv}) (cf. Eqs.~(\ref{eq:sig_0_tranf-loc}--\ref{eq:sig_a_tranf-loc})):
{}
\begin{eqnarray}
\sigma_b(y_a(x))&=& \Big(1 -
\frac{2}{c^2}v_{a_0\epsilon}v^\epsilon_{a_0}\Big)\hat \sigma_b(x)+
\frac{4}{c^2}v_{a_0\epsilon} \hat \sigma^\epsilon_b(x)
+\frac{2}{c^2}
\big(\frac{\partial \hat{\cal K}_a}{\partial x^0}+
{\textstyle\frac{1}{2}}v_{a_0\epsilon}v^\epsilon_{a_0}\big)
\hat \sigma_b(x)+{\cal O}(c^{-4}),~~~
\label{eq:sig_0_tranf-loc-inverse}\\[3pt]
{}
\sigma^\alpha_b(y_a(x))&=& \hat \sigma^{\alpha}_b(x)-
v^\alpha_{a_0}\hat \sigma_b(x)+{\cal O}(c^{-2}),
\label{eq:sig_a_tranf-loc-inverse}
\end{eqnarray}
where $\hat \sigma_b(x)$ and $\hat \sigma^\alpha_b(x)$ are given by (\ref{eq:sig0_tranf})--(\ref{eq:siga_tranf}), correspondingly.

The metric tensor (\ref{(B1a+*)})--(\ref{eq:(B1+*)}) and the partial-differential form of the gauge conditions (\ref{eq:harm-gauge-0})--(\ref{eq:harm-gauge-a}) together with condition (\ref{eq:(DeDgGa)OKsa-ctv}) allow one to simplify the expressions for the Ricci tensor and to present its contravariant components $R^{mn}=g^{mk}g^{nl}R_{kl}$ in the following form (similar to that of Eqs.~(\ref{eq:(1R00)})--(\ref{eq:(1Rab)})):
{}
\begin{eqnarray}
R^{00}(y_a(x)) &=& {1\over2} \Box_{y_a}
\Big(c^{-2}g^{[2]00}+ c^{-4}\Big\{g^{[4]00}-\frac{1}{2}\big(g^{[2]00}\big)^2\Big\}\Big)+
c^{-4}{1\over2}\Big(g^{[2]\epsilon\lambda}+\gamma^{\epsilon\lambda}g^{[2]00}\Big)\partial_\epsilon\partial_\lambda g^{[2]00}
+{\cal O}(c^{-6}), \label{eq:(1R00)-loc_ctv}\\
{}
R^{0\alpha}(y_a(x)) &=& c^{-3}{1\over2}
\Delta_{y_a} g^{[3]0\alpha}
+ {\cal O}(c^{-5}),
\label{eq:(1R0a)-loc_ctv}\\
{}
R^{\alpha\beta}(y_a(x)) &=& c^{-2}{1\over2}
\Delta_{y_a} g^{[2]\alpha\beta}
+{\cal O}(c^{-4}). \label{eq:(1Rab)-loc_ctv}
\end{eqnarray}

These expressions allow one to express the gravitational field equations as below:
{}
\begin{eqnarray}
\Box_{y_a}\Big[\sum_b\Big\{\Big(1-\frac{2}{c^2}v^{}_{a_0\epsilon} v_{a_0}^\epsilon\Big)\hat w_b(x)&+&
\frac{4}{c^2}{v^{}_{a_0}}_{\!\epsilon} \hat w^\epsilon_b(x)\Big\}
+
\nonumber\\[0pt]
{}
+\frac{\partial \hat {\cal K}_a}{\partial x^0}+
{\textstyle\frac{1}{2}}v^{}_{a_0\epsilon} v_{a_0}^\epsilon
+\frac{1}{c^2}\Big\{ \frac{\partial \hat {\cal L}_a}{\partial x^0}
&+&
{\textstyle\frac{1}{2}}
\Big(\frac{\partial \hat {\cal K}_a}{\partial x^0}\Big)^{2}
+
\frac{v_{a_0}^\epsilon}{c}\frac{\partial \hat {\cal L}_a}{\partial x^\epsilon}-
\Big(\frac{\partial \hat {\cal K}_a}{\partial x^0}+
{\textstyle\frac{1}{2}}
v^{}_{a_0\epsilon} v_{a_0}^\epsilon\Big)^{\!2}\Big\}+{\cal O}(c^{-4})\Big]=\nonumber\\[0pt]
{}
=
{4\pi G}\sum_b\Big\{\Big(1 -
\frac{2}{c^2}v^{}_{a_0\epsilon}v^\epsilon_{a_0}\Big)\hat\sigma_b(x)&+&\frac{4}{c^2}v^{}_{a_0\epsilon} \hat\sigma^\epsilon_b(x))+
 \frac{2}{c^2}\big(\frac{\partial \hat{\cal K}_a}{\partial x^0}+
{\textstyle\frac{1}{2}}v^{}_{a_0\epsilon}v^\epsilon_{a_0}\big)
\hat\sigma_b(x)+{\cal O}(c^{-4})\Big\},
\label{eq:GR00_loc-ctv}\\[3pt]
{}
\Delta_{y_a} \Big[
\sum_b \Big(\hat w^\alpha_b(x)-v_{a_0}^\alpha \hat w_b(x)\Big)
&+&\frac{1}{4}\Big\{\gamma^{\alpha\epsilon}\frac{1}{c}\frac{\partial \hat {\cal L}_a}{\partial x^\epsilon}+
c\frac{\partial \hat {\cal Q}^\alpha_a}{\partial x^0}+v_{a_0}^\epsilon\frac{\partial \hat {\cal Q}^\alpha_a}{\partial x^\epsilon}-
v_{a_0}^\alpha\frac{\partial \hat {\cal K}_a}{\partial x^0}\Big\}+{\cal O}(c^{-2})\Big]=\nonumber\\[0pt]
{}
&=& {4\pi G}\sum_b\Big\{\hat \sigma^{\alpha}_b(x)-
v^\alpha_{a_0}\hat \sigma_b(x)\Big\}+ {\cal O}(c^{-2}),
\label{eq:GR0a_loc-ctv}\\
{}
\Delta_{y_a} \Big[\sum_b \hat w_b(x)+
\frac{\partial \hat {\cal K}_a}{\partial x^0}+ {\textstyle\frac{1}{2}}v^{}_{a_0\epsilon} v_{a_0}^\epsilon
&+&{\cal O}(c^{-2})\Big]
= 4\pi G \sum_b\hat \sigma_b(x)+{\cal O}(c^{-2}).~~~
\label{eq:GRab_loc-ctv}
\end{eqnarray}
Remembering Eq.~(\ref{eq:Box_y-w}) and noticing that $ \Delta_{y_a}= \Delta_{x}+{\cal O}(c^{-2})$, with help of Eqs.~(\ref{eq:DeDoGa-K-hat})--(\ref{eq:DeDoGa-Q-hat}), we can verify that Eqs.~(\ref{eq:GR00_loc-ctv})--(\ref{eq:GRab_loc-ctv}) are satisfied.

Similarly to Eqs.~(\ref{eq:g00-cov*})--(\ref{eq:gab-cov*}), the  contra-variant components of the metric tenor $g^{mn}(y_a(x))$ given by Eqs.~(\ref{eq:g00-loc_ctv2})--(\ref{eq:gab-loc_ctv2}) can be expressed using two harmonic potentials as given below:
{}
\begin{eqnarray}
g^{00}(y_a(x))&=& 1+\frac{2}{c^2}w(x)+\frac{2}{c^4}w^2(x)+O(c^{-6}),
\label{eq:g00-ctv*}
\\[3pt]
{}
g^{0\alpha}(y_a(x))&=& \frac{4}{c^3}w^\alpha(x)+O(c^{-5}),
\label{eq:g0a-ctv*}\\[3pt]
{}
g^{\alpha\beta}(y_a(x))&=& \gamma^{\alpha\beta}-\gamma^{\alpha\beta}\frac{2}{c^2} w(x)+O(c^{-4}),
\label{eq:gab-ctv*}
\end{eqnarray}
where the total (gravitation plus inertia) local scalar $w(x)$ and vector $w^\alpha(x)$ potentials have the following form:
{}
\begin{eqnarray}
w(x)&=& \sum_b {w}_b(x)+
\frac{\partial \hat {\cal K}_a}{\partial x^0}+
{\textstyle\frac{1}{2}}v^{}_{a_0\epsilon} v_{a_0}^\epsilon +
\frac{1}{c^2}\Big\{ \frac{\partial \hat {\cal L}_a}{\partial x^0}+
\frac{v_{a_0}^\epsilon}{c}\frac{\partial \hat {\cal L}_a}{\partial x^\epsilon}+
{\textstyle\frac{1}{2}}\Big(\frac{\partial \hat {\cal K}_a}{\partial x^0}\Big)^2-
\Big(\frac{\partial \hat {\cal K}_a}{\partial x^0}+
{\textstyle\frac{1}{2}}v^{}_{a_0\epsilon} v_{a_0}^\epsilon\Big)^2\Big\}+{\cal O}(c^{-4}),~~~~~
\label{eq:pot_loc_tot-w_0-ctv*}\\[3pt]
{}
w^\alpha(x)&=& \sum_b {w}^\alpha_b(x)+{\textstyle\frac{1}{4}}\Big\{\gamma^{\alpha\epsilon}\frac{1}{c}\frac{\partial \hat {\cal L}_a}{\partial x^\epsilon}+
c\frac{\partial \hat {\cal Q}^\alpha_a}{\partial x^0}+v_{a_0}^\epsilon\frac{\partial \hat {\cal Q}^\alpha_a}{\partial x^\epsilon}-
v_{a_0}^\alpha\frac{\partial \hat {\cal K}_a}{\partial x^0}\Big\}+{\cal O}(c^{-2}),
\label{eq:pot_loc_tot-w_a-ctv*}
\end{eqnarray}
with ${w}_b(x)$ and ${w}^\alpha_b(x)$ being the scalar and vector potentials of body $b$ transformed to the local frame of body $a$, but expressed as functions of the global coordinates as
{}
\begin{eqnarray}
{w}_b(x)&=& \Big(1-\frac{2}{c^2}v^{}_{a_0\epsilon} v_{a_0}^\epsilon\Big)\hat w_b(x)+\frac{4}{c^2}{v^{}_{a_0}}_{\!\epsilon}
\hat w^\epsilon_b(x)+{\cal O}(c^{-4}),
\label{eq:pot_loc_grav-w_0-ctv**}\\[3pt]
{w}^\alpha_b(x)&=& \hat w^\alpha_b(x)-v_{a_0}^\alpha \hat w_b(x)+{\cal O}(c^{-2}).
\label{eq:pot_loc_grav-w_a-ctv**}
\end{eqnarray}

Clearly, the potentials ${w}(x)$ and ${w}^\alpha(x)$ introduced by Eqs.~(\ref{eq:pot_loc_tot-w_0-ctv*})--(\ref{eq:pot_loc_tot-w_a-ctv*}) satisfy the following harmonic equations in the local coordinates $\{y^m_a\}$:
{}
\begin{align}
\Box_{y_a} {w}(x)&={4\pi G}\sum_b\Big\{\Big(1 -
\frac{2}{c^2}v^{}_{a_0\epsilon}v^\epsilon_{a_0}\Big)\hat\sigma_b(x)+
\frac{4}{c^2}v^{}_{a_0\epsilon} \hat\sigma^\epsilon_b(x)
+\frac{2}{c^2}
\big(\frac{\partial \hat{\cal K}_a}{\partial x^0}+
\textstyle{\frac{1}{2}}v^{}_{a_0\epsilon}v^\epsilon_{a_0}\big)
\hat\sigma_b(x)+{\cal O}(c^{-4})\Big\}, ~~~~
\label{eq:eq-w+}\\[3pt]
{}
\Delta_{y_a} {w}^\alpha(x)&=4\pi G\sum_b\Big\{\hat\sigma^{\alpha}_b(x)-v^\alpha_{a_0}\hat\sigma_b(x)+{\cal O}(c^{-2})\Big\}.
\label{eq:eq-w^a+}
\end{align}
With the help of Eqs.~(\ref{eq:DeDoGa-K-hat})--(\ref{eq:DeDoGa-Q-hat}) and (\ref{eq:D'Alemb_y+==}) one can verify that both Eqs.~(\ref{eq:eq-w+}) and (\ref{eq:eq-w^a+}) are identically satisfied.

Finally, using the relations $c\partial/\partial y^0_a=c\partial/\partial x^0+v^\epsilon_{a_0}\partial/\partial x^\epsilon+{\cal O}(c^{-2})$ and $\partial/\partial y^\alpha_a=\partial/\partial x^\alpha+{\cal O}(c^{-2})$ in Eq.~(\ref{eq:(DeDgGa)OKs0-ctv}), one can verify that potentials ${w}_b(x)$ and ${w}^\alpha_b(x)$ satisfy the following continuity equation:
\begin{equation}
\Big(c\frac{\partial }{\partial x^0}+v^\epsilon_{a_0}\frac{\partial }{\partial x^\epsilon}\Big) {w}(x)+
\frac{\partial }{\partial x^\epsilon} {w}^\epsilon(x)=
{\cal O}(c^{-2}).  \label{eq:(DeDgGa)cont-eq+}
\end{equation}

These expressions allow us to present the scalar and vector potentials given by Eqs.~(\ref{eq:pot_loc_tot-w_0-ctv*})--(\ref{eq:pot_loc_tot-w_a-ctv*}) entirely as functions of the global coordinates. Remembering Eq.~(\ref{eq:D'Alemb_y+==}) and also the fact that
{}
\begin{eqnarray}
\Box_{x} \hat w_b(x) &=&{4\pi G}\hat \sigma_b(x) +{\cal O}(c^{-4}), \qquad
\Delta_{x} \hat w^\alpha_b(x) =
{4\pi G}\hat\sigma^\alpha_b(x) +{\cal O}(c^{-2}),
\label{eq:Delta_x}
\end{eqnarray}
where solutions are given as:
{}
\begin{eqnarray}
{\hat w}_b(x)&=&G\int d^3x'\frac{\hat\sigma_b(x')}{|\vec{x}-\vec{x}'|}+\frac{1}{2c^2}G\frac{c^2\partial^2}{\partial {x^0}^2}\int d^3x'
{\hat\sigma_b(x')}{|\vec{x}-\vec{x}'|}+{\cal O}(c^{-4}),
\label{eq:w_0_ctv_bar+0}\\
{\hat w}^\alpha_b(x)&=&G\int d^3x'\frac{\hat\sigma^\alpha_b(x')}{|\vec{x}-\vec{x}'|}
+{\cal O}(c^{-2}),
\label{eq:w_a_ctv_bar+0}
\end{eqnarray}
we can use these results to present the gravitational potentials (\ref{eq:w_0_ctv_bar+0})--(\ref{eq:w_a_ctv_bar+0}) in coordinates of the global frame in the form of the integrals over the body's volume as below (where $x^m$ is understood as $x^m=x^m(y_a)$):
{}
\begin{eqnarray}
{w}_b(x)&=& \big(1 -
\frac{2}{c^2}v^{}_{a_0\epsilon}v^\epsilon_{a_0}\big)
G\int d^3x'\frac{\hat\sigma_b(x')}{|\vec{x}-\vec{x}'|}
+\frac{4}{c^2}v^{}_{a_0\epsilon}
G\int d^3x'\frac{\hat\sigma^\epsilon_b
(x')}{|\vec{x}-\vec{x}'|}+
\nonumber\\
&&\hskip 15pt +~
\frac{1}{2c^2}G\frac{c^2\partial^2}{\partial {x^0}^2}\int d^3x'
{\hat\sigma_b(x')}{|\vec{x}-\vec{x}'|}+
{\cal O}(c^{-3}),
\label{eq:w_0_ctv_bar+fin02*}\\
{w}^\alpha_b(x)&=& G\int d^3x'\frac{\hat\sigma^\alpha_b(x')}{|\vec{x}-\vec{x}'|}-
v^\alpha_{a_0}G\int d^3x'\frac{\hat\sigma_b(x')}{|\vec{x}-\vec{x}'|}
+{\cal O}(c^{-2}).
\label{eq:w_a_ctv_bar+fin02*}
\end{eqnarray}

Although the expressions for the scalar inertial potentials have different functional dependence on the transformation functions (i.e., (${\cal K},{\cal L},{\cal Q}^\alpha$) vs. ($\hat{\cal K},\hat{\cal L},\hat{\cal Q}^\alpha$)), it is clear that both expressions for $w(y_a)$ given by Eqs.~(\ref{eq:pot_loc-w_0-cov*}) and (\ref{eq:pot_loc_tot-w_0-ctv*}) are identical. The same is true for the inertial vector potentials $w^\alpha(y_a)$ given by (\ref{eq:pot_loc-w_a-cov*}) and (\ref{eq:pot_loc_tot-w_a-ctv*}).

\subsection{Determining the functional form of the inverse transformation}

Using the harmonic gauge conditions, we determined that functions $\hat{\cal K}_a,\hat{\cal L}_a,\hat{\cal Q}_a^\alpha$ must satisfy Eq.~(\ref{eq:DeDoGa-K-hat})--(\ref{eq:DeDoGa-Q-hat}). Below, we will establish the functional form of these harmonic functions in a way, analogous to that presented in Sec.~\ref{sec:harm-fun_struct}.

\subsubsection{Determining the structure of $\hat{\cal K}_a$}

The general solution to Eq.~(\ref{eq:DeDoGa-K-hat}) can be sought in the following form: \begin{equation}
\hat{\cal K}_a (x) = \hat\kappa_{a_0}+ \hat\kappa_{a_0\mu} r^\mu_a + \delta \hat\kappa_a(x)+ {\cal O}(c^{-4}), \qquad {\rm where} \qquad
\delta \hat\kappa_a(x)=\sum_{k=2}\frac{1}{k!}\hat\kappa_{a_0\mu_1...\mu_k} (r^0)r^{\mu_1...}_ar^{\mu_k}_a + {\cal O}(c^{-4}),
\label{eq:K-gen-ctv}
\end{equation}
where $r^\alpha_a$ is defined as below:
{}
\begin{equation}
r^\alpha_a=x^\alpha-x^\alpha_{a_0}, \qquad x^\alpha_{a_0}=z^\alpha_{a_0}+c^{-2}{\hat q}^\alpha_{a_0}+{\cal O}(c^{-4}),
\label{eq:x0a}
\end{equation}
where $x^\alpha_{a_0}(x^0)$ is the position vector of the body $a$, complete to ${\cal O}(c^{-4})$, and expressed as a function of the global time-like coordinate $x^0$, given by Eq.~(\ref{eq:x_0Q}) and $\hat\kappa_{a_0\mu_1...\mu_k} (x^0)$ being Cartesian STF tensors \cite{Thorne-1980} which depend only on the time taken at the origin of the coordinate system on the observer's world-line. Substituting this form of the function $\hat{\cal K}_a$ in the equation (\ref{eq:form-inv_0a+}), we find solution for $\hat\kappa_{a_0\mu}$ and $\delta \hat\kappa_a$:
{}
\begin{equation}
\hat\kappa_{a_0\mu} =c v_{a_0\mu}  + {\cal O}(c^{-4}),
\qquad \delta \hat\kappa_a={\cal O}(c^{-4}).
\label{eq:Ka-ctv}
\end{equation}

As a result, the function $\hat{\cal K}_a$ that satisfies the harmonic gauge conditions was determined to be
{}
\begin{equation}
\hat{\cal K}_a (x) = \hat\kappa_{a_0} + c (v_{a_0\mu}r^\mu_a)  + {\cal O}(c^{-4}).
\label{eq:Ka-ctv+}
\end{equation}
This expression completely fixes the spatial dependence of the function $\hat{\cal K}_a$, but still has unknown dependence on time via function $\hat\kappa_{a_0} (x^0)$. This dependence determines the proper time on the observers world-line which will be done later.

\subsubsection{Determining the structure of $\hat{\cal Q}^\alpha_a$}

The general solution for the function $\hat{\cal Q}^\alpha_a$ that satisfies Eq.~(\ref{eq:DeDoGa-Q-hat}) may be presented as a sum of a solution of the inhomogeneous Poisson equation and a solution of the homogeneous Laplace equation. Furthermore, the part of that solution with regular behavior in the vicinity of the world-line may be given in the following form:
\begin{equation}
\hat{\cal Q}^\alpha_a (x) = - \hat q^\alpha_{a_0}+ \hat q^\alpha_{a_0\mu} r^\mu_a +
{\textstyle\frac{1}{2}}\hat q^\alpha_{a_0\mu\nu} r^\mu_a r^\nu_a + \delta\hat\xi^\alpha_a(x)+ {\cal O}(c^{-2}),
\label{eq:K-gen-ctv+}
\end{equation}
where $\hat q^\alpha_{a_0\mu\nu}$ can be determined directly from Eq.~(\ref{eq:DeDoGa-Q-hat}) and the function $\delta\hat\xi^\alpha_a$ satisfies the Laplace equation
\begin{eqnarray}
\gamma^{\epsilon\lambda}\frac{\partial^2}{\partial x^\epsilon \partial x^\lambda} \delta\hat\xi^\alpha_a &=& {\cal O}(c^{-2}).
\label{eq:DeDoGa-a_xi-ctv}
\end{eqnarray}

We can see that Eq.~(\ref{eq:DeDoGa-Q-hat}) can be used to determine $\hat q^\alpha_{a_0\mu\nu}$, but would leave the other terms in the equation unspecified. To determine these terms, we use Eq.~(\ref{eq:form-inv_ab+}) (which is equivalent to Eq.~(\ref{eq:DeDoGa-Q-hat})) together with Eq.~(\ref{eq:form-inv_0a+}), and obtain:
{}
\begin{eqnarray}
v_{a_0}^\alpha v_{a_0}^\beta+\gamma^{\alpha\lambda}
\frac{\partial \hat {\cal Q}^\beta_a}{\partial x^\lambda}+\gamma^{\beta\lambda}\frac{\partial \hat {\cal Q}^\alpha_a}{\partial x^\lambda}
+2\gamma^{\alpha\beta}\Big(\frac{\partial \hat{\cal K}_a}{\partial x^0}+
{\textstyle\frac{1}{2}}v^{}_{a_0\epsilon} v_{a_0}^\epsilon\Big)&=&{\cal O}(c^{-2}).
\label{eq:Q-form-inv-ctv}
\end{eqnarray}

Using the intermediate solution for the function $\hat{\cal K}_a$ from Eq.~(\ref{eq:Ka-ctv+}) in Eq.~(\ref{eq:Q-form-inv-ctv}), we obtain the following equation for $\hat{\cal Q}^\alpha_a$:
{}
\begin{eqnarray}
v_{a_0}^\alpha v_{a_0}^\beta+\gamma^{\alpha\lambda}
\frac{\partial \hat {\cal Q}^\beta_a}{\partial x^\lambda}+\gamma^{\beta\lambda}\frac{\partial \hat {\cal Q}^\alpha_a}{\partial x^\lambda}
+2\gamma^{\alpha\beta}\Big(\frac{\partial \hat\kappa_{a_0}}{\partial x^0}-
{\textstyle\frac{1}{2}}v^{}_{a_0\epsilon} v_{a_0}^\epsilon+a^{}_{0\epsilon} r^\epsilon_a\Big)&=&{\cal O}(c^{-2}).
\label{eq:form-inv_exp-ctv}
\end{eqnarray}

A trial solution to Eq.~(\ref{eq:form-inv_exp-ctv}) may be given in the following general from:
{}
\begin{eqnarray}
\hat{\cal Q}^\alpha_a (x) &=& -\hat q^{\alpha}_{a_0}+
c_1 v_{a_0}^\alpha v^{}_{a_0\epsilon} r^{\epsilon}_a +
c_2 v^{}_{a_0\epsilon} v_{a_0}^\epsilon r^{\alpha}_a +
c_3 a^{\alpha}_{a_0} r^{}_{a\epsilon} r^\epsilon_a+
c_4 a^{}_{a_0\epsilon} r^\epsilon_a r^{\alpha}_a +
c_5\big(\frac{\partial \hat\kappa_{a_0}}{\partial x^0}-
{\textstyle\frac{1}{2}}v^{}_{a_0\epsilon} v_{a_0}^\epsilon\big)r^\alpha_a
-
\nonumber\\
&&-~
\hat\omega_{a_0}^{\alpha\epsilon}r^{}_{a\epsilon}+\delta\hat\xi^\mu_a(x)+ {\cal O}(c^{-2}),
\label{eq:delta-Q-c-ctv}
\end{eqnarray}
where $\hat q^{\alpha}_{a_0}$ and  antisymmetric matrix $\hat \omega_{a_0}^{\alpha\epsilon}=-\hat\omega_{a_0}^{\epsilon\alpha}$ are some functions of time, and $c_1,..., c_5$ are constants and $\delta\hat\xi^\mu_a(x)$ is given by Eq.~(\ref{eq:DeDoGa-a_xi-ctv}) and is at least of the third order in spatial coordinates $r^\mu_a$, namely $\delta\hat\xi^\mu_a(x)\propto{\cal O}(|r^\mu_a|^3)$. Direct substitution of Eq.~(\ref{eq:delta-Q-c-ctv}) into Eq.~(\ref{eq:form-inv_exp-ctv}) results in the following unique solution for these coefficients:
{}
\begin{equation}
c_1 = -\frac{1}{2},\quad
c_2 = 0,\quad
c_3 = \frac{1}{2}, \quad
c_4 = -1,\quad
c_5 = -1.
\label{eq:delta-Q-c-sol-hat}
\end{equation}
As a result, function $\hat{\cal Q}^\alpha_a$ has the following structure
\begin{eqnarray}
\hat{\cal Q}^\alpha_a(x)&=& -\hat q^{\alpha}_{a_0} -
\Big({\textstyle\frac{1}{2}}v^\alpha_{a_0} v^\epsilon_{a_0}+
\hat\omega_{a_0}^{\alpha\epsilon}+\gamma^{\alpha\epsilon}\big(\frac{\partial \hat\kappa_{a_0}}{\partial x^0}-
{\textstyle\frac{1}{2}}v_{a_0}{}_\lambda v_{a_0}^\lambda\big)\Big)r_{a\epsilon} -
a^{}_{a_0\epsilon}\Big(r^\alpha_a r^\epsilon_a-
{\textstyle\frac{1}{2}}\gamma^{\alpha\epsilon}r^{}_{a\lambda} r^\lambda_a\Big)
+\delta\hat\xi^\alpha_a(x) + {\cal O}(c^{-2}),
\hskip 25pt
\label{eq:d-Q-hat}
\end{eqnarray}
where $\hat q^{\alpha}_{a_0}$ and $\hat\omega_{a_0}^{\alpha\epsilon}$ are yet to be determined.

By substituting Eq.~(\ref{eq:d-Q-hat}) into Eq.~(\ref{eq:form-inv_exp-ctv}), we see that the function $\delta\hat \xi^\alpha_a(x)$ in Eq.~(\ref{eq:d-Q-hat}) must satisfy the equation:
{}
\begin{equation}
\frac{\partial}{\partial x_\alpha} \delta\hat \xi^\beta_a + \frac{\partial}{\partial x_\beta} \delta\hat \xi^\alpha_a ={\cal O}(c^{-2}).
\label{eq:form-inv_xi-ctv}
\end{equation}
We keep in mind that the function $\delta\hat \xi^\alpha_a(x)$ must also satisfy Eq.~(\ref{eq:DeDoGa-a_xi-ctv}). The solution to the partial differential equation (\ref{eq:DeDoGa-a_xi-ctv}) with regular behavior on the world-line (i.e., when $|\vec r_a|\rightarrow 0$) can be given in powers of $r^\mu_a$ as
\begin{equation}
\delta\hat \xi^\alpha_a(x)=\sum_{k\ge3}^K\frac{1}{k!}\delta\hat \xi^\alpha_{a_0\,\mu_1...\mu_k}(x^0)r^{\mu_1}_a{}^{...}r^{\mu_k}_a+{\cal O}(|r^{\mu}_a|^{K+1}) + {\cal O}(c^{-2}),
\label{eq:delxi-hat}
\end{equation}
where $\delta\hat \xi^\alpha_{a_0\,\mu_1...\mu_k} (x^0)$ are STF tensors that depend only on time-like coordinate $x_0$. Using the solution (\ref{eq:delxi-hat}) in Eq.~(\ref{eq:form-inv_xi-ctv}), we can see that $\delta\hat \xi^\alpha_{a_0\,\mu_1...\mu_n}$ is also antisymmetric with respect to the index $\alpha$ and any of the spatial indices $\mu_1...\mu_k$. Combination of these two conditions suggests that $\delta\hat \xi^\alpha_{a_0\,\mu_1...\mu_n}=0$ for all $k\ge 3$, thus
{}
\begin{equation}
\delta\hat \xi^\alpha_a(x)={\cal O}(c^{-2}).
\end{equation}

Therefore, application of the harmonic gauge conditions leads to the following structure of function $\hat{\cal Q}^\alpha_a$:
\begin{eqnarray}
\hat {\cal Q}^\alpha_a(x)&=& -\hat q^{\alpha}_{a_0} -
\Big({\textstyle\frac{1}{2}}v^\alpha_{a_0} v^\epsilon_{a_0}+
\hat \omega_{a_0}^{\alpha\epsilon}+\gamma^{\alpha\epsilon}\big(\frac{\partial \hat \kappa_{a_0}}{\partial x^0}-
{\textstyle\frac{1}{2}}v^{}_{a_0\lambda} v_{a_0}^\lambda\big)\Big)r^{}_{a\epsilon} -
a^{}_{a_0\epsilon}\Big(r^\alpha_a r^\epsilon_a-
{\textstyle\frac{1}{2}}\gamma^{\alpha\epsilon}r^{}_{a\lambda} r^\lambda_a\Big)+ {\cal O}(c^{-2}),
\label{eq:d-Q-hat+}
\end{eqnarray}
where $\hat q^{\alpha}_{a_0},\hat \omega_{a_0}^{\alpha\epsilon}$ and $\hat \kappa_{a_0}$  are yet to be determined.

\subsubsection{Determining the structure of $\hat {\cal L}_a$}

We now turn our attention to the second gauge condition on the temporal coordinate transformation, Eq.~(\ref{eq:DeDoGa-L-hat}). Using the intermediate solution (\ref{eq:Ka-ctv+}) for the function $\hat{\cal K}_a$, we obtain the following equation for $\hat{\cal L}_a$:
{}
\begin{eqnarray}
\gamma^{\epsilon\lambda}\frac{\partial^2 \hat{\cal L}_a}{\partial x^\epsilon \partial x^\lambda}
&=& -c^2\frac{\partial^2 \hat{\cal K}_a}{\partial {x^0}^2}+
{\cal O}(c^{-2})=c\big(2v^{}_{a_0\epsilon} a_{a_0}^\epsilon-
\dot a^{}_{a_0\epsilon} r^\epsilon_a\big)-c^2\frac{\partial}{\partial x^0}\Big(\frac{\partial \hat\kappa_{a_0}}{\partial x^0}-
{\textstyle\frac{1}{2}}v^{}_{a_0\epsilon} v_{a_0}^\epsilon\Big)+{\cal O}(c^{-2}).
\label{eq:DeDoGa-0_LK-hat}
\end{eqnarray}

The general solution of Eq.~(\ref{eq:DeDoGa-0_LK-hat}) for $\hat{\cal L}_a$  may be presented as a sum of a solution $\delta \hat{\cal L}_a$ for the inhomogeneous Poisson equation and a solution $\delta{\hat{\cal L}}_{a0}$ of the homogeneous Laplace equation.
A trial solution of the  inhomogeneous equation to this equation, $\delta\hat{\cal L}_a$, is sought in the following form:
{}
\begin{equation}
\delta\hat{\cal L}_a(x) =
ck_1 (v^{}_{a_0\epsilon} a_{a_0}^\epsilon) (r^{}_{a\mu} r^\mu_a)+ck_2 (\dot{a}^{}_{a_0\epsilon}r^\epsilon_a)(r^{}_{a\nu} r^\nu_a)-
k_3c^2\frac{\partial}{\partial x^0}\Big(\frac{\partial \kappa_{a_0}}{\partial x^0}-
{\textstyle\frac{1}{2}}v^{}_{a_0\epsilon} v_{a_0}^\epsilon\Big)(r^{}_{a\nu} r^\nu_a) +{\cal O}(c^{-2}),
\label{eq:delta-L}
\end{equation}
where $k_1, k_2, k_3$ are some constants. Direct substitution of Eq.~(\ref{eq:delta-L}) into Eq.~(\ref{eq:DeDoGa-0_LK-hat}) yields the following values for these coefficients:
{}
\begin{equation}
k_1 = \frac{1}{3},\quad
k_2 = -\frac{1}{10}, \quad
k_3 = \frac{1}{6}.
\label{eq:delta-Q-k-sol-hat}
\end{equation}
As a result, the solution for $\delta\hat{\cal L}_a$ that satisfies the harmonic gauge conditions has the following form:
{}
\begin{eqnarray}
\delta\hat{\cal L}_a(x)&=&
{\textstyle\frac{1}{3}}c
\big(v^{}_{a_0\epsilon} a_{a_0}^\epsilon\big)(r^{}_{a\nu} r^\nu_a)-
{\textstyle\frac{1}{10}}c(\dot{a}^{}_{a_0\epsilon}r^\epsilon_a)
(r^{}_{a\nu} r^\nu_a) -
{\textstyle\frac{1}{6}}c^2\frac{\partial}{\partial x^0}\Big(\frac{\partial \kappa_{a_0}}{\partial x^0}-
{\textstyle\frac{1}{2}}v^{}_{a_0\epsilon} v_{a_0}^\epsilon\Big)(r^{}_{a\nu} r^\nu_a) +{\cal O}(c^{-2}).
\end{eqnarray}

The solution for the homogeneous equation with regular behavior on the world-line (i.e., when $|r^\mu_a|\rightarrow0$) may be presented as follows:
{}
\begin{eqnarray}
\hat{\cal L}_{a0}(x)&=&\hat\ell_{a_0}(x^0)+\hat\ell_{a_0\lambda}(x^0)\,r^\lambda_a+
{\textstyle\frac{1}{2}}\hat\ell_{a_0\lambda\mu}(x^0)\,r^\lambda_a r^\mu_a
+\delta\hat\ell_a(x)+{\cal O}(c^{-2}),
\end{eqnarray}
where $\hat\ell_{a_0\lambda\mu}$ is the STF tensor of second rank and $\delta\hat\ell_a$ is a function formed from the similar STF tensors of higher order
\begin{equation}
\delta\hat\ell_a(x)=\sum_{k\ge3}^K\frac{1}{k!}\delta\hat\ell_{a_0\,\mu_1...\mu_k}(x^0)r^{\mu_1}_a{}^{...}r^{\mu_k}_a+{\cal O}(|r^{\mu}_a|^{K+1})+
{\cal O}(c^{-2}).
\end{equation}

Finally, the general solution of the Eq.~(\ref{eq:DeDoGa-0_LK-hat}) may be presented as a sum of the special solution of inhomogeneous equation $\delta\hat{\cal L}_a$ and solution, $\hat{\cal L}_{a0}$, to the homogeneous equation, $\Delta \hat{\cal L}_a=0$.
Therefore, the general solution for the gauge equations for function $\hat{\cal L}_a(y)= \hat{\cal L}_{a0}+\delta\hat{\cal L}_a$ has the following form:
{}
\begin{eqnarray}
\hat{\cal L}_a(x)&=&\hat\ell_{a_0}+\hat\ell_{a_0\lambda}\,r^\lambda_a+
{\textstyle\frac{1}{2}}\hat\ell_{a_0\lambda\mu}\,r^\lambda_a r^\mu_a
+{\textstyle\frac{1}{3}}c
\big(v^{}_{a_0\epsilon} a_{a_0}^\epsilon\big)(r^{}_{a\nu} r^\nu_a)-
{\textstyle\frac{1}{10}}c(\dot{a}^{}_{a_0\epsilon}r^\epsilon_a)
(r^{}_{a\nu} r^\nu_a) -
\nonumber\\[3pt]
&&~~~~-~{\textstyle\frac{1}{6}}c^2\frac{\partial}{\partial x^0}\Big(\frac{\partial \hat\kappa_{a_0}}{\partial x^0}-
{\textstyle\frac{1}{2}}v^{}_{a_0\epsilon} v_{a_0}^\epsilon\Big)(r^{}_{a\nu} r^\nu_a)+\delta\hat\ell_a(x) +{\cal O}(c^{-2}).
\label{eq:L-gen-ctv}
\end{eqnarray}

We successfully determined the structure of the transformation functions ${\cal K}_a, {\cal Q}^\alpha_a$ and ${\cal L}_a$, which are imposed by the harmonic gauge conditions. Specifically, the harmonic structure for ${\cal K}_a$ is given by Eq.~(\ref{eq:Ka-ctv+}), the function  ${\cal Q}^\alpha_a$ was determined to have the structure given by Eq.~(\ref{eq:d-Q-hat+}), and the structure for ${\cal L}_a$ is given by  Eq.~(\ref{eq:L-gen-ctv}). The structure of the functions $\hat{\cal K}_a, \hat{\cal Q}^\alpha_a$ and $\hat{\cal L}_a$ is given in the following form:
{}
\begin{eqnarray}
\hat{\cal K}_a(x) &=& \hat\kappa_{a_0} + c (v_{a_0\mu}r^\mu_a)  + {\cal O}(c^{-4}),
\label{eq:K-sum-hat}\\
\hat{\cal Q}^\alpha_a(x)&=& -\hat q^{\alpha}_{a_0} -
\Big({\textstyle\frac{1}{2}}v^\alpha_{a_0} v^\epsilon_{a_0}+
\hat\omega_{a_0}^{\alpha\epsilon}+\gamma^{\alpha\epsilon}\big(\frac{\partial \hat\kappa_{a_0}}{\partial x^0}-
{\textstyle\frac{1}{2}}v^{}_{a_0\lambda} v_{a_0}^\lambda\big)\Big)r_{a\epsilon} -
a^{}_{a_0\epsilon}\Big(r^\alpha_a r^\epsilon_a-
{\textstyle\frac{1}{2}}\gamma^{\alpha\epsilon}r_{a\lambda} r^\lambda_a\Big)+ {\cal O}(c^{-2}),
\label{eq:Q-sum-hat}\\
\hat{\cal L}_a(x)&=&\hat\ell_{a_0}+\hat\ell_{a_0\lambda}\,r^\lambda_a+
{\textstyle\frac{1}{2}}\hat\ell_{a_0\lambda\mu}\,r^\lambda_a r^\mu_a
+{\textstyle\frac{1}{6}}c\Big(
2\big(v^{}_{a_0\epsilon} a_{a_0}^\epsilon\big)-
c\frac{\partial}{\partial x^0}\big(\frac{\partial \hat\kappa_{a_0}}{\partial x^0}-
{\textstyle\frac{1}{2}}v^{}_{a_0\epsilon} v_{a_0}^\epsilon\big)\Big)(r^{}_{a\lambda} r^\lambda_a)-\nonumber\\[3pt]
&&~~~~-~
{\textstyle\frac{1}{10}}c(\dot{a}^{}_{a_0\epsilon}r^\epsilon_a)
(r^{}_{a\nu} r^\nu_a) +\delta\hat\ell_a(x)+{\cal O}(c^{-2}).
\label{eq:L-sum-hat}
\end{eqnarray}

Note that the harmonic gauge conditions allow us to reconstruct the structure of the functions  with respect to spatial coordinate $x^\mu$. The time-dependent functions $\hat\kappa_{a_0}, \hat q^{\alpha}_{a_0},\hat\omega_{a_0}^{\alpha\epsilon}$, $\hat\ell_{a_0}, \hat\ell_{a_0\lambda}, \hat\ell_{a_0\lambda\mu}$, and $\delta\hat\ell_{a_0\,\mu_1...\mu_k}$ cannot be determined from the gauge conditions alone. Instead, just as we did in Sec.~\ref{sec:dyn}, we resort to a set of dynamic conditions to define these functions unambiguously by introducing the proper reference frame of a moving observer.

\subsection{Finding the form of the coordinate transformation functions}

In the case of the metric tensor given by the expressions Eqs.~(\ref{eq:g00-ctv*})--(\ref{eq:gab-ctv*}) and the total potentials Eqs.~(\ref{eq:pot_loc_tot-w_0-ctv*})--(\ref{eq:pot_loc_tot-w_a-ctv*}) representing gravity and inertia, the conditions Eqs.~(\ref{eq:fermi-cov_pot-0})--(\ref{eq:fermi-cov_pot-a}) taken at the world-line of body $a$ for which $r_a\equiv |r^\epsilon_a| = 0$, with $r^\epsilon_a=x^\epsilon-x^\epsilon_{a_0}$ given by Eq.~(\ref{eq:x0a}),  lead to the following set of equations:
{}
\begin{eqnarray}
\Big(\sum_{b\not=a}w_b(x)+ u_a(x)\Big)\Big|_{r_a=0}&=&{\cal O}(c^{-4}), \qquad\qquad  ~
\Big(\sum_{b\not=a}\frac{\partial w_b(x)}{\partial x^\beta}
+ \frac{\partial u_a(x)}{\partial x^\beta}\Big)\Big|_{r_a=0}=\,
{\cal O}(c^{-4}),
\label{eq:fermi-ctv_0}\\
\Big(\sum_{b\not=a}w^\alpha_b(x)+ u^\alpha_a(x)\Big)\Big|_{r_a=0} &=&
{\cal O}(c^{-2}), \qquad\qquad
\Big(\sum_{b\not=a}\frac{\partial w^\alpha_b(x)}{\partial x^\beta}+ \frac{\partial u^\alpha_a(x)}{\partial x^\beta}\Big)\Big|_{r_a=0}=\,
{\cal O}(c^{-2}),
\label{eq:fermi-ctv_a}
\end{eqnarray}
where, as seen in Eqs.~(\ref{eq:pot_loc_tot-w_0-ctv*})--(\ref{eq:pot_loc_tot-w_a-ctv*}), the scalar, $u_a(x)$, and vector, $u^\alpha_a(x)$, inertial potentials in global coordinates $\{x^m\}$ are given as (see also discussion in \cite{TMT:2011}):
{}
\begin{eqnarray}
u_a(x)&=& \frac{\partial \hat {\cal K}_a}{\partial x^0}+
{\textstyle\frac{1}{2}}v^{}_{a_0\epsilon} v_{a_0}^\epsilon +
\frac{1}{c^2}\Big\{ \frac{\partial \hat {\cal L}_a}{\partial x^0}+
\frac{v_{a_0}^\epsilon}{c}\frac{\partial \hat {\cal L}_a}{\partial x^\epsilon}+
{\textstyle\frac{1}{2}}\Big(\frac{\partial \hat {\cal K}_a}{\partial x^0}\Big)^2-
\Big(\frac{\partial \hat {\cal K}_a}{\partial x^0}+
{\textstyle\frac{1}{2}}v^{}_{a_0\epsilon} v_{a_0}^\epsilon\Big)^2\Big\}+{\cal O}(c^{-4}),~~~~~
\label{eq:inert_pot-u_0}\\[3pt]
{}
u^\alpha_a(x)&=& {\textstyle\frac{1}{4}}\Big\{\gamma^{\alpha\epsilon}\frac{1}{c}\frac{\partial \hat {\cal L}_a}{\partial x^\epsilon}+
c\frac{\partial \hat {\cal Q}^\alpha_a}{\partial x^0}+v_{a_0}^\epsilon\frac{\partial \hat {\cal Q}^\alpha_a}{\partial x^\epsilon}-
v_{a_0}^\alpha\frac{\partial \hat {\cal K}_a}{\partial x^0}\Big\}+{\cal O}(c^{-2}).
\label{eq:inert_pot-u_a}
\end{eqnarray}

The inertial potentials (\ref{eq:inert_pot-u_0}) and (\ref{eq:inert_pot-u_a}) depend on the first derivatives of the ``hatted'' transformation functions $\hat{\cal K}_a$, $\hat{\cal L}_a$, and $\hat{\cal Q}^\alpha_a$, which are determined from the harmonic gauge conditions up to several time-dependent functions as shown by (\ref{eq:K-sum-hat})--(\ref{eq:L-sum-hat}). To determine these yet-unknown functions and, thus, to completely identify $\hat{\cal K}_a$, $\hat{\cal L}_a$, and $\hat{\cal Q}^\alpha_a$, we impose the dynamical conditions (\ref{eq:fermi-ctv_0})--(\ref{eq:fermi-ctv_a}), which results in the following set of partial differential equations on the world-line of the local observer expressed via global coordinates $\{x^m\}$:
{}
\begin{eqnarray}
\sum_{b\not=a} \overline{w_b}+\frac{\partial \hat \kappa_{a_0}}{\partial x^0}-
{\textstyle\frac{1}{2}}v^{}_{a_0\epsilon} v_{a_0}^\epsilon
+\hskip 260pt &&
\nonumber\\
+~
\frac{1}{c^2}\Big\{-v^{}_{a_0\epsilon} \dot{\hat q}_{a_0}^\epsilon+
\frac{\partial \hat {\cal L}_a}{\partial x^0}+
\frac{v_{a_0}^\epsilon}{c}\frac{\partial \hat {\cal L}_a}{\partial x^\epsilon}+
{\textstyle\frac{1}{2}}
\Big(\frac{\partial \hat \kappa_{a_0}}{\partial x^0}-v^{}_{a_0\epsilon} v_{a_0}^\epsilon\Big)^2-
\Big(\frac{\partial \hat \kappa_{a_0}}{\partial x^0}-
{\textstyle\frac{1}{2}}v^{}_{a_0\epsilon} v_{a_0}^\epsilon\Big)^2\Big\}\Big|_{r_a=0}&=&
{\cal O}(c^{-4}),\hskip24pt
\label{eq:match-w_0-ctv}\\
\sum_{b\not=a}
\overline{\frac{\partial{w}_b}{\partial x^\beta}}+a^{}_{a_0\beta} +
\frac{1}{c^2}\Big\{ \frac{\partial^2 \hat{\cal L}_a}{\partial x^\beta \partial x^0}+\frac{v^\epsilon_{a_0}}{c}\frac{\partial^2 \hat{\cal L}_a}{\partial x^\beta\partial x^\epsilon}-
a_{a_0\beta}\frac{\partial \hat\kappa_{a_0}}{\partial x^0}\Big\}\Big|_{r_a=0}&=&{\cal O}(c^{-4}),\hskip24pt
\label{eq:match-w_0dir-ctv}\\
\sum_{b\not=a} \overline{w^\alpha_b}+
{\textstyle\frac{1}{4}}\Big\{ \gamma^{\alpha\epsilon}\frac{1}{c}\frac{\partial \hat {\cal L}_a}{\partial x^\epsilon}+
c\frac{\partial \hat {\cal Q}^\alpha_a}{\partial x^0}+v_{a_0}^\epsilon\frac{\partial \hat {\cal Q}^\alpha_a}{\partial x^\epsilon}-
v_{a_0}^\alpha\big(\frac{\partial \hat \kappa_{a_0}}{\partial x^0}-v^{}_{a_0\epsilon} v_{a_0}^\epsilon\big)\Big\}\Big|_{r_a=0}&=&{\cal O}(c^{-2}),
\label{eq:match-w_a-ctv}\\
\sum_{b\not=a} \overline{\frac{\partial{w}_\alpha^b}{\partial x^\beta}}+
{\textstyle\frac{1}{4}}\Big\{ \frac{1}{c}\frac{\partial^2 \hat{\cal L}_a}{\partial x^\alpha\partial x^\beta}+
c\gamma_{\alpha\lambda}\frac{\partial^2 \hat{\cal Q}^\lambda_a}{\partial x^0\partial x^\beta}+
\gamma_{\alpha\lambda}v^\epsilon_{a_0}\frac{\partial^2 \hat{\cal Q}^\lambda_a}{\partial x^\epsilon\partial x^\beta}-v_{a_0\alpha} a_{a_0\beta}\Big\}\Big|_{r_a=0}&=&{\cal O}(c^{-2}),
\label{eq:match-w_adir-ctv}
\end{eqnarray}
with all the quantifies above now being the functions of the global time-like coordinate $x^0$.

The equations above can now be used to determine uniquely the form of the coordinate transformation functions. From the first two conditions (\ref{eq:match-w_0-ctv}) and (\ref{eq:match-w_0dir-ctv}) above, we immediately obtain
{}
\begin{eqnarray}
\sum_{b\not=a} \overline{U_b}+\frac{\partial \hat \kappa_{a_0}}{\partial x^0}-
{\textstyle\frac{1}{2}}v^{}_{a_0\epsilon} v_{a_0}^\epsilon
&=&{\cal O}(c^{-4}), \label{eq:dir0-K-ctv}\\
\sum_{b\not=a} \overline{\frac{\partial{U}_b}{\partial x^\beta}}+a^{}_{a_0\beta}&=&{\cal O}(c^{-4}),
\label{eq:a-Newton-ctv}
\end{eqnarray}
where, similarly to (\ref{eq:w-U-cov}), we split the scalar potential $w_b(x)$ defined in Eq.~(\ref{eq:pot_loc_grav-w_0-ctv**}) into Newtonian ($U_b$) and post-Newtonian ($\delta w_b$) parts, as in Eq.~(\ref{eq:w-U-cov}). This allows us to present gravitational scalar and vector potentials defined by Eqs.~(\ref{eq:pot_loc_grav-w_0-ctv**})--(\ref{eq:pot_loc_grav-w_a-ctv**}) in the following form:
\begin{eqnarray}
{w}^{}_b(x)&=& U_b(x) +\frac{1}{c^2}\delta w_b(x)
+{\cal O}(c^{-4}),
\label{eq:pot_loc_grav-w_0-ctv*}\\[3pt]
{w}^\alpha_b(x)&=& \hat w^\alpha_b(x)-v_{a_0}^\alpha U_b(x)+{\cal O}(c^{-2}).
\label{eq:pot_loc_grav-w_a-ctv*}
\end{eqnarray}
where the post-Newtonian part of the scalar potential $\delta{w}^{}_b(x)$ is given as below:
\begin{equation}
\delta w_b(x)=\delta \hat w^{}_b(x)- 2(v^{}_{a_0\epsilon} v_{a_0}^\epsilon)U_b(x)+4{v^{}_{a_0}}_{\!\epsilon} \hat w^\epsilon_b(x)+{\cal O}(c^{-2}).
\label{eq:w-U}
\end{equation}

Eqs.~(\ref{eq:dir0-K-ctv}) and (\ref{eq:K-sum-hat}) allow us to determine function $\hat\kappa_{a_0}$ and to present the solution for $\hat{\cal K}_a$ as below:
{}
\begin{equation}
\hat{\cal K}_a(x)= -\int_{x^0_0}^{x^0}\!\!\!
\Big(\sum_{b\not=a} \overline{U_b}-
{\textstyle\frac{1}{2}}v^{}_{a_0\epsilon} v_{a_0}^{\epsilon}
\Big)dx'^0+
c(v^{}_{a_0\epsilon} r_a^\epsilon) + {\cal O}(c^{-4}).
\label{eq:Ka-sol-ctv}
\end{equation}

Substituting this expression into Eq.~(\ref{eq:Q-sum-hat}), we can determine $\hat{\cal Q}^\alpha_a$:
\begin{eqnarray}
\hat{\cal Q}^\alpha_a(x)&=& -\hat q^{\alpha}_{a_0} +
\Big(\gamma^{\alpha\epsilon}\sum_{b\not=a} \overline{U_b}-{\textstyle\frac{1}{2}}v^\alpha_{a_0} v^\epsilon_{a_0}-
\hat\omega_{a_0}^{\alpha\epsilon}\Big)r_{a\epsilon} -
a^{}_{a_0\epsilon}\Big(r^\alpha_a r^\epsilon_a-
{\textstyle\frac{1}{2}}\gamma^{\alpha\epsilon}r_{a\lambda} r^\lambda_a\Big)+ {\cal O}(c^{-2}).
\label{eq:d-Q-hat=}
\end{eqnarray}

Finally, the general solution, given by Eq.~(\ref{eq:L-sum-hat}), for the function $\hat{\cal L}_a$ takes the following form:
{}
\begin{eqnarray}
\hat{\cal L}_a(x)&=&\hat\ell_{a_0}+\hat\ell_{a_0\lambda}\,r^\lambda_a+
{\textstyle\frac{1}{2}}\Big(\hat\ell_{a_0\lambda\mu}
+c\gamma_{\lambda\mu}\big(
{\textstyle\frac{2}{3}}\big(v^{}_{a_0\epsilon} a_{a_0}^\epsilon\big)+
{\textstyle\frac{1}{3}}c\frac{\partial}{\partial x^0}\sum_{b\not=a} \overline{U_b}\big)
\Big)r^\lambda_a r^\mu_a-
{\textstyle\frac{1}{10}}c(\dot{a}^{}_{a_0\epsilon}r^\epsilon_a)
(r^{}_{a\nu} r^\nu_a) +
\nonumber\\
&&~~~~+~
\delta\hat\ell_a(x) +{\cal O}(c^{-2}).
\label{eq:L-gen-ctv++}
\end{eqnarray}

This equation allows us to proceed onto finding the function $\hat{\cal L}_a$. To do this, we rely on the yet-unused parts of Eqs.~(\ref{eq:match-w_0-ctv})--(\ref{eq:match-w_adir-ctv}) which take the following form
{}
\begin{eqnarray}
\sum_{b\not=a} \overline{\delta{w}_b} -v^{}_{a_0\epsilon} \dot{\hat q}_{a_0}^\epsilon+
\Big\{\frac{\partial \hat {\cal L}_a}{\partial x^0}+
\frac{v_{a_0}^\epsilon}{c}\frac{\partial \hat {\cal L}_a}{\partial x^\epsilon}+
{\textstyle\frac{1}{2}}
\Big(\frac{\partial \hat \kappa_{a_0}}{\partial x^0}-v^{}_{a_0\epsilon} v_{a_0}^\epsilon\Big)^2-
\Big(\frac{\partial \hat \kappa_{a_0}}{\partial x^0}-
{\textstyle\frac{1}{2}}v^{}_{a_0\epsilon} v_{a_0}^\epsilon\Big)^2\Big\}\Big|_{r_a=0}&=&
{\cal O}(c^{-2}),\hskip24pt
\label{eq:match-w_0-ctv-L}\\
\sum_{b\not=a} \overline{\frac{\partial \delta{w}_b}{\partial x^\beta}}+
\Big\{ \frac{\partial^2 \hat{\cal L}_a}{\partial x^\beta \partial x^0}+\frac{v^\epsilon_{a_0}}{c}\frac{\partial^2 \hat{\cal L}_a}{\partial x^\beta\partial x^\epsilon}-
a_{a_0\beta}\frac{\partial \hat\kappa_{a_0}}{\partial x^0}\Big\}\Big|_{r_a=0}&=&{\cal O}(c^{-2}),\hskip24pt
\label{eq:match-w_0dir-ctv-L}\\
\sum_{b\not=a} \overline{w^\alpha_b}+
{\textstyle\frac{1}{4}}\Big\{ \gamma^{\alpha\epsilon}\frac{1}{c}\frac{\partial \hat {\cal L}_a}{\partial x^\epsilon}+
c\frac{\partial \hat {\cal Q}^\alpha_a}{\partial x^0}+v_{a_0}^\epsilon\frac{\partial \hat {\cal Q}^\alpha_a}{\partial x^\epsilon}-
v_{a_0}^\alpha\big(\frac{\partial \hat \kappa_{a_0}}{\partial x^0}-v^{}_{a_0\epsilon} v_{a_0}^\epsilon\big)\Big\}\Big|_{r_a=0}&=&{\cal O}(c^{-2}),
\label{eq:match-w_a-ctv-L}\\
\sum_{b\not=a} \overline{\frac{\partial{w}_\alpha^b}{\partial x^\beta}}+
{\textstyle\frac{1}{4}}\Big\{ \frac{1}{c}\frac{\partial^2 \hat{\cal L}_a}{\partial x^\alpha\partial x^\beta}+
c\gamma_{\alpha\lambda}\frac{\partial^2 \hat{\cal Q}^\lambda_a}{\partial x^0\partial x^\beta}+
\gamma_{\alpha\lambda}v^\epsilon_{a_0}\frac{\partial^2 \hat{\cal Q}^\lambda_a}{\partial x^\epsilon\partial x^\beta}-v_{a_0\alpha} a_{a_0\beta}\Big\}\Big|_{r_a=0}&=&{\cal O}(c^{-2}).
\label{eq:match-w_adir-ctv-L}
\end{eqnarray}

Using the expressions for the functions $\hat{\cal K}_a$, $\hat{\cal Q}^\alpha_a$ given by Eqs.~(\ref{eq:Ka-sol-ctv}) and (\ref{eq:d-Q-hat=}) correspondingly together with the expression (\ref{eq:L-gen-ctv++}) for $\hat{\cal L}_a$, and also using transformation rule for the gravitational potentials given by Eqs.~(\ref{eq:pot_loc_grav-w_0-ctv*}) and (\ref{eq:pot_loc_grav-w_a-ctv*}) together with the Newtonian equations of motion (\ref{eq:a-Newton-ctv}), we can now obtain the equations needed to determine the remaining unknown functions $\hat\ell_{a_0}, \hat\ell_{a_0\lambda}$, $\hat\ell_{a_0\lambda\mu}$, $\hat\omega_{a_0}^{\alpha\beta}$ and ${\hat q}^\alpha_{a_0}$.

Eq.~(\ref{eq:match-w_0-ctv-L}) leads to the following equation for $\hat\ell_{a_0}$:
{}
\begin{eqnarray}
{\textstyle\frac{1}{c}}\dot{\hat\ell}_{a_0}&=&v^{}_{a_0\epsilon} \dot{\hat q}_{a_0}^\epsilon
-{\textstyle\frac{1}{8}}(v^{}_{a_0\epsilon}v^\epsilon_{a_0})^2+
{\textstyle\frac{3}{2}}(v^{}_{a_0\epsilon}v^\epsilon_{a_0})\sum_{b\not=a} \overline{U_b}-
4v^{}_{a_0\epsilon}\!\sum_{b\not=a} \bar{\hat w}^\epsilon_b+
{\textstyle\frac{1}{2}}(\sum_{b\not=a} \overline{U_b})^2-
\sum_{b\not=a} \overline{\delta{\hat w}_b}+{\cal O}(c^{-2}).
\label{eq:ell_0-hat}
\end{eqnarray}

Next, Eq.~(\ref{eq:match-w_0dir-ctv-L}) results in the equation for $\dot{\hat\ell}^\alpha_{a_0}$:
{}
\begin{eqnarray}
{\textstyle\frac{1}{c}}\dot{\hat\ell}^\alpha_{a_0}&=&
-\sum_{b\not=a} \overline{\frac{\partial \delta{\hat w}_b}{\partial x_\alpha}}-
{\textstyle\frac{3}{2}}(v^{}_{a_0\epsilon}v^\epsilon_{a_0})a^\alpha_{a_0}-
a^\alpha_{a_0}\!\sum_{b\not=a} \overline{U_b}-4v^{}_{a_0\epsilon}\sum_{b\not=a} \overline{\partial^\alpha {\hat w}^\epsilon_b}
+{\cal O}(c^{-2}).
\label{eq:ell_a-dot-hat}
\end{eqnarray}

From Eq.~(\ref{eq:match-w_a-ctv-L}) we determine ${\hat\ell}^\alpha_{a_0}$:
{}
\begin{eqnarray}
{\textstyle\frac{1}{c}}{\hat\ell}^\alpha_{a_0}&=&\dot {\hat q}^\alpha_{a_0}-
{\textstyle\frac{1}{2}} v^\alpha_{a_0}(v^{}_{a_0\epsilon}v^\epsilon_{a_0})+3v^\alpha_{a_0}\sum_{b\not=a} \overline{U_b}-4\sum_{b\not=a} \overline{{\hat w}^\alpha_b}
+{\cal O}(c^{-2}).
\label{eq:ell_a-hat}
\end{eqnarray}

Furthermore, Eq.~(\ref{eq:match-w_adir-ctv-L}) leads us to the following solution for $\hat\ell^{\alpha\beta}_{a_0}$:
{}
\begin{eqnarray}
{\textstyle\frac{1}{c}}{\hat\ell}^{\alpha\beta}_{a_0}&=&-4\sum_{b\not=a}
\overline{\frac{\partial {\hat w}^{\alpha}_b}{\partial x_{\beta}}}-{\textstyle\frac{4}{3}}\gamma^{\alpha\beta}\sum_{b\not=a}
c\overline{\frac{\partial U_b}{\partial x^0}}-
{\textstyle\frac{5}{2}}v^\alpha_{a_0}a^\beta_{a_0}+
{\textstyle\frac{1}{2}}v^\beta_{a_0}a^\alpha_{a_0}-
{\textstyle\frac{2}{3}}\gamma^{\alpha\beta}v^{}_{a_0\epsilon}a_{a_0}^\epsilon+
{\dot{\hat\omega}}_{a_0}^{\alpha\beta}
+{\cal O}(c^{-2}).
\label{eq:ell_ab-hat0}
\end{eqnarray}

Using the same argument concerning the symmetry properties of $\hat\ell^{\alpha\beta}_{a_0}$ that led to Eq.~(\ref{eq:omega+}), we find the following unique choice for the anti-symmetric matrix ${\dot{\hat \omega}}_{a_0}^{\alpha\beta}$ that symmetrizes Eq.~(\ref{eq:ell_ab-hat0}):
{}
\begin{eqnarray}
{\dot{\hat\omega}}_{a_0}^{\alpha\beta}&=&{\textstyle\frac{3}{2}}(v^\alpha_{a_0} a^\beta_{a_0}-v^\beta_{a_0} a^\alpha_{a_0})-
2\sum_{b\not=a}\Big(
\overline{\frac{\partial{\hat w}^\beta_b}{\partial x_\alpha}}-
\overline{\frac{\partial{\hat w}^\alpha_b}{\partial x_\beta}}\Big)+
{\cal O}(c^{-2}),
\label{eq:omega-hat}
\end{eqnarray}
which, if we now account for the Newtonian equation of motion (\ref{eq:a-Newton-ctv}), actually has the following form:
{}
\begin{eqnarray}
{\dot{\hat\omega}}_{a_0}^{\alpha\beta}&=&-{\textstyle\frac{1}{2}}(v^\alpha_{a_0} a^\beta_{a_0}-v^\beta_{a_0} a^\alpha_{a_0})-
2\sum_{b\not=a}\Big(v^\alpha_{a_0}\overline{\frac{\partial{U}^{}_b}{\partial x_\beta}}-v^\beta_{a_0}\overline{\frac{\partial{U}^{}_b}{\partial x_\alpha}}\Big)-
2\sum_{b\not=a}(\overline{\partial^\alpha{\hat w}^\beta_b}-
\overline{\partial^\beta{\hat w}^\alpha_b})+
{\cal O}(c^{-2}).
\label{eq:omega-hat2}
\end{eqnarray}
The first term in the equation above is the special relativistic Thomas precession, the second term is the geodetic precession and the last term is the Lense-Thirring precession (compare this to its analog in a local frame, Eq.~(\ref{eq:omega+})).

The result (\ref{eq:omega-hat}) allows us to present Eq.~(\ref{eq:ell_ab-hat0}) for the function ${\hat\ell}^{\alpha\beta}_{a_0}$ in the following form:
{}
\begin{eqnarray}
{\textstyle\frac{1}{c}}{\hat\ell}^{\alpha\beta}_{a_0}&=&-2\sum_{b\not=a}\Big(
\overline{\frac{\partial {\hat w}^{\beta}_b}{\partial x_{\alpha}}}+\overline{\frac{\partial {\hat w}^{\alpha}_b}{\partial x_{\beta}}}\Big)-{\textstyle\frac{4}{3}}\gamma^{\alpha\beta}\sum_{b\not=a}
c\overline{\frac{\partial U_b}{\partial x^0}}-
\Big(v^\alpha_{a_0}a^\beta_{a_0}+v^\beta_{a_0}a^\alpha_{a_0}+{\textstyle\frac{2}{3}}\gamma^{\alpha\beta}v^{}_{a_0\epsilon}a_{a_0}^\epsilon\Big)+{\cal O}(c^{-2}).
\label{eq:ell_ab-hat}
\end{eqnarray}

Finally, Eqs.~(\ref{eq:ell_a-dot-hat}) and (\ref{eq:ell_a-hat}) allow us to determine the equation for $\hat q_{a_0}^{\alpha}$. Indeed, differentiating  Eq.~(\ref{eq:ell_a-hat}) with respect to time and subtracting the result from Eq.~(\ref{eq:ell_a-dot-hat}), we obtain:
{}
\begin{eqnarray}
\ddot {\hat q}_{a_0}^{\alpha}(x^0)
&=& -\gamma^{\alpha\epsilon}\sum_{b\not=a}
\overline{\frac{\partial\delta{\hat w}_b}{\partial x^\epsilon}}+
4\sum_{b\not=a} c\overline{\frac{\partial{\hat w}^\alpha_b}{\partial x^0}}-
4v^{}_{a_0\lambda}\gamma^{\alpha\epsilon}\sum_{b\not=a}
\overline{\frac{\partial{\hat w}^\lambda_b}{\partial x^\epsilon}}-
4a^\alpha_{a_0}\sum_{b\not=a} \overline{U_b}-\nonumber\\
&&-
3v^\alpha_{a_0}\sum_{b\not=a} c\overline{\frac{\partial{U}_b}{\partial x^0}}-
(v^{}_{a_0\epsilon}v_{a_0}^\epsilon)a^\alpha_{a_0}+v^\alpha_{a_0}(v^{}_{a_0\epsilon}a_{a_0}^\epsilon)+
{\cal O}(c^{-2}).
\label{eq:q-ddot-hat}
\end{eqnarray}

By combining Eqs.~(\ref{eq:a-Newton-ctv}) and (\ref{eq:q-ddot-hat}), we obtain the equations of motion of the body $a$ with respect to the inertial reference frame $\{x^m\}$:
{}
\begin{eqnarray}
\ddot x^\alpha_{a_0}(x^0)&=&\ddot z^\alpha_{a_0}+c^{-2}\ddot {\hat q}_{a_0}^{\alpha}+
{\cal O}(c^{-4})= \nonumber\\
&=&
-\gamma^{\alpha\epsilon}\sum_{b\not=a} \overline{\frac{\partial{\hat w}_b}{\partial x^\epsilon}}+\frac{1}{c^2}
\Big(
4\sum_{b\not=a} c\overline{\frac{\partial {\hat w}^\alpha_b}{\partial x^0}}-
4v^{}_{a_0\lambda}\gamma^{\alpha\epsilon}\sum_{b\not=a}
\overline{\frac{\partial{\hat w}^\lambda_b}{\partial x^\epsilon}}-
4a^\alpha_{a_0}\sum_{b\not=a} \overline{U_b}-\nonumber\\
&&\hskip 140pt -~
3v^\alpha_{a_0}\sum_{b\not=a} c\overline{\frac{\partial {U}_b}{\partial x^0}}-
(v^{}_{a_0\epsilon}v_{a_0}^\epsilon)a^\alpha_{a_0}+
v^\alpha_{a_0}(v^{}_{a_0\epsilon}a_{a_0}^\epsilon)\Big)+
{\cal O}(c^{-4}).
\label{eq:geod_eq-local-hat}
\end{eqnarray}
The equations of motion (\ref{eq:geod_eq-local-hat}) establish the corresponding barycentric equations of motion of the body $a$.

\subsection{Summary of results for the inverse transformation}
\label{sec:inverse-summary}

In this section, we sought to write the transformation between global coordinates $\{x^k\}$ of the barycentric reference system and local coordinates $\{y^k_a\}$ introduced in the proper body-centric coordinate reference system associated with the body $a$  in the form:
{}
\begin{eqnarray}
y^0_a&=& x^0+c^{-2}\hat{\cal K}_a(x^k)+c^{-4}\hat{\cal L}_a(x^k)+{\cal O}(c^{-6})x^0,\\[3pt]
y^\alpha_a&=& x^\alpha-z^\alpha_{a_0}(x^0)+c^{-2}\hat{\cal Q}^\alpha_a(x^k)+{\cal O}(c^{-4}).
\end{eqnarray}

Imposing the same conditions discussed in Sec.~\ref{sec:sum-direct}, we were able to find explicit forms for the transformation functions $\hat{\cal K}_a, \hat{\cal L}_a$ and $\hat{\cal Q}^\alpha_a$:
{}
\begin{eqnarray}
\hat{\cal K}_a(x)&=& -\!\int_{x^0_{0}}^{x^0}\!\!\!
\Big(\sum_{b\not=a} \overline{U_b}-
{\textstyle\frac{1}{2}}v^{}_{a_0\epsilon} v_{a_0}^\epsilon
\Big)dx'^0+
c(v^{}_{a_0\epsilon} r_a^\epsilon) +{\cal O}(c^{-4}),\\
{}
\hat{\cal Q}^\alpha_a(x)&=& -{\hat q}^{\alpha}_{a_0} +
\Big(\gamma^{\alpha\epsilon}\sum_{b\not=a} \overline{U_b}-
{\textstyle\frac{1}{2}}v^\alpha_{a_0} v^\epsilon_{a_0}-
\hat\omega_{a_0}^{\phantom{0}\alpha\epsilon}\Big)r_{a\epsilon} -
{a_{a_0}}_\epsilon\Big(r^\alpha_a r^\epsilon_a-{\textstyle\frac{1}{2}}
\gamma^{\alpha\epsilon}{r_a}_\lambda r_a^\lambda\Big)+{\cal O}(c^{-2}),
\end{eqnarray}
where the anti-symmetric relativistic precession matrix ${\hat\omega}_{a_0}^{\alpha\beta}$ in the barycentric reference frame is given as:
\begin{equation}
\dot{\hat\omega}_{a_0}^{\alpha\beta}=
{\textstyle\frac{3}{2}}(v^\alpha_{a_0} a^\beta_{a_0}-v^\beta_{a_0} a^\alpha_{a_0})-
2\sum_{b\not=a}\Big(
\overline{\frac{\partial{\hat w}^\beta_b}{\partial x_\alpha}}-
\overline{\frac{\partial{\hat w}^\alpha_b}{\partial x_\beta}}\Big)+
{\cal O}(c^{-2}).
\end{equation}

Also, the function $\hat{\cal L}_a$ was determined in the following form:
\begin{eqnarray}
\hat{\cal L}_a(x)&=& \hat\ell_{a_0}+\hat\ell_{a_0\lambda}\,r_a^\lambda-
{\textstyle\frac{1}{2}}c \,r^{}_{a\lambda} r^{}_{a\mu}
\Big[
v^\lambda_{a_0} a_{a_0}^\mu + v^\mu_{a_0} a_{a_0}^\lambda+\gamma^{\lambda\mu} \sum_{b\not=a}\dot{\overline{U_b}}+
2\sum_{b\not=a}\Big(\overline{\frac{\partial {\hat w}^{\mu}_b}{\partial x_{\lambda}}}+
\overline{\frac{\partial {\hat w}^{\lambda}_b}{\partial x_{\mu}}}\Big)\Big]-\nonumber\\
&&-
{\textstyle\frac{1}{10}}c(\dot{a}^{}_{a_0\epsilon}r^\epsilon_a)(r^{}_{a\nu}r^\nu_a)+ \delta\hat\ell_a(x)+{\cal O}(c^{-2}),
\end{eqnarray}
{}
with the functions ${\hat\ell}_{a_0}$ and ${\hat\ell}^\alpha_{a_0}$ given by
{}
\begin{eqnarray}
{\textstyle\frac{1}{c}}\dot{\hat\ell}_{a_0}&=&
v^{}_{a_0\epsilon} \dot{\hat q}_{a_0}^\epsilon
-{\textstyle\frac{1}{8}}(v^{}_{a_0\epsilon}v^\epsilon_{a_0})^2+
{\textstyle\frac{3}{2}}(v^{}_{a_0\epsilon}v^\epsilon_{a_0})\sum_{b\not=a} \overline{U_b}-
4v^{}_{a_0\epsilon}\!\sum_{b\not=a} \overline{{\hat w}^\epsilon_b}+
{\textstyle\frac{1}{2}}\big(\sum_{b\not=a} \overline{U_b}\big)^2-
\sum_{b\not=a} \overline{\delta{\hat w}_b}+{\cal O}(c^{-2}),\\
{}
{\textstyle\frac{1}{c}}{\hat\ell}^\alpha_{a_0}&=&\dot {\hat q}^\alpha_{a_0}-
{\textstyle\frac{1}{2}} v^\alpha_{a_0}(v^{}_{a_0\epsilon}v^\epsilon_{a_0})+3v^\alpha_{a_0}\sum_{b\not=a} \overline{U_b}-
4\sum_{b\not=a} \overline{{\hat w}^\alpha_b}
+{\cal O}(c^{-2}).
\end{eqnarray}

Finally, the function ${\hat q}_{a_0}^{\alpha}$ is the post-Newtonian part of the position vector of the body $a$ given by Eq.~(\ref{eq:q-ddot-hat}). Together with the Newtonian part of the acceleration, $\ddot{z}_{a_0}=a^\alpha_{a_0}$, the post-Newtonian part $\ddot {\hat q}_{a_0}^{\alpha}$ completes the barycentric equation of motions of the body $a$ given by Eq.~(\ref{eq:geod_eq-local-hat}). This equation essentially is the geodetic equation of motion written for the body $a$ in the global reference frame.

Substituting these solutions for the functions $\hat{\cal K}_a, \hat{\cal Q}^\alpha_a$ and $\hat{\cal L}_a$ into the expressions for the potentials $w$ and $w^\alpha$ given by Eqs.~(\ref{eq:pot_loc_tot-w_0-ctv*})--(\ref{eq:pot_loc_tot-w_a-ctv*}), we find the following form for these potentials:
{}
\begin{eqnarray}
w(x)&=&\sum_b w_b(x)-\sum_{b\not=a}\Big(\overline{\hat w_b} +r^\epsilon_a \overline{\frac{\partial{\hat w}_b}{\partial x^\epsilon}} \Big)-
\nonumber\\
&&\hskip 35pt ~-
\frac{1}{c^2}\Big\{{\textstyle\frac{1}{2}}
r^{\epsilon}_{a}r^{\lambda}_{a}\Big[
3a^{}_{a_0\epsilon} a^{}_{a_0\lambda}+
{\dot a}^{}_{a_0\epsilon} v^{}_{a_0\lambda}+
v^{}_{a_0\epsilon}{\dot a}^{}_{a_0\lambda}
+
\gamma_{\epsilon\lambda}\sum_{b\not=a}
\ddot{\overline{U_b}}+2\sum_{b\not=a}\Big(
\frac{\dot{\overline{\partial\hat{\hskip -8pt w\hskip 9pt}_{b\lambda}}}}{\partial x^\epsilon} +
\frac{\dot{\overline{\partial\hat{\hskip -7pt w\hskip 8pt}_{b\epsilon}}}}{\partial x^\lambda}\Big)\Big]
+\nonumber\\
&&\hskip 90pt ~+
{\textstyle\frac{1}{10}}
(\ddot a^{}_{a_0\epsilon} r^\epsilon_a)(r^{}_{a\mu} r_a^\mu)-
\Big(c\frac{\partial}{\partial x^0}+ v^\epsilon_{a_0}\frac{\partial}{\partial x^\epsilon}\Big) {\textstyle\frac{1}{c}}\delta\hat\ell_a\Big\}+
{\cal O}(c^{-4}),
\label{eq:pot_loc-w_0-ctv+*}\\[3pt]
w^\alpha(x)&=&\sum_b w^\alpha_b(x)-\sum_{b\not=a}\Big(\overline{\hat w^\alpha_b} +r_a^\epsilon \overline{\frac{\partial{\hat w}^\alpha_b}{\partial x^\epsilon}} \Big) -
{\textstyle\frac{1}{10}}\big\{3r^\alpha_a r^\epsilon_a-
\gamma^{\alpha\epsilon}r^{}_{a\mu} r^\mu_a\big\}{\dot a}^{}_{a_0\epsilon}+
{\textstyle\frac{1}{4c}}\frac{\partial \delta\hat\ell_a}{\partial x_\alpha} +
{\cal O}(c^{-2}).
\label{eq:pot_loc-w_a-ctv+*}
\end{eqnarray}

The same argument that allowed us to eliminate $\delta\ell_a$ in Sec.~\ref{sec:sum-direct} works here, allowing us to omit $\delta\hat\ell_a$. Therefore, we can now present the metric of the moving observer expressed using the global coordinates $\{x^n\}$. Substituting the expressions for the potentials given by (\ref{eq:pot_loc-w_0-ctv+*}) and (\ref{eq:pot_loc-w_a-ctv+*}) into Eqs.~(\ref{eq:g00-ctv*})--(\ref{eq:gab-ctv*}) we obtain the following expressions
{}
\begin{eqnarray}
g^{00}(x)&=& 1+\frac{2}{c^2}\Big\{\sum_b w_b(x)-\sum_{b\not=a}\Big(\overline{\hat w_b} +r_a^\epsilon \overline{\frac{\partial{\hat w}_b}{\partial x^\epsilon}} \Big)\Big\}+\frac{2}{c^4}\Big\{
\Big[\sum_b U_b(x)-\sum_{b\not=a}\Big(\overline{U_b} +r_a^\epsilon \overline{\frac{\partial{U}_b}{\partial x^\epsilon}} \Big)\Big]^2-\nonumber\\
&&\hskip 30pt ~-{\textstyle\frac{1}{2}}
r^{\epsilon}_{a}r^{\lambda}_{a}\Big(
3a^{}_{a_0\epsilon} a^{}_{a_0\lambda}+
{\dot a}^{}_{a_0\epsilon} v^{}_{a_0\lambda}+
v^{}_{a_0\epsilon}{\dot a}^{}_{a_0\lambda}+
\gamma_{\epsilon\lambda}\sum_{b\not=a}
\ddot{\overline{U_b}}+2\sum_{b\not=a}(
\dot{\overline{\partial_\epsilon\hat{\hskip -8pt w\hskip 9pt}_{b\lambda}}} +
\dot{\overline{\partial_\lambda\hat{\hskip -7pt w\hskip 8pt}_{b\epsilon}}})\Big)-\nonumber\\[-3pt]
&&\hskip 65pt ~-
{\textstyle\frac{1}{10}}
(\ddot a^{}_{a_0\epsilon} r^\epsilon_a)(r^{}_{a\mu} r_a^\mu)\Big\}+{\cal O}(c^{-6}),
\label{eq:g00-ctv-fin*}\\[3pt]
g^{0\alpha}(x)&=& \frac{4}{c^3}\Big\{\sum_b w^\alpha_b(x)-\sum_{b\not=a}\Big(\overline{\hat w^\alpha_b}
+r_a^\epsilon \overline{\frac{\partial{\hat w}^\alpha_b}{\partial x^\epsilon}} \Big)-
{\textstyle\frac{1}{10}}\big\{3r^\alpha_a r^\epsilon_a-
\gamma^{\alpha\epsilon}r^{}_{a\mu} r^\mu_a\big\}{\dot a}^{}_{a_0\epsilon}\Big\}+{\cal O}(c^{-5}),
~~
\label{eq:g0a-ctv-fin*}\\
g^{\alpha\beta}(x)&=& \gamma^{\alpha\beta}-\gamma^{\alpha\beta}\frac{2}{c^2} \Big\{\sum_b w_b(x)-\sum_{b\not=a}\Big(\overline{\hat w_b} +r_a^\epsilon \overline{\frac{\partial{\hat w}_b}{\partial x^\epsilon}} \Big)\Big\}+{\cal O}(c^{-4}).
\label{eq:gab-ctv-fin*}
\end{eqnarray}

The coordinate transformations that put the observer in this reference frame are given by:
{}
\begin{eqnarray}
y^0_a&=& x^0+c^{-2}\Big\{\!\int_{x^0_{0}}^{x^0}\!\!\! \Big(
{\textstyle\frac{1}{2}}\big(v^{}_{a_0\epsilon} +c^{-2}\dot{\hat q}_{a_0\epsilon}\big)\big(v_{a_0}^\epsilon+c^{-2}\dot{\hat q}^\epsilon_{a_0}\big)-\sum_{b\not=a} \overline{{\hat w}_b}\Big)dx'^0+\nonumber\\
&&~~~{}\hskip 50pt{}
+c\,\Big(\big(v^{}_{a_0\epsilon} + c^{-2}\dot{\hat q}_{a_0\epsilon}\big)\big(1+c^{-2}\big(3\sum_{b\not=a} \overline{U_b} -{\textstyle\frac{1}{2}} v^{}_{a_0\mu}v^\mu_{a_0}- 
4\sum_{b\not=a} \overline{{\hat w}_{b\epsilon}}\big)\big)\Big) r_a^\epsilon\Big\}+\nonumber\\[-7pt]
&&~~~{}+
c^{-4}\Big\{
\!\int_{x^0_{0}}^{x^0}\!\!\!
\Big(-{\textstyle\frac{1}{8}}
(v^{}_{a_0\epsilon}v^\epsilon_{a_0})^2+
{\textstyle\frac{3}{2}}(v^{}_{a_0\epsilon}v^\epsilon_{a_0})\sum_{b\not=a} \overline{U_b}-
4v^{}_{a_0\epsilon}\sum_{b\not=a} \overline{{\hat w}^\epsilon_b}+
{\textstyle\frac{1}{2}}\big(\sum_{b\not=a} \overline{U_b}\big)^2
\Big)dx'^0 -\nonumber\\
&-&
{\textstyle\frac{1}{2}}c \,r^{}_{a\lambda} r^{}_{a\mu}
\Big(v^\lambda_{a_0} a_{a_0}^\mu + v^\mu_{a_0} a_{a_0}^\lambda+
\gamma^{\lambda\mu} \sum_{b\not=a}\dot{\overline{U_b}}
+2\sum_{b\not=a}(
\overline{\partial^{\lambda} {\hat w}^{\mu}_b}+
\overline{\partial^{\mu} {\hat w}^{\lambda}_b})\Big)-
{\textstyle\frac{1}{10}}c(\dot{a}^{}_{a_0\epsilon}r^\epsilon_a)(r^{}_{a\nu}r^\nu_a)
\Big\}+{\cal O}(c^{-6})x^0,~~~
\label{eq:transfrom-fin-0-ctv*}\\[3pt]
y^\alpha_a&=& r^\alpha_a-
c^{-2}\Big\{\Big({\textstyle\frac{1}{2}}v^\alpha_{a_0} v^\epsilon_{a_0}+
\hat\omega_{a_0}^{\alpha\epsilon}-\gamma^{\alpha\epsilon}\sum_{b\not=a} \overline{U_b}\Big)r_{a\epsilon} +
a^{}_{a_0\epsilon}\Big(r^\alpha_a r^\epsilon_a-
{\textstyle\frac{1}{2}}\gamma^{\alpha\epsilon}r^{}_{a\lambda} r^\lambda_a\Big)\Big\}+{\cal O}(c^{-4}).
\label{eq:transfrom-fin-a-ctv*}
\end{eqnarray}

The inverse coordinate transformations that we obtained are new and extend previously obtained results (for reviews, see \cite{Brumberg-book-1991,Brumberg-Kopeikin-1989}, and also \cite{Soffel:2003cr}). This set of results concludes our derivation of the coordinate transformations from a local accelerated reference frame to the global inertial frame. Generalization to the case of $N$ extended bodies can be done in a complete analogy with Eqs.~(\ref{eq:Ka-sol-cov!!})--(\ref{eq:a-Newton+!!}), which is sufficient for the solar system applications.

\subsection{Alternative form of the inverse transformations}
\label{sec:note}

As we scrutinize the results for the inverse transformations that were summarized in Sec.~\ref{sec:inverse-summary}, we notice that the two expressions in the transformation of the temporal coordinate (\ref{eq:transfrom-fin-0-ctv*}) look very familiar \cite{Brumberg-book-1991}. Indeed, the term under the integral sign in this expression is the Lagrangian, $\hat L_a$, of the inertial spacetime in the presence of a fictitious  frame-reaction force, which is needed to keep a test particle $a$ on its world-line. Following the derivation of the Lagrangian in the coordinates $\{y^m_a\}$ of the local frame (\ref{eq:lagr}) of the body $a$ and using the metric tensor Eqs.~(\ref{eq:g00-ctv*})--(\ref{eq:gab-ctv*}), while keeping only the inertial potentials (\ref{eq:inert_pot-u_0})--(\ref{eq:inert_pot-u_a}) on the world-line of body $a$, we find that the dynamics of body $a$ in the coordinates $\{x^m\}$ of the global frame are governed by the following Lagrangian $\hat L_{a}$:
\begin{eqnarray}
\hat L_{a}&=&-m_ac^2\frac{ds}{dx^0}=-m_ac^2\Big(g_{mn}\frac{dx^m_{a_0}}{dx^0}\frac{dx^n_{a_0}}{dx^0}\Big)^{1/2}=\nonumber\\
&=&-m_ac^2\Big\{ 1+c^{-2}\Big(
{\textstyle\frac{1}{2}}\big(v^{}_{a_0\epsilon} +c^{-2}\dot{\hat q}_{a_0\epsilon}\big)\big(v_{a_0}^\epsilon+c^{-2}\dot{\hat q}^\epsilon_{a_0}\big)-\sum_{b\not=a} \overline{\hat w_b}\Big)+\nonumber\\
&&\hskip 40pt +~
c^{-4}\Big( -{\textstyle\frac{1}{8}}
(v^{}_{a_0\epsilon}v^\epsilon_{a_0})^2+
{\textstyle\frac{3}{2}}(v^{}_{a_0\epsilon}v^\epsilon_{a_0})\sum_{b\not=a} \overline{U_b}-4v^{}_{a_0\epsilon}\sum_{b\not=a} \overline{\hat w^\epsilon_b}+
{\textstyle\frac{1}{2}}\big(\sum_{b\not=a} \overline{U_b}\big)^2
\Big)+{\cal O}(c^{-6})\Big\}.
\label{eq:transfrom-fin-0-ctv+}
\end{eqnarray}
As the frame-reaction force must compensate for the effect of the external gravity measured by an observer that is co-moving with body $a$, we can introduce the frame-reaction Lagrangian per unit of energy: $\hat{\mathfrak L}_{a_0}=-\hat L_{a}/m_ac^2$. This quantity represents the time transformation between the origins of the global and local frames.

We also observe that the term linearly proportional to $r^\alpha_a$ in (\ref{eq:transfrom-fin-0-ctv*})  is the momentum per mass unit of the test particle $a$ in the global reference frame (cf. Eq.~(\ref{eq:canmon})):
\begin{equation}
p^\alpha_{a}=c^2\frac{\partial{\hat {\mathfrak L}}_{a_0}}{\partial v^\alpha_{a_0}}=\big(v^\alpha_{a_0} + c^{-2}\dot{\hat q}^\alpha_{a_0}\big)\Big(1+c^{-2}\big(3\sum_{b\not=a} \overline{U_b} -{\textstyle\frac{1}{2}} v^{}_{a_0\mu}v^\mu_{a_0}\big)\Big)- c^{-2}4\sum_{b\not=a} \overline{\hat w^\alpha_b}+{\cal O}(c^{-4}).
\label{eq:momentum_a0}
\end{equation}

Similarly to \cite{Ashby-Bertotti-1986}, we introduce a $4\times4$  matrix, ${\cal S}_{a_0}{}^m_n$, with the following components:
{}
\begin{eqnarray}
{\cal S}_{a_0}{}^0_0&=& \frac{dx^0}{ds}=
\hat{\mathfrak L}^{-1}_{a_0}+{\cal O}(c^{-6}),
\label{eq:S-00}
%\\
\qquad \qquad
{\cal S}_{a_0}{}^\alpha_0= \frac{dx^\alpha_{a_0}}{ds}=
\hat {\mathfrak L}^{-1}_{a_0}\frac{v^\alpha_{a_0}}{c}+{\cal O}(c^{-5}),
\label{eq:S-a0}\\
{\cal S}_{a_0}{}^0_\alpha&=& c^{-1}p^{}_{a_0\alpha}=c\frac{\partial\hat{\mathfrak L}_{a_0}}{\partial v^\alpha_{a_0}}+{\cal O}(c^{-5}),
\label{eq:S-0a}\\
{\cal S}_{a_0}{}^\alpha_\beta&=& \gamma_{\beta\epsilon}\Big(\gamma^{\alpha\epsilon}-
c^{-2}\big({\textstyle\frac{1}{2}}v^\alpha_{a_0} v^\epsilon_{a_0}+
\hat\omega_{a_0}^{\alpha\epsilon}-\gamma^{\alpha\epsilon}\sum_{b\not=a} \overline{U_b}\big)+{\cal O}(c^{-4})\Big),
\label{eq:S-ab}
\end{eqnarray}
where the vector ${\cal S}_{a_0}{}^m_0$ is defined with respect to the four-velocity of the base geodesic of body $a$, $u^m=dx^m_{a_0}/ds$. As a result, expressions (\ref{eq:transfrom-fin-0-ctv*})--(\ref{eq:transfrom-fin-a-ctv*}) can be re-written as
{}
\begin{eqnarray}
y^0_a&=& \int_{x^0_{0}}^{x^0}\! \hat{\mathfrak L}_{a_0} dx'^0+{\cal S}^{\phantom{~~\,}0}_{a_0\epsilon} r^{\epsilon}_{a}+
{\textstyle\frac{1}{2}}\Gamma^0_{mn}{\cal S}^{\phantom{~~\,}m}_{a_0\lambda}{\cal S}^{\phantom{~~\,}n}_{a_0\mu}
\,r^{\lambda}_{a} r^{\mu}_{a}+
{\textstyle\frac{1}{10}}\Gamma_{\epsilon\mu\nu,0}\,r^\epsilon_ar^{\mu}_{a}r^\nu_a+{\cal O}(c^{-6})x^0,~~~~
\label{eq:transform-geod-01}\\[3pt]
y^\alpha_a&=& {\cal S}^{\phantom{~~\,}\alpha}_{a_0\epsilon}\,r^{\epsilon}_{a} +{\textstyle\frac{1}{2}}\Gamma^\alpha_{mn}{\cal S}^{\phantom{~~\,}m}_{a_0\lambda}{\cal S}^{\phantom{~~\,}n}_{a_0\mu}
\,r^{\lambda}_{a} r^{\mu}_{a}+{\cal O}(c^{-4}),
\label{eq:transform-geod-a1}
\end{eqnarray}
where the important Christoffel symbols on the world-line of body $a$ were calculated to be
{}
\begin{eqnarray}
\Gamma^{0}_{0\alpha}\big|_{r_a=0} &=& \frac{1}{c^2}a^{}_{a_0\alpha}  +  {\cal O}(c^{-4}), \label{eq:Gamma-00a)}\\
{}
\Gamma^{0}_{\alpha\beta}\big|_{r_a=0} &=& -\frac{1}{c^3}\gamma_{\alpha\lambda}\gamma_{\beta\mu}\Big\{\gamma^{\lambda\mu} \sum_{b\not=a}\dot{\overline{U_b}}+2\sum_{b\not=a}(\overline{\partial^{\lambda} {\hat w}^{\mu}_b}+\overline{\partial^{\mu}{\hat w}^{\lambda}_b})+
2(v^\lambda_{a_0} a_{a_0}^\mu + v^\mu_{a_0} a_{a_0}^\lambda)
\Big\}  +  {\cal O}(c^{-5}), \label{eq:Gamma-0ab)}\\[-3pt]
{}
\Gamma^{\alpha}_{\beta\nu}\big|_{r_a=0} &=& -\frac{1}{c^2}\Big\{
\delta^\alpha_\nu a^{}_{a_0\beta}+
\delta^\alpha_\beta a^{}_{a_0\nu}-
\gamma_{\beta\nu}a^{\alpha}_{a_0}\Big\} +
 {\cal O}(c^{-4}).\label{eq:Gamma-abg)}
\end{eqnarray}

Eqs.~(\ref{eq:transform-geod-01})--(\ref{eq:transform-geod-a1}) represent an alternative form of the coordinate transformations to those given by Eqs.~(\ref{eq:transfrom-fin-0-ctv*})--(\ref{eq:transfrom-fin-a-ctv*}). Both sets of equations describe the transformations from the local coordinates of the proper reference system $\{y^m_a\}$ centered on the world-line of body $a$ to the global coordinates $\{x^m\}$ centered at the barycenter of the $N$-body system. The result that was obtained is quite interesting, as it emphasizes the connection of the coordinate transformations to the dynamics of the body as described by the Lagrange functional (\ref{eq:transfrom-fin-0-ctv+}). A similar result was derived in \cite{Ashby-Bertotti-1986,DSX-I} under a set of completely different assumptions. The fact that our results agree with those derived earlier further demonstrates the validity of our approach to the development of a theory of relativistic coordinate reference systems. Note that similar expressions could also be derived for the direct transformations (\ref{eq:transfrom-fin-0+})--(\ref{eq:transfrom-fin-a+}).

The presence of Lagrangian-derived quantities in the coordinate transformations Eqs.~(\ref{eq:transform-geod-01})--(\ref{eq:transform-geod-a1}) can be seen as a direct consequence of the fact that the dynamics of a test particle were used to identify a `good' proper reference frame. It is also a strong indication of the self-consistency of the approach that we followed in these derivations. Beyond such formal considerations, the connection between the dynamical equations governing the motion of a test particle and the coordinate transformations between its reference frame and a global inertial frame may be worthy of further study. The results of any such study which we may undertake in the future will be reported elsewhere.

\section{Discussion and Conclusions}
\label{sec:end}

In this paper we introduced a new approach to find a solution for the $N$-body problem in the general theory of relativity in the weak-field and slow motion approximation. The approach is based on solving all three problems discussed in Sec.~\ref{sec:introduction}, including finding solutions to global and local problems, as well as developing a theory of relativistic reference frames. Our objective was to establish the properties of a local frame associated with the world-line of one of the bodies in the system  and to find a suitable form of the metric tensor corresponding to this frame; we achieved this goal by imposing a set of clearly defined coordinate and physical conditions.

Specifically, we combined a new perturbation theory ansatz introduced in Sec.~\ref{sec:ansatz-N-body} for the gravitational $N$-body system. In this approach, the solution to the gravitational field equations in any reference frame is presented as a sum of three terms:
\begin{inparaenum}[i)]
\item the inertial flat spacetime in that frame,
\item unperturbed solutions for each body in the system transformed to the coordinates of this frame, and
\item the gravitational interaction term.
\end{inparaenum}
We use the harmonic gauge conditions that allow reconstruction of a significant part of the structure of the post-Galilean coordinate transformation functions relating non-rotating global coordinates in the inertial reference frame to the local coordinates of the non-inertial frame associated with a particular body. The remaining parts of these functions are determined from dynamical conditions, obtained by constructing the relativistic proper reference frame associated with a particular body. In this frame, the effect of external forces acting on the body is balanced by the fictitious frame-reaction force that is needed to keep the body at rest with respect to the frame, conserving its relativistic three-momentum. We find that this is sufficient to determine explicitly all the terms of the coordinate transformation. Exactly the same method is then used to develop the inverse transformations.

Our approach naturally incorporates properties of dynamical coordinate reference systems into the hierarchy of relativistic reference systems in the solar system and relevant time scales, accurate to the $c^{-4}$ order. The results obtained enable us to address the needs of practical astronomy and allow us to develop adequate models for high-precision experiments. In our approach, an astronomical reference system is fully defined by a metric tensor in a particular coordinate frame, a set of gravitational potentials describing the gravitational field in that frame, a set of direct, inverse, and mutual coordinate transformations between the frames involved, ephemerides, and the standard physical constants and algorithms. We anticipate that the results presented here may find immediate use in many areas of modern geodesy, astronomy, and astrophysics.

The new results reported in this paper agree well with those previously obtained by other researchers. In fact, the coordinate transformations that we derived here are in agreement with the results established for both direct and inverse coordinate transformations given in  Refs~\cite{DSX-I,DSX-II,DSX-III,DSX-IV} and  \cite{Brumberg-Kopeikin-1989,Kopeikin-1988,Brumberg-Kopeikin-1989-2,Kopeikin:2004ia}, correspondingly. However, our approach allows one to consistently and within the same framework develop both direct and inverse transformations, the corresponding equations of motion, and explicit forms of  the metric tensors and gravitational potentials in the various reference frames involved. The difficulty of this task was mentioned in Ref.~\cite{Soffel:2003cr} when the post-Newtonian motion of a gravitational $N$-body system was considered; our proposed formulation successfully resolves this important issue. As an added benefit, the new approach provides one with a good justification to eliminate the functions $\delta \kappa$, $\delta \xi$ and $\delta \ell$, yielding a complete form for the transformation functions ${\cal K}$, ${\cal L}$ and ${\cal Q}^\alpha$ involved  in the transformations (as well as their ``hatted'' inverse counterparts).

The significance of our result is that for the first time, a formalism for the coordinate transformation between relativistic reference frames is provided, presenting both the direct and inverse transformations in explicit form. By combining inverse and direct transformations, the transformation rules between arbitrary accelerating frames can be obtained. Furthermore, it is possible to combine direct (or inverse) transformations, and obtain a complete set of transformations that can be represented by our formalism, as shown explicitly in \cite{Turyshev-96,TMT:2011}. This leads to an {\it approximate} finite group structure that extends the Poincar\'e group of global transformations to accelerating reference frames. In fact, we explicitly verified that the two coordinate transformations presented in this work are inverses of each other, the fact that further supports our method for solving the $N$-body problem in general theory of relativity.

The results obtained in this paper are designed to facilitate the analysis of relativistic phenomena with ever increasing accuracy. We should note that the approach we presented can be further developed in an iterative manner: if greater accuracy is desired, the coordinate transformations (\ref{eq:trans-0})--(\ref{eq:trans-a}) can be expanded to include higher-order terms. Furthermore, the same approach relying on the functional ${\cal KLQ}$-parameterization may be successfully applied to the case of describing the gravitational dynamics of an astronomical $N$-body system and dynamically rotating reference frames. This work has begun and the results, when available, will be reported elsewhere.

There are some problems that remain to be solved.  One of these is the problem of relativistic rotation. In particular, it is known that the rotational motion of extended bodies in general relativity is a complicated problem that has no complete solution up to now. On the other hand, modern observational accuracy of the geodynamical observations makes it necessary to have a rigorous relativistic model of the Earth's rotation. The method presented in this paper may now be used to study the rotation of extended bodies to derive a theory of relativistic rotation within the general theory of relativity. Our ${\cal KLQ}$ formalism presented here provides one with the necessary conceptual basis to study this  problem from a very general position and could serve as the foundation for future work. Also, a complete formal treatment of the case of a system of $N$ extended bodies should be presented. Although this treatment would result in corrections much smaller to those anticipated in the solar system experiments, they may be important to a stronger gravitational regime, such as in binary pulsars. This work had been initiated. The results will be reported elsewhere.

Our current efforts are directed towards the practical application of the results obtained in this paper (similar to those discussed in \cite{Turyshev:2012nw} and \cite{ref:Turyshev:2014dea} in the context of the GRAIL and GRACE-FO missions, correspondingly). We aim to establish the necessary relativistic models of measurement, equations of motion for the celestial bodies and spacecraft, as well as the light propagation equations in the different frames involved. We also plan to implement these results in the form of computer codes for the purpose of high-precision spacecraft navigation and will perform the relevant scientific data analysis. The analysis of the  above-mentioned problems  from the standpoint of the new theory of relativistic astronomical reference systems will be the subject of specific studies and future publications.

\begin{acknowledgments}
We thank Sami W. Asmar, William M. Folkner, Michael M. Watkins, and James G. Williams of JPL for their interest, support and encouragement during the work and preparation of this manuscript. It is our pleasure to acknowledge fruitful conversations with Olivier L. Minazzoli. We would also like to thank the anonymous referee for valuable suggestions that improved the manuscript. This work was performed at the Jet Propulsion Laboratory, California Institute of Technology, under a contract with the National Aeronautics and Space Administration.

\end{acknowledgments}
%
%\bibliography{ref-frames}

\end{document}